\documentclass[aps,reprint,twocolumn,superscriptaddress,prd,nofootinbib]{revtex4-2}

\usepackage[utf8]{inputenc}
\usepackage{url} 
\usepackage{float}
\usepackage{amsmath, amssymb}

\usepackage{array}
\usepackage{tcolorbox}
\usepackage{hyperref} 
\hypersetup{colorlinks=true, citecolor=blue}

\usepackage{graphicx} 
\usepackage{subcaption} 
\usepackage{cleveref} 
\usepackage[toc,page]{appendix} 

\usepackage{natbib} 

\usepackage{tikz} 
\usetikzlibrary{shapes.geometric, arrows}
\tikzstyle{fkksflow} = [rectangle,rounded corners,text centered,fill = blue!30, text width = 7cm]
\tikzstyle{arrow} = [thick,->,>=stealth]

\Crefname{figure}{figure}{Figs.} 
\Crefname{equation}{Eq.}{Eqs.} 

\def\ProcaBounds{{$[1.8 \times 10^{-17}, 4.47 \times 10^{-16}]$}}

\begin{document}
	\title{Detecting Fundamental Vector Fields with LISA}
	
	\author{Shaun Fell}
	\email{fell@thphys.uni-heidelberg.de}
	\affiliation{Institute for Theoretical Physics, Universit{\"a}t Heidelberg , Philosophenweg 16, 69120 Heidelberg, Germany}
	
	\author{Lavinia Heisenberg}
	\email{lavinia.heisenberg@phys.ethz.ch}
	\affiliation{Institute for Theoretical Physics, Universit{\"a}t Heidelberg , Philosophenweg 16, 69120 Heidelberg, Germany}
	\affiliation{Institute for Theoretical Physics, ETH Z{\"u}rich, Wolfgang-Pauli-Strasse 27, 8093, Z{\"u}rich, Switzerland}
    \affiliation{Perimeter Institute for Theoretical Physics, 31 Caroline St N, Waterloo, Ontario, N2L 6B9, Canada}

    \author{Do\u{g}a Veske}
	\email{veske@thphys.uni-heidelberg.de}
	\affiliation{Institute for Theoretical Physics, Universit{\"a}t Heidelberg , Philosophenweg 16, 69120 Heidelberg, Germany}
	
	\begin{abstract}
		The advent of gravitational wave astronomy has seen a huge influx of new predictions for potential discoveries of beyond the Standard Model fields. The coupling of all fundamental fields to gravity, together with its dominance on large scales, makes gravitational physics a rich laboratory to study fundamental physics. This holds especially true for the search for the elusive dark photon, a promising dark matter candidate. The dark photon is predicted to generate instabilities in a rotating black hole spacetime, birthing a macroscopic Bose-Einstein condensate. These condensates can especially form around super massive black holes, modifying the dynamical inspiralling process. This then opens another window to leverage future space-borne gravitational wave antennas to join the hunt for the elusive dark matter particle. This study builds a preliminary model for the gravitational waveform emitted by such a dressed extreme mass-ratio inspiral. Comparing these waveforms to the vacuum scenario allows projections to the potential constrainability on the dark photon mass by space-borne gravitational wave antennas. The superradiant instability of a massive vector field on a Kerr background is calculated and, under reasonable approximations, the backreaction on the orbit of an inspiralling solar mass-scale compact object due to the secular evolution of the resulting boson cloud is determined. The end result is the projection that the LISA mission should be able to constrain the dark photon mass using extreme mass ratio inspirals in the range \ProcaBounds eV.
	\end{abstract}
	
	\maketitle

	\section{Introduction}
	
	The advent of gravitational wave (GW) astronomy~\cite{PhysRevLett.116.061102} has brought forth a plethora of avenues to study a wide range of physical phenomena, including fundamental physics. The coupling of gravity to all fundamental particles allows for the study of fundamental fields in the proximity of a strong gravitational field, such as coalescing compact objects. These studies have already placed stringent limits on the allowed parameter space of some physical theories~\cite{LIGO1,LIGOdarkphoton,An_Pospelov_Pradler_Ritz_2020,D_Eramo_2022,Pitjev_2013,Polisensky_2011}. Current GW observatories such as LIGO~\cite{2015}, Virgo~\cite{virgo1, virgo2}, KAGRA~\cite{kagra1,kagra2}; future missions such as LISA~\cite{lisa1, lisa2}, Einstein Telescope~\cite{ET}, Cosmic Explorer~\cite{reitze2019cosmic,PhysRevD.91.082001}, DECIGO~\cite{decigo1, decigo2}; and indirect detectors such as pulsar timing arrays, for example EPTA~\cite{EPTA1, EPTA2}, NANOGrav~\cite{nanograv1, nanograv2}, and the PPTA~\cite{ppta} are thus invaluable tools to study fundamental physics. They shed light on the accuracy of theoretical models of strong gravity, but they can also potentially ameliorate current big problems in physics such as the nature of dark matter and dark energy. 
	
	The LISA mission in particular, operating in the millihertz frequency range, offers a unique capability to answer fundamental questions. These range from probing the dynamics of extreme mass ratio inspirals (EMRIs), to studying the expansion of the universe~\cite{lisawhite}. The EMRI scenario has particular interest due to the long timescale for which the system remains in the strong gravity regime. Moreover, the wide difference in their respective masses translates to a wide difference in the curvature of spacetime they generate. The substantial gap in masses implies the central super massive black hole (SMBH) can be treated as generating a Kerr geometry on which the secondary compact object acts as a perturbing particle. This approximation drastically simplifies estimation techniques for the gravitational waveform, without requiring an appeal to full numerical relativity~\cite{Babak_Fang_Gair_Glampedakis_Hughes_2008,Barack_Cutler_2004,AAK}. 
 
    The most common formation mechanism of EMRIs is thought to occur by gravitational capture of a stellar mass compact object from a cusp onto a highly eccentric orbit~\cite{Sigurdsson_1997,Alexander_2005,Merritt_2006,Bortolas_2019}. These captures are thought to happen as a result of two body relaxations of the objects where an object is deflected to an orbit around the SMBH with a small pericenter distance. Eventually, emission of GWs will reduce the eccentricity to more circular values. Only compact objects such as stellar mass black holes (BHs), neutron stars, white dwarfs or Helium cores of giant stars can produce detectable extragalactic EMRI signals. Less compact or lighter objects such as main sequence stars either cannot withstand the tidal forces of the SMBH or are not massive enough to produce sufficiently strong GWs. Other less common formation scenarios exist such as tidal separation of binaries, Bondi-like capture of passing objects, separation of a massive stars core from its envelope, and compact object formation in accretion disks around the SMBH~\cite{Amaro_Seoane_2007}.
    
    A typical EMRI system will complete $ \sim 10^4 - 10^5$ orbits, with most of the orbits exhibiting relativistic velocities~\cite{lisaemri}. Thanks to the large number of orbits, despite the low amplitude of the signal, a considerable amount of signal-to-noise ratio (SNR) is obtainable. However, such a large parameter space of the system makes a fully coherent matched filtering computationally infeasible due to the required large number of templates ($\gtrsim10^{35}$)~\cite{Gair_2004,Amaro_Seoane_2007}. Different approaches such as semi-coherent matched filtering, time-frequency methods or stochastic methods are under consideration. The required amplitude SNR threshold seems to be at least around 15-20~\cite{Babak_2017,Babak_2010}. 
    
    The timescale with which the secondary BH experiences relativistic orbital velocities suggests EMRIs will be a prime target for performing a litany of tests of general relativity (GR), study environmental effects around SMBH's, and enables high precision estimation of the physical parameters of the system, such as redshifted masses and black hole spins~\cite{Gair_Vallisneri_Larson_Baker_2013, lisawhite,Gair_2017}. The high precision measurements of massive black holes (MBHs) and EMRI properties allows a remarkably precise test of many theories, including the predictions of new fundamental fields outside the standard model. 
	
	Such fundamental fields include the predicted dark matter particle. Dark matter candidates cover a wide range of predicted mass values, depending on the underlying theory. Some models predict the dark matter particle to have a mass as low as $10^{-22}$eV~\cite{Hui_2017}, so called fuzzy dark matter, and as high as (sub-) solar masses in the form of primordial black holes~\cite{Miller_Clesse_De}. Due to the feeble (or vanishing) interaction with the other forces, dark matter has thus far evaded direct detection. However, since gravity propagates on extremely large scales, large dark matter structures are detectable through their gravitational influence\footnote{For example, the Bullet Cluster or MACS J0025.4-1222, Baryon acoustic oscillations, structure formation, etc. }. One model for dark matter includes the dark photon, which can be extremely light, $ m \gtrsim 10^{-22}$eV~\cite{Chen_Schive_Chiueh_2017}, and behaves as non-relativistic matter. The relic abdunance of such dark matter particles may be produced by a misalignment mechanism \footnote{Though, this requires a non-minimal coupling to gravity}~\cite{Nelson_Scholtz_2011} and also arise naturally in certain string theories~\cite{Goodsell_Jaeckel_Redondo_Ringwald_2009}. Other production mechanisms include tachyonic couplings to a misaligned axion~\cite{Co_2019} and quantum fluctuations during inflation~\cite{Graham_2016}. Constraints on the couplings of such particles to the standard model come from equivalence principle tests, such as of the Eöt-Wash group~\cite{PhysRevD.50.3614, Schlamminger_2008} and Lunar Laser Ranging groups~\cite{PhysRevLett.93.261101, TURYSHEV_2007}. Near future constraints will come from GW measurements, for example from black hole superradiance~\cite{Baryakhtar_Lasenby_Teo_2017, East_Pretorius_2017, East_2017, Pierce_Riles_Zhao_2018} \footnote{See~\cite{Maselli_Franchini_Gualtieri_Sotiriou_2020, Maselli_Franchini_Gualtieri_Sotiriou_Barsanti_Pani_2022} for the case of a scalar-charged secondary BH in an EMRI and the associated effects on the GW waveform.}. Current constraints in the literature coming from superradiance suggest LISA would be able to constrain the mass of the vector field in the range $1 \times 10^{-16}$ eV to $6 \times 10^{-16}$ eV~\cite{superrad}. The superradiance phenomenon, and more generally GW measurements, will be a powerful tool in probing the potential vector nature of dark matter.
	
	The purpose of this paper is to extend these superradiant-based constraints for the dark photon mass by considering the interaction of an EMRI system with a superradiantly-generated Proca cloud. Here, the massive vector field describing the dark photon is evolved according to the Proca equations, describing a massive vector field decoupled from the standard model. For the purposes of a preliminary study on detecting superradiant Proca clouds, at this time we don't consider bound-bound or bound-unbound state transitions in the Proca cloud induced by the secondary black hole \footnote{See~\cite{Baumann_Chia_Porto_Stout_2020} for a study on the transitions induced in a superradiant Proca cloud. The effect on the waveform due to these transitions will be studied in the future.}. We defer an estimation of the effect on the waveform due to state transitions (as well as dynamical friction, accretion, and self-gravity) to appendix \ref{app:comp}. We also don't consider dynamical friction or accretion effects on the secondary black hole. Both of these effects on the waveform are expected to be significant and will be studied in the near future. Section \ref{sec:theory} describes the theory of a superradiant Proca field on a Kerr background and the interaction with an orbiting secondary black hole. The method used to solve the Proca equations of motion are also described, along with the theory of superradiance. Section \ref{sec:results} presents the results of this study, the Proca mass range potentially reached with the LISA mission. A brief discussion on the obtained Proca solutions is also made. As the linearized Proca equations of motion have been solved in many other studies, a more detailed analysis is deferred to the literature. Section \ref{sec:conclusion} concludes the study, detailing future work of the authors.

	\section{Theory} \label{sec:theory}
	\subsection{Curve Spacetime With Proca}
	
	The starting point is the specification of the relevant fields, via the action functional
	\begin{equation}
	S[g,A,\psi] = S_0 [g,A] + S_m[g,A,\psi]
	\end{equation}
	where $g$ is the metric tensor, $A$ is the Proca field, $S_0[g,A]$ describes the background,  and $S_m[g,A, \psi]$ is the action for the matter field $\psi$. Here, the "skeletonized" approach is adopted. The matter action for the generic matter field $\psi$ is replaced by the action for the point particle. This is a phenomenological reduction of the description of the secondary black hole to that of a "probe" particle following the geodesics set by the background. This is achieved via the replacement
	\begin{equation}
	S_m[g,A,\psi] \rightarrow S_p[g,A,\{x\}]
	\end{equation}
	where $\{x\}$ are the co-ordinates of the secondary black hole. 
	
	Specifying the background to be that of a Proca field minimally coupled to a supermassive Kerr black hole, the background action becomes
	\begin{equation}
	S_0[g,A] = \int d^4x \sqrt{-g} \left[ \frac{1}{\kappa} \mathfrak{R} - \frac{1}{4}F_{\mu \nu} F^{\mu \nu} - \frac{1}{2} \mu^2 A_{\mu} A^{\mu} \right]
	\end{equation}
	where $\kappa = \frac{16 \pi G}{c^4}$. Due to the lack of an obvious separability of the point particle current in the chosen ansatz (see below), we consider no coupling between the secondary BH and the Proca field\footnote{More generally, we don't consider any direct coupling between the Proca field and the Standard Model fields.}
	\begin{align}
	S_p[g,A,\psi] &\equiv - \int m_p d\tau +   q\int A_{\mu}J^{\mu}\\ & \rightarrow  - m_p \int \sqrt{ -g_{\mu \nu} \frac{dx^{\mu}}{dt\tau} \frac{dx^{\mu}}{d\tau}} d\tau
	\end{align}
	The equations of motion (EOM) associated to the action functional are
    \begin{widetext}
	\begin{align} \label{eq:einsteinproca}
		G^{\rho \sigma} &= 8 \pi \left( \frac{-1}{4} F_{\mu \nu}F^{\mu \nu} g^{\rho \sigma} + F^{\rho \nu} F^{\sigma}_{\nu} - \frac{1}{2} \mu^2 g^{\rho \sigma} A_{\mu} A^{\mu} +  \mu^2 A^{\rho} A^{\sigma} \right) + 8\pi \mathfrak{T}^{\rho \sigma}_{p}\\
		0&= \nabla_{\rho} F^{\rho \sigma} - \mu^2 A^{\sigma}
	\end{align}
    \end{widetext}
	where $\mathfrak{T}^{\rho \sigma}_{p}$ is the energy-momentum tensor of the point particle. Note that, due to the source terms on the RHS of the Einstein equations, the Proca equations cannot be written in a Klein-Gordon-type form. Instead, using the Lorentz constraint from the conserved current, one finds
	\begin{equation}
	\nabla^2 A^{\nu} - R^{\nu}_{\sigma} A^{\sigma} - \mu^2 A^{\nu} = 0
	\end{equation}
	However, this statement is true for the full non-linear system. If we work in a perturbative regime in which the Proca amplitude is small and the mass ratio between the secondary and the SMBH is large, then the right hand side of the Einstein equations vanish, so we can write
	\begin{equation}
	\nabla^2 A^{\nu}  - \mu^2 A^{\nu} = 0
	\end{equation}
	So the Proca equations become, in the linearized regime,
	\begin{align}
		\nabla^2 A^{\nu} - \mu^2 A^{\nu} &= 0\\
		\nabla_{\sigma}A^{\sigma} &=0
	\end{align}

	\paragraph*{Decomposition in FKKS ansatz:}
	The first step in solving the Proca equations is to decompose the Proca EOM in the Frolov-Krtouš-Kubizňák-Santos (FKKS) ansatz~\cite{fkks}. We define the ansatz for the Proca field as
	\begin{equation}
	A^{\mu} = B^{\mu \nu} \nabla_{\nu}Z
	\end{equation}
	where $B^{\mu \nu}$ is implicitly defined through the complex-valued algebraic equation
	\begin{equation}
	B^{\mu \nu} (g_{\nu \gamma} + \frac{i}{\lambda} h_{\nu \gamma}) = \delta^{\mu}_{\gamma}
	\end{equation}
	and $h_{\mu \nu}$ is the so-called principle tensor of the Kerr spacetime. In Boyer-Lindquist co-ordinates, it is defined as

    \begin{widetext}
	\begin{equation}
	h_{\mu \nu} = 
	\begin{bmatrix}
		0 & r & a^2 \cos(\theta) \sin(\theta) & 0\\
		-r & 0 & 0 & a r \sin(\theta)^2\\
		-a^2 \cos(\theta) \sin(\theta) & 0 & 0 & a \cos(\theta) (a^2 + r^2) \sin(\theta) \\
		0 & -a r \sin(\theta)^2 & -a \cos(\theta) \sin(\theta) (a^2 + r^2) & 0
	\end{bmatrix}
	\end{equation}
    \end{widetext}

	We then perform a separation of variables in the FKKS ansatz via
	\begin{equation}
	Z = R(r) S(\theta) e^{-i \omega t} e^{i \mathfrak{m} \phi}
	\end{equation}
	This is a multiplicative separation into 2 arbitrary single-coordinate-dependent functions and 2 eigenfunctions of the spacetime killing vectors \footnote{We write the mode number as $\mathfrak{m}$ in order to distinguish it from the mass of the secondary black hole, which is denoted as $m$.}
	\begin{align}
		\mathfrak{L}_{T} Z &= -i \omega Z\\
		\mathfrak{L}_{\Phi} Z &= i \mathfrak{m} Z
	\end{align}
	where $\mathfrak{L}$ is the Lie derivative, and $T$,$\Phi$ are the temporal and azimuthal Killing vectors, respectively.
	After inserting this separated form into the Proca EOM, one finds the following coupled second-order system of differential equations
	\begin{align} \label{eq:ProcaFKKSradial}
		\frac{d}{dr} \left( \frac{\Delta}{q_r} \frac{dR}{dr} \right) + \left( \frac{K_r^2}{q_r \Delta} + \frac{2-q_r}{q_r^2} \frac{\sigma}{\nu} - \frac{\mu^2}{\nu^2} \right) R &= 0 \\ \label{eq:ProcaFKKStheta}
		\frac{1}{\sin{\theta}}\frac{d}{d \theta} \left( \frac{\sin{\theta}}{q_{\theta}} \frac{dS}{d \theta}\right) - \left( \frac{K_{\theta}^2}{q_{\theta} \sin^2{\theta}} + \frac{2-q_{\theta}}{q_{\theta}^2} \frac{\sigma}{\nu} - \frac{\mu^2}{\nu^2} \right)S &= 0
	\end{align}
	where 
	\begin{align}
		K_r &= a \mathfrak{m} - (a^2 + r^2)\omega  & q_r &= 1+\nu^2 r^2\\
		\sigma &= a \nu^2 (\mathfrak{m} - a \omega) + \omega & K_{\theta} &= \mathfrak{m} - a \omega \sin{\theta}^2 \\
		q_{\theta} &= 1 - \nu^2 a^2 \cos{\theta}^2 & \Delta &= r^2 + a^2 - 2Mr
	\end{align}

    Henceforth, we focus only on a single mode specified by the tuple $(\nu, \omega, \mathfrak{m})$. A generic solution to the Proca EOM will be a linear combination of the single mode solutions. See appendix \ref{app:decomp} for a more indepth discussion of the method to solving the coupled eigenvalue problem \Cref{eq:ProcaFKKSradial,eq:ProcaFKKStheta}.
    Some additional quantities that are important for the later analysis are the total energy and the normalization of the Proca field. \Cref{eq:ProcaFKKSradial,eq:ProcaFKKStheta} determines the Proca field only up to an overall normalization constant. This constant must be determined as part of a full description of the state of the cloud. We choose to normalize the field by the requirement that the total energy of the Proca field matches the reduction in energy of the black hole (energy conservation), utilizing the clear separation of timescales between the superradiant instability and the gravitational radiation from the cloud. The cloud generally builds up much more rapidly than it depletes via gravitational radiation, hence we can safely neglect the depletion during the instability~\cite{Baumann_Chia_Porto_Stout_2020}. 
    The total energy of the cloud at a particular instance is defined by
    \begin{equation} \label{eq:energyintegral}
    E_{c} = - \int \mathfrak{T}^{t}_{t} \sqrt{-g} dr d\theta d\phi
    \end{equation}
    where $g$ is the metric determinant and $\mathfrak{T}$ is the stress-energy tensor of the Proca field, \Cref{eq:einsteinproca}. Normalization of the Proca field then follows from the requirement $E_{c} = M_{0, bh} - M_{f, bh}$, where $M_{0,bh}$ and $M_{f,bh}$ are the masses of the black hole before and immediately after the superradiant instability, respectively.

	\paragraph*{Systematics of Solving Radial and Angular Equations:} \label{sec:solvingProca}
	First define the parameters and eigenvalues of the problem as $(\mathfrak{m}, S, n, a, \mu, M) = \mathcal{P}$ and $(\nu, \omega) = \mathcal{E}$, respectively. The system of equations can then be represented schematically as
	\begin{align} \label{eq:abstractsystem}
		L[r;\mathcal{P}, \mathcal{E}]R(r) &=0 \\ \nonumber
		O[\theta; \mathcal{P}, \mathcal{E}] S(\theta) &= 0
	\end{align}
	where $L$ and $O$ are linear operators defined in Eq. \eqref{eq:ProcaFKKSradial} and \eqref{eq:ProcaFKKStheta} and which are coupled only through the set of eigenvalues $\mathcal{E}$. In the non-relativistic limit $\mu M \ll 1$, the real and imaginary parts of the frequency read (see~\cite{Dolan_2018, Baumann_Chia_Porto_2019,gravatom} for the definitions of $C_{l \mathfrak{m} Sn}$ and $g_{j \mathfrak{m}}$)

    \begin{widetext}
	\begin{align}\label{eq:nonrelfreq}
		\frac{\omega_R}{\mu} &\approx 1 - \frac{\mu^2 M^2}{2( |\mathfrak{m}| + n + S + 1)^2} + O((\mu M)^4)\\
		M \omega_I &\approx 2 r_+ C_{l \mathfrak{m} Sn} g_{j \mathfrak{m}}(a, \mu M, \omega) (\mathfrak{m} \Omega_H - \omega_R) (M \mu)^{4 |\mathfrak{m}| + 5 + 2S}
	\end{align}
    \end{widetext}
 
	These, together with the non-relativistic limits for the eigenvalue $\nu$ in appendix \ref{sec:angularequation}, provide good starting guesses for iteratively solving the system Eq. \eqref{eq:abstractsystem}.
	The algorithm we employ to numerically solve the Proca field in the FKKS ansatz follows similarily to~\cite{Siemonsen_2020} and goes as follows:
	\begin{itemize}
        \setlength{\itemsep}{15pt}
        \setlength{\parskip}{0pt}
        \setlength{\parsep}{0pt}
		\item After specifying the initial parameters to be considered (BH spin, Proca spin, mode number, overtone number, etc.), an initial guess for the $\omega$ and $\nu$ eigenvalues are formed by the non-relativistic limit.
		\item Solve the determinant of the angular equation matrix and pick the eigenvalue that is nearest to either the non-relativistic limit or the previous result for a different mass.
		\item Solve the Proca radial equation using the Frobenius method and find the initial conditions from evaluating the Frobenius solution at a starting radius, close to the outer horizon.
		\item Numerically solve the radial equation with the previously obtained $\omega$ and $\nu$ eigenvalue and boundary conditions.
		\item Find the logarithmic minimum of the radial equation at the outer boundary of the radial integration over $\omega$-space. Minimization is carried out in $\omega$-space, recalculating $\nu$ for each $\omega$-value, using a native Nelder-Mead algorithm in the software system Mathematica~\cite{Mathematica}.
		\item With the found value of $\nu$ and $\omega$, the angular matrix can be solved for the expansion of the angular function in terms of the spherical harmonics
		\item With the radial and angular functions in hand, the Proca EOM are solved in the FKKS ansatz.
		\item This process can be repeated for varying choices of the Proca mass parameter, overtone number, and mode number. Initial guesses for $\omega$ and $\nu$ switch from using the non-relativistic limit to using a 4th-order polynomial fit to previous results in $\mu - \omega$ space. These fits perform much better than the non-relativistic limit for higher mass parameters, typically $\mu \gtrapprox 0.3$
	\end{itemize}

	The flow of the algorithm proceeds as follows:
	\begin{center}
		\begin{tikzpicture}[node distance=1.5cm]
			
			\node  (N1) [fkksflow] {Setup Iteration};
			\node (N2) [fkksflow, below of=N1] {Fix branch of Frobenius solution};
			\node (N3) [fkksflow, below of=N2] {Construct guess for $\omega$ from either previous results or $\omega_{non rel}$};
			\node (N4) [fkksflow, below of=N3] {Walk $\omega$-space until slope $\frac{dR}{d \omega}$ changes sign. \newline $\rightarrow$ Indicates rough location of $\omega_{true}$};
			\node (N5) [fkksflow, below of=N4] {Minimize $R(r_{max})$ using Nelder-Mead simplicies};
			\node (N6) [fkksflow, below of=N5] {Solve angular equation for kernel vectors};
			
			\draw [arrow] (N1) -- (N2);
			\draw [arrow] (N2) -- (N3);
			\draw [arrow] (N3) -- (N4);
			\draw [arrow] (N4) -- (N5);
			\draw [arrow] (N5) -- (N6);
			
		\end{tikzpicture}
	\end{center}
		
	\paragraph*{Computing the Asymptotic Flux:}
	We used the package \textsc{superrad}~\cite{superrad} for the calculation of the asymptotic fluxes from the Proca cloud. It uses a combination of analytic and numerical results to compute the asymptotic energy flux from a Proca cloud, assuming all the energy of the cloud resides in a single mode. The asymptotic angular momentum flux can then be computed from the Teukolsky formalism~\cite{Teukolsky3} as
	\begin{equation}
		\langle \frac{d J}{dt} \rangle = \frac{\mathfrak{m}}{\omega} \langle \frac{d E}{dt} \rangle
	\end{equation}
	
	\subsection{Superradiance}
	Black hole superradiance is a dissipative phenomenon which involves the unstable growth of field amplitudes due to the collection of negative energy states by the ergoregion \footnote{For a review of superradiance, see~\cite{Brito_Cardoso_Pani_2015}}. This superradiant instability of matter fields around spinning black holes can lead to, under certain conditions, a quasibound state. In fact, a quite general argument for the existence of superradiance can be shown to follow from the black hole area theorem, which states
    \begin{equation}
		\delta M = \frac{T_H}{4} \delta A + \Omega_H \delta J
	\end{equation}
	for an uncharged black hole, where $T_H$ is the Hawking temperature, $A_H$ is the area of the horizon, $\Omega_H = \frac{a}{r_+^2 + a^2}$, $a$ is the spin of the black hole, and $r_+$ is the radius of the outer event horizon. For a matter wave of frequency $\omega$ and azimuthal number $\mathfrak{m}$, the ratio of angular momentum to energy is
	
 \begin{equation}
		\frac{L}{E} = \frac{\mathfrak{m}}{\omega}
	\end{equation}
	Hence, an interaction of the matter wave with the black hole causes the latter to change its angular momentum by
	
 \begin{equation} \label{eq:fracchange}
		\frac{\delta J}{\delta M} = \frac{\mathfrak{m}}{\omega}
	\end{equation}
 
	The first law then tells us
	
 \begin{equation}
		\delta M = \frac{\omega T_H}{4} \frac{\delta A_H}{\omega - \mathfrak{m} \Omega_H}
	\end{equation}
	The second law of black hole thermodynamics, $\delta A_H>0$, implies waves impinging on the event horizon with frequency 
	
 \begin{equation} 
		\omega < \mathfrak{m} \Omega_H
  \label{eq:superradcondition}
	\end{equation}
	causes the black hole to lose mass and hence energy is extracted by the wave, increasing its own energy. This wave can become trapped by the potential well of the black hole, causing the wave to again impinge on the black hole. 
	
	This is the mechanism of superradiance. A small amplitude wave initially impinging from past infinity will be continuously excited in a runaway process until the black hole loses enough angular momentum and mass to turn off the superradiant condition Eq. \eqref{eq:superradcondition}. This is a purely classical description. A quantum description, in which vacuum states at past and future infinity contain different particles numbers, has also been formulated~\cite{unruh,starobinsky,BALAKUMAR2020135904}. The process is reminiscent of the well-known Penrose process, though they are distinct phenomena~\cite{richartz2009generalised}.

    \subsection{Modified Gravitational Waves}
	
	After the superradiant instability has turned off, saturating the superradiant threshold \Cref{eq:superradcondition}, the system exists in a quasi-static state consisting of a black hole surrounded by a quasibound Proca condensate. It is not in an eternal bound state due to gravitational emission from the cloud itself, which manifests as a long-duration depletion of the condensate \cite{Baumann_Chia_Porto_Stout_2020}. This long-timescale depletion of the cloud provides a secular change in the mass and angular momentum of the background, in addition to any other emissions from the system. Hence, the presence of a Proca environment surrounding the primary black hole in an EMRI system modifies the inspiraling dynamics of the secondary black hole, resulting in a modification to the measured waveform at the detector \footnote{Here, we neglect higher order effects such as resonant depletion of the Proca cloud (see~\cite{Baumann_Chia_Porto_2019, Berti_Brito_Macedo_Raposo_Rosa_2019} for resonant depletion of a superradiant scalar cloud in an EMRI system). We also neglect dynamical friction effects on the secondary black hole. These effects have been shown to have  dramatic effects in the scalar field case and concievably will also have large effects in the vector case~\cite{Baumann_Chia_Porto_2019, Traykova_2021}, though a study of this nature has yet to be performed. However, for the preliminary and simplified analysis considered here, we relegate these effects to studies that will be performed in the future.}. In an EMRI system, the inspiralling dynamics is well approximated by assuming the trajectory follows a sequential evolution of geodesics of the Kerr spacetime. In the Kerr spacetime, geodesics are determined by three constants of motion: the energy $E$, the projection of the angular momentum along the spin axis $L$, and the Carter constant $C$. Assuming an equatorial orbit, the Carter constant vanishes and receives no evolution. Hence, the geodesic motion is determined by only two constants, $E$ and $L$. Sequential evolution along a series of geodesics corresponds to an adiabatic evolution of the orbital constants. This adiabatic change in the integrals of motion arise due to the asymptotic flux of energy and angular momentum from the system, sourced by either environmental effects or gravitational emission. 
	
	In particular, for the case of an EMRI system immersed in a superradiantly-generated Proca cloud, the evolution of the integrals of motion is given by
	\begin{align}
		\frac{dE_{geo}}{dt} &= - \left( \frac{dE_{GW}}{dt} + \left (\frac{dE_{geo}}{dt}\right)_{Proca} \right)\\
		\frac{dL_{geo}}{dt} &= - \left( \frac{dL_{GW}}{dt} + \left (\frac{dL_{geo}}{dt} \right)_{Proca}\right) 
	\end{align}
	where $I_{geo}$ represents the integral of motion for the geodesic and $\left (\frac{dI_{geo}}{dt}\right)_{Proca}$ represents the change in the orbital constants due to the flux of energy and momentum from the quasibound Proca cloud. Since the secondary black hole is minimally coupled to the energy-momentum of the Proca field, via the Einstein equations, the change in the orbital constants will not be the same as the change in the energy and angular momentum of the Proca cloud. Instead, at the linear level, the presence of the Proca cloud modifies the energy and angular momentum of the background spacetime, which enters as an additional change in the orbital constants. In particular, the change in the integrals of motion, due to the presence of an uncoupled Proca cloud, arise due to the change in the energy and angular momentum of the Kerr background:
	\begin{align}
		dE_{geo} &= \frac{\partial E_{geo}}{\partial L_{Kerr}}dL_{Kerr} + \frac{\partial E_{geo}}{\partial E_{Kerr}} dE_{Kerr}\\
		dL_{geo} &= \frac{\partial L_{geo}}{\partial L_{Kerr}}dL_{Kerr} + \frac{\partial L_{geo}}{\partial E_{Kerr}} dE_{Kerr}
	\end{align}
	
	such that
	\begin{align} \label{eq:modifiedfluxes}
		\left (\frac{dE_{geo}}{dt}\right)_{Proca} &= \frac{dE_{Proca}}{dt} \Gamma(r) \left [ \frac{dE_{geo}}{dE_{Kerr}} + \frac{\mathfrak{m}}{\omega} \frac{dE_{geo}}{dL_{Kerr}} \right ] \\
        \left (\frac{dL_{geo}}{dt}\right)_{Proca} &= \frac{dE_{Proca}}{dt} \Gamma(r) \left [ \frac{dL_{geo}}{dE_{Kerr}} + \frac{\mathfrak{m}}{\omega} \frac{dL_{geo}}{dL_{Kerr}} \right ]
	\end{align}
	where we introduced a radially dependent prefactor that accounts for the fraction of the Proca cloud within the orbital radius, and $E_{Kerr}$ and $L_{Kerr}$ are the total mass and angular momentum of the Kerr spacetime, respectively. $E_{Kerr}$ and $L_{Kerr}$ are calculated prior to the superradiant instability and hence represent the total mass and angular momentum of the black hole-cloud system after the instability has turned off. It follows from the Teukolsky equation and our choice of normalization of the Proca field that $\frac{dE_{Proca}}{dt} \propto E_{Proca}^2$, and hence $\Gamma(r) = \left( \frac{E(r<r_{orbit})}{E_{total}} \right)^2 $ (see \Cref{app:proccloud}). This prefactor accounts for the portion of the radiating Proca cloud that modifies the orbital trajectory. $\Gamma(r)$ asymptotes to unity at asymptotic infinity, meaning all of the Proca cloud is within the orbital radius and contributes to the trajectory modification. At the other extreme, near the horizon, $\Gamma(r)$ approaches zero since all of the cloud is external to the orbital radius. This prefactor hence represents the fraction of the Proca cloud the inspiralling black hole "sees". At infinity, the mass the secondary black hole "sees" is the total mass of the black hole-cloud system; At the horizon, it's the mass of the central black hole which is "seen" by the secondary black hole and hence none of the radiating Proca cloud modifies the trajectory at this point. 
 
    This is an approximation in several respects. First, the angular structure of the cloud is integrated out to produce a purely radial function. Secondly, the energy integral \Cref{eq:energyintegral} is calculated using the stress-energy tensor from the perturbative calculation of the Proca field and not from the full Einstein-Proca system. Thirdly, an additional averaging of the radial distance over an orbital period is performed when calculating $\Gamma(r)$, due to limitations of the waveform generator. 
	
	For the gravitational terms in the flux functions, 5PN accurate analytic expressions for the energy and momentum fluxes (Hence, semi-latus rectum and eccentricity evolution. See below.) are employed. 
	
	For our purposes, its more convenient to express adiabatic evolution of the integrals of motion in terms of the eccentricity and semi-latus rectum, from which the integrals of motion can be expressed. The asymptotic fluxes can be transformed into rates of change of the orbital parameters by inverting $\dot{I}_{geo} = \frac{dI_{geo}}{dp} \dot{p} + \frac{dI_{geo}}{de} \dot{e}$. This then gives us the rate of change of the geometry of the trajectory which the secondary black hole follows.
	
	The full trajectory is calculated by integrating the flux equations, after choosing suitable initial conditions, using an 8th-order explicit Runge-Kutta integrator. The trajectory is integrated to within 0.2 gravitational radii of the separatrix, calculated using the previous iteration loop of the integration. Initial conditions for the integration, namely $(p_0, e_0, \Phi_{\theta,_0}, \Phi_{\phi,_0}, \Phi_{r,0})$, are chosen such that coalescence occurs approximately after 5 years. This gives the greatest possible chance a Proca cloud will be detected during the mission lifetime of the LISA observatory.
	
	The trajectory, once computed, is then fed into a waveform model. The model currently employed is the Fast EMRI Waveforms (few) Augmented Analytic Kludge (AAK) model~\cite{AAK, few}. The AAK model is built using Keplerian ellipses for the orbital trajectory, and evolves the inspiral, periapsis precession, and Lense-Thirring precession using PN fluxes. The difference to the original Analytic Kludge model is that the orbital frequencies and 2 precession rates are enforced to be the original Kerr values, which is achieved by solving an algebraic expression for some unphysical values of the mass, spins, and semi-latus rectum. This defines a map $(M,a,p) \rightarrow (\tilde{M}, \tilde{a}, \tilde{p})$ which maps the frequencies of the Keplerian orbit onto the frequencies for the Kerr geodetic motion. This greatly improves the accuracy of the original AK model and agrees remarkably well with Teukolsky-based waveforms. The few version, the version employed in this study, removes this mapping and instead directly calculates the fundamental frequencies and converts them into the basis for the AAK model:
	\begin{align}
		\dot{\Phi} &= \Omega_r \\
		\dot{\gamma} &= \Omega_{\theta} - \Omega_{r} \\
		\dot{\alpha} &= \Omega_{\phi} - \Omega_{\theta}
	\end{align}
	where $\dot{\Phi}$ is the variation of the quasi-Keplerian mean anomaly, $\dot{\alpha}$ is the Lense-Thirring precession, and $\dot{\gamma} + \dot{\alpha}$ is the periapsis precession.  These phase evolutions are then fed into the Peters-Matthew formula for the gravitational strain amplitudes~\cite{PeterMatthew} \footnote{The custom code used to generate these trajectories and waveforms is available at \url{https://github.com/Shaun-F/GWGenerator.git}}.

    The states of the cloud in this analysis are restricted to the $m=1$ mode and $n=0$ overtone. This is for several reasons. First, the asymptotic flux values from the numerical solver are numerically unstable for larger mode and overtone values. Secondly, gravitational emission from higher modes are $\alpha$-suppressed~\cite{Yoshino_Kodama_2014, superrad}, with higher modes being suppressed by powers of $\alpha^4$. Hence, the secular variation in the cloud largely comes from the $m=1$ mode. Moreover, the total mass contained in the higher modes is less than that in the $m=1$ mode. Reducing the analysis to the single choice of these parameters is thus reasonable within the approximations already employed.

	\section{Results} \label{sec:results}

	\subsection{Proca Clouds}
\begin{figure*}
		\centering
		\begin{subfigure}{.49\textwidth}
			\centering
			\includegraphics[width=\linewidth]{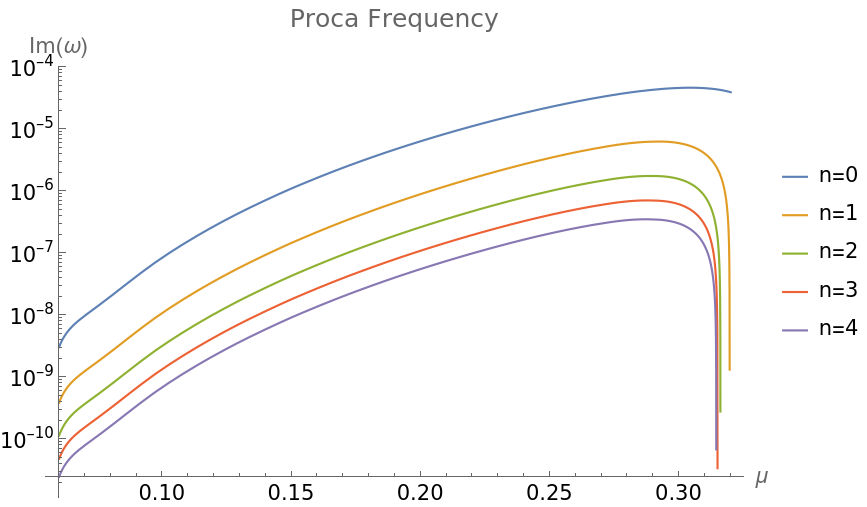}
		\end{subfigure}
		\begin{subfigure}{.49\textwidth}
			\centering
			\includegraphics[width=\linewidth]{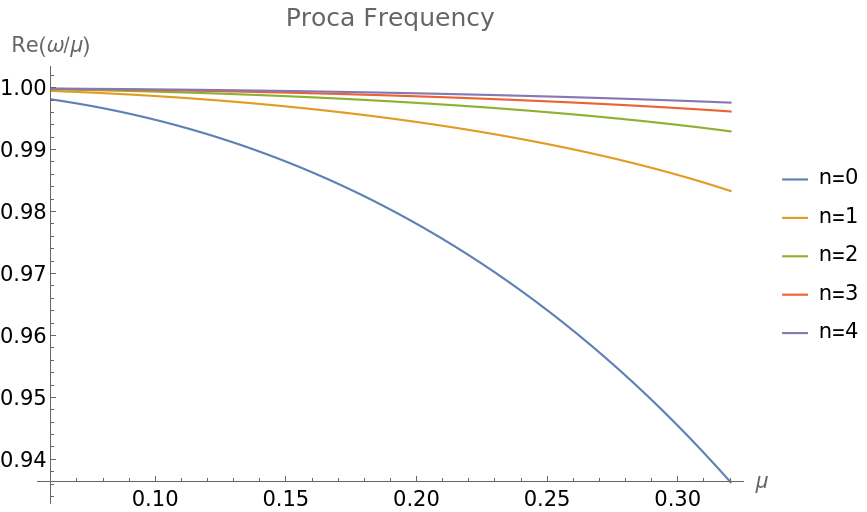}
		\end{subfigure}
		\caption{Superradiant Proca data for $\mathfrak{m}=1$ mode, dimensionless spin $\chi=0.9$, and Proca spin $S=-1$. The left plot displays the imaginary part of the Proca field frequency, which directly yields the instability rate of the cloud. The right plot displays the real part of the frequency, which yields the oscillatory part of the field. As expected, the $\mathfrak{m}=1$, $n=0$ mode is the most unstable. Here we take $G=c=\hbar=1$, so that $\alpha = \mu M \frac{G}{c \hbar} \rightarrow \mu M$.}
		\label{fig:ProcaFreq}
	\end{figure*}

\begin{figure*}
		\centering
		\includegraphics[width=\linewidth]{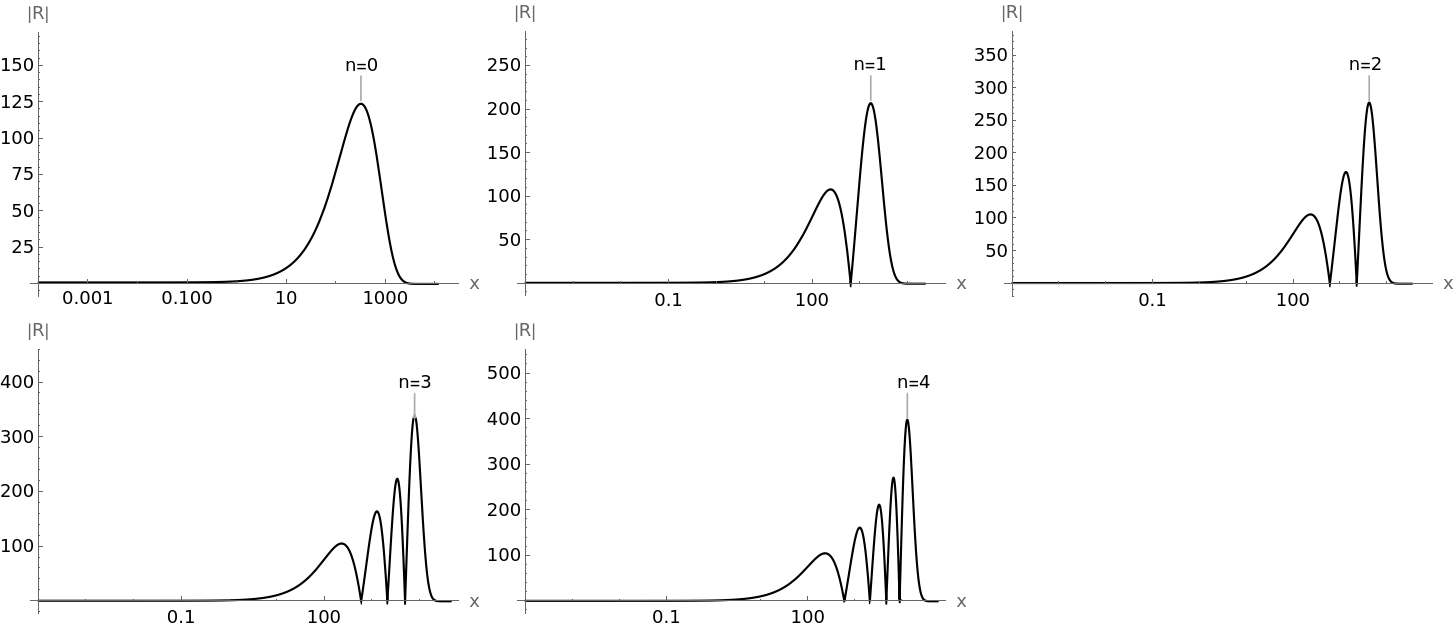}
		\caption{Example radial functions for $\mathfrak{m}=1$ mode with dimensionless spin $\chi=0.9$ and gravitational coupling $\alpha = \frac{11}{100}$ for various overtone numbers.}
		\label{fig:ProcaRadial1}
	\end{figure*}

\begin{figure*}
		\centering
		\includegraphics[width=\linewidth]{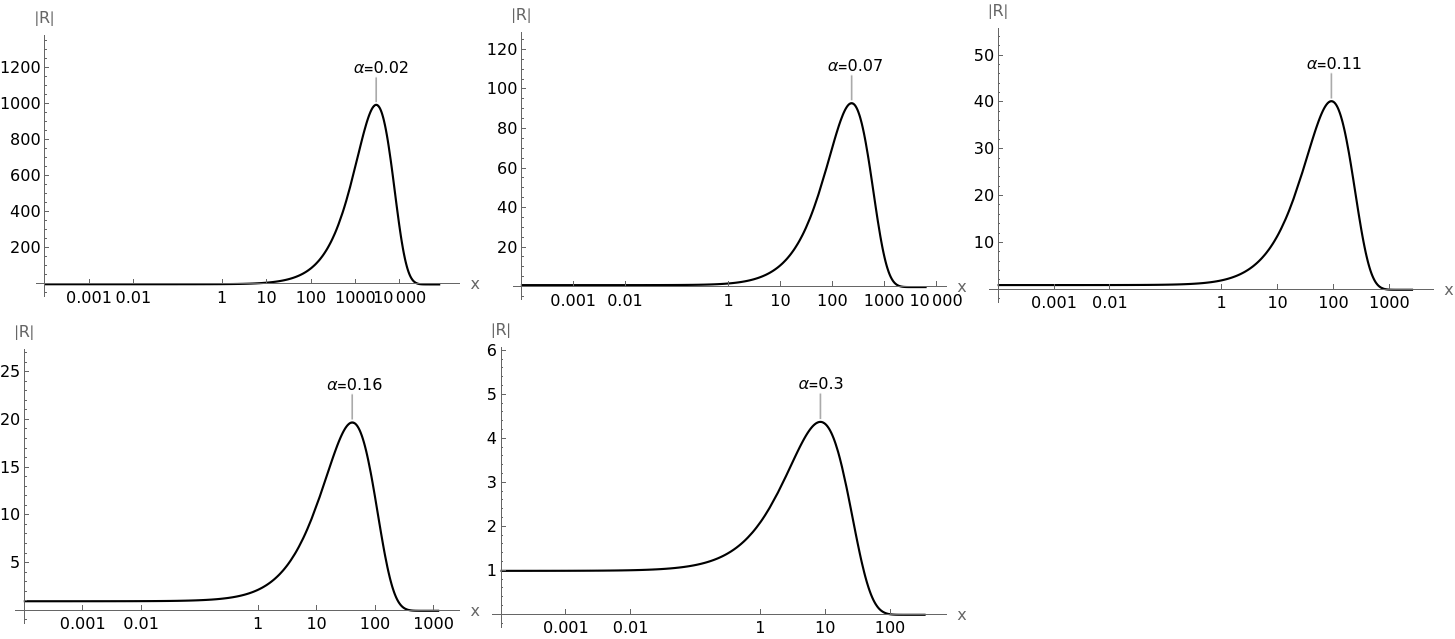}
		\caption{Example radial functions for $\mathfrak{m}=1$ mode with dimensionless spin $\chi=0.9$ and overtone number $n=0$ for various gravitational couplings. As expected, higher gravitational couplings translate to a more dense cloud, with most of the mass being concentrated closer to the horizon.}
		\label{fig:ProcaRadial2}
	\end{figure*}

	Following the procedure layed out in Section \ref{sec:solvingProca}, the Proca field equations were solved for mode numbers $\mathfrak{m}=\{1,2,3,4\}$, overtone numbers $n=\{0,1,2,3,4\}$, SMBH dimensionless spin $\chi \in [0.6,0.9]$ and Proca spin $S=-1$ \footnote{The code used to generate this dataset is available at \url{https://github.com/Shaun-F/KerrDressedWithProca.git}}. For our purposes, we restrict to $S=-1$ as this is the most unstable~\cite{Dolan_2018}. Examples of our generated data are shown in \Cref{fig:ProcaFreq,fig:ProcaRadial1,fig:ProcaRadial2}. figure \ref{fig:ProcaFreq} shows the evolution of the Proca field frequency as a function of the gravitational coupling for the $\mathfrak{m}=1$ mode and SMBH dimensionless spin $\chi=0.9$, for various overtone numbers. The imaginary part of the frequency gives the instability rate of the cloud, while the real part yields the oscillation frequency. As can be seen, the $n=0$ overtone number is the most unstable. The maximum instability occurs at, for $\mathfrak{m}=1$, $n=0$, and $\chi=0.9$, $\alpha=0.304$ with an instability rate of $\tau = 2.1\times10^4 \frac{\text{GM}}{\text{c}^3} = 0.105$ s. Compared to the maximum instability of the corresponding scalar superradiant cloud, this is $\sim 2500$ times faster~\cite{Dolan_2018}.
	
	\Cref{fig:ProcaRadial1,fig:ProcaRadial2} show example radial functions for the $\mathfrak{m}=1$ mode and dimensionless spin $\chi=0.9$. The overtone structure of the Proca field is clearly displayed. The number of roots of the radial function is given by the overtone number, which also specifies the number of maxima and minima. The compactness of the cloud is also apparent, being directly given by the gravitational coupling, as expected. A higher gravitational coupling translates to a more compact Proca cloud. Lower values of the gravitational coupling yield a Proca cloud that can span thousands of gravitational radii, as expected from the rough scaling of the radial function as $\sim \frac{1}{\alpha}$. Higher values of the gravitational coupling yield Proca clouds that span tens of gravitational radii. Hence, higher values of the gravitational coupling are expected to have the greatest effect on an EMRI system.
	
	For a more in-depth analysis of superradiant Proca fields on Kerr backgrounds, see~\cite{Dolan_2018, Siemonsen_2020, East_2018, East_2017,fkks,Pani_Cardoso_Gualtieri_Berti_Ishibashi_2012}
	
    \subsection{Identifying Modified GWs}
    To assess the detectability and measureability of a superradiantly-generated Proca cloud around an EMRI system from the emitted GWs, a couple of figures of merit are leveraged. We first perform a simple analysis based on generated SNR and traditionally used faithfulness criteria to have an idea about the distinguishable mass region of the Proca particles. Then we perform a custom statistical study to have a more precise estimate.
    \vspace{0.3\baselineskip}
    
    \subsubsection{A simple rough estimate based on SNR}
    
	First, we define a noise-weighted inner product between two waveforms as
    \begin{equation} \label{eq:wvinner}
		\langle h_1, h_2 \rangle = 4\times \mathfrak{R}\mathfrak{e} \sum_{\alpha = I, II} \int_{f_{min}}^{f_{max}} \frac{\tilde{h}_{1,\alpha}(f) \tilde{h}_{2,\alpha}^*(f)}{S_n(f)} df
	\end{equation}
	where $\tilde{h}_{1/2,\alpha}$ are the Fourier-transforms of the detector response signals, $\tilde{h}_{2,{\alpha}}^*$ the latters complex conjugate, and $S_n(f)$ the one-sided noise power spectral density (PSD) of LISA~\cite{lisapsd,Barack_Cutler_2004}. The LISA PSD model receives contributions from 3 noise sources. The analytic expression for the PSD is
    \begin{widetext}
	\begin{equation}
		S_n(f) = \frac{10}{3L^2}\left(P_{OMS}(f) + \frac{4 P_{acc}(f)}{(2\pi f)^4}\right) \left (1+\frac{6}{10}\left ( \frac{f}{f_*}\right)^2 \right) + S_c(f)
	\end{equation}
	where $P_{OMS}$ is the single-link optical metrology noise, $P_{acc}$ is the single test mass acceleration noise, and $S_c$ is the galactic confusion noise, accounting for unresolved galactic sources that manifest in the noise. The analytic expressions for the three noise contributions are
	\begin{align}
		P_{OMS} &= (1.5\times10^{-11}~{\rm m})^2 \left(1 + \left( \frac{2~{\rm mHz}}{f}\right)^4 \right) {\rm Hz}^{-1}\\
		P_{acc} &= \left(3\times10^{-15} \frac{\rm m}{\rm s^2}\right) \left( 1 + \left( \frac{0.4~ {\rm mHz}}{f}\right)^2 \right) \left( 1 + \left( \frac{f}{8~ {\rm mHz}} \right)^4 \right) {\rm Hz}^{-1}\\
		S_c &= A f^{-\frac{7}{3}} e^{-f^{\alpha} + \beta f \sin{(\kappa f)}} \left(1 + \tanh{(\gamma(f_k - f))} \right) {\rm Hz}^{-1}
	\end{align}
    \end{widetext}
    
    where the parameters of the galactic confusion noise are fit to simulations for a 4 year data run. These fit values are $(A, \alpha, \beta, \kappa, \gamma, f_k) = (1.8\times10^{-44}, 0.138, -221, 521, 1680, 0.00113)$.
	
\begin{figure*}
		\centering
		\includegraphics[width=\linewidth]{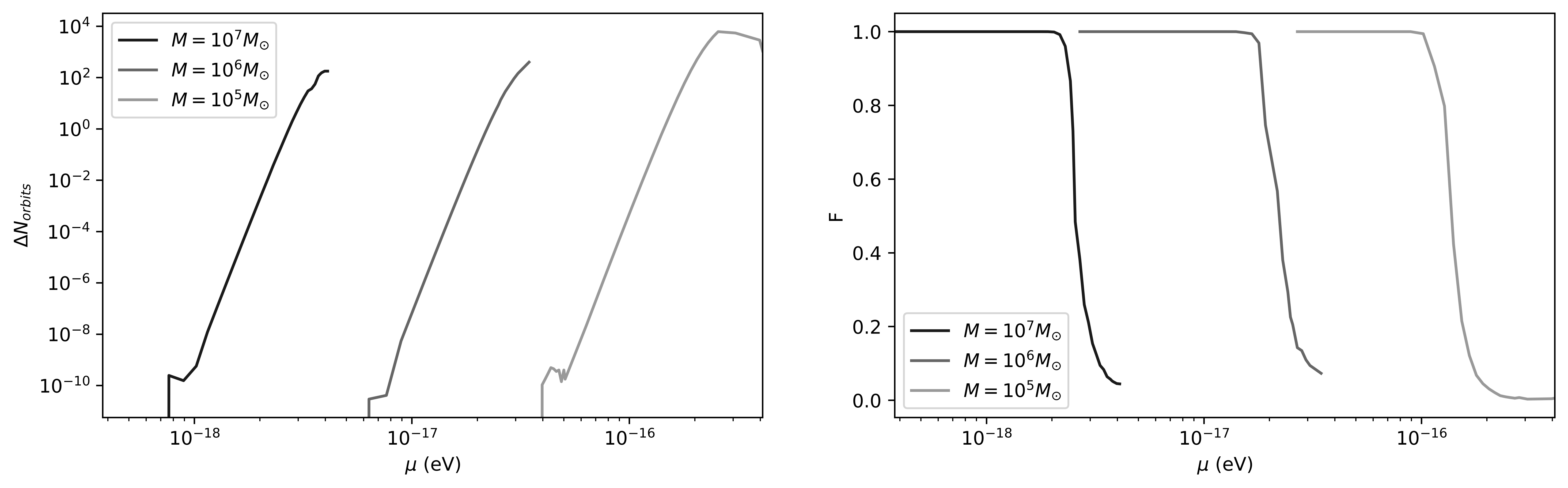}
		\caption{The difference in number of orbits and faithfulness as a function of Proca mass, respectively. The total spacetime dimensionless spin is $\chi = 0.9$ and the initial eccentricity was chosen to be 0.2. The difference in number of orbits is the absolute difference between the number of orbits completed by the dressed and undressed waveform at separatrix. This data suggests LISA should be able to distinguish GR-in-vacua waveforms and Proca-modified waveforms, for the given initial data, for Proca masses above $\sim 2 \times 10^{-18}$ eV. The upper limit on the mass is enforced by the superradiant threshold cutoff. Above this threshold, no bound state exists.  For the $M=10^{5}M_{\odot}$ data, this upper limit is $\mu_{max} = 4.47\times10^{-16}$ eV.}
		\label{fig:faith_Norbits}
	\end{figure*}

	Using the noise-weighted inner product, the SNR of a particular signal is defined as ${\rm SNR}^2 = \left \langle h | h \right \rangle$. Further, we also define the faithfulness between two signals as~\cite{Finn_1992,Damour_Iyer_Sathyaprakash_1998}
	\begin{equation}
		F \equiv \underset{t_c, \phi_c}{max} \frac{ \langle h_1 | h_2 \rangle}{\sqrt{ \langle h_1 | h_1 \rangle \langle h_2 | h_2 \rangle}}
	\end{equation}
	which is maximized over time and phase offsets of the two signals and takes values between -1 and 1, where the latter indicates perfect agreement between the waveforms. The maximization translates to maximizing over the variables $T$ in 
	\begin{equation}
		\langle h_{1,T} | h_2 \rangle = \mathfrak{R}\mathfrak{e} \sum_{\alpha = I, II} \int \frac{\tilde{h}_{1, \alpha}(f) \tilde{h}_{2,\alpha}^*(f)}{S_{n,2}(f)} e^{-2 \pi f T} df
	\end{equation}
	where $h_{1,T}$ is the time-offset version of the original waveform $h_1$ by time T, $S_{n,2}$ is now the two-sided noise power spectral density, and we have extended the integration domain to the entire reals using the fact that the noise PSD forces the integrand to vanishing values outside the range $[-f_{max}, -f_{min}] \cup [f_{min}, f_{max}]$ and that its an even function of $f$. Using the convolution theorem, this translates into the convolution
	\begin{equation}
		\langle h_{1,T} | h_2 \rangle = \mathfrak{R}\mathfrak{e}\sum_{\alpha = I, II}(H_{1,\alpha} \ast H_{2, \alpha})
	\end{equation}
	where $H_{1, \alpha} = \mathfrak{F}^{-1}(\frac{\tilde{h}_{1, \alpha}(f)}{S_{n,2}(f)})$, $H_{2, \alpha} = \mathfrak{F}^{-1}(\tilde{h}_{2, \alpha}^*(f))$, and $\mathfrak{F}$ denotes the Fourier transform. The faithfulness then becomes \footnote{Maximization over the coalescence phase $\phi_c$ is achieved by iteration over phase offsets of one of the signals. See~\cite{HPBB} for further discussion of the maximization procedure.}
	\begin{equation}
		F \equiv max \left \{ \left (\frac{\mathfrak{R}\mathfrak{e}\sum_{\alpha = I, II}(H_{1,\alpha} \ast H_{2, \alpha})}{\sqrt{ \langle h_1 | h_1 \rangle \langle h_2 | h_2 \rangle}} \right) (T, \phi_c) \right \}
	\end{equation}
	The detectability requirement places a threshold on the faithfulness statistic. Under Gaussian likelihoods for the parameters, this threshold arises from the requirement that a systematic mismodeling error, i.e. the error between the true waveform and the model waveform, should be smaller than the statistical measurement error. If the mismodeling error were larger than the measurement error, then the signals would be measurably different in the LISA data. Thus, if the faithfulness between the bare EMRI and the dressed EMRI waveforms are below this critical threshold, LISA should likely be able to distinguish between the two EMRI systems. This threshold assumes a Gaussian distribution for the model parameters around the true values. This prerequisite does not hold for the Proca mass, since the distribution is asymmetric and the waveforms are being compared against the vacuum case (c.f. figure \ref{fig:faith_Norbits}). However, it provides a rough estimate for the threshold of distinguishability. For a more rigorous statistical analysis, see the next section. The expression for this threshold is
	\begin{equation}
		F_c = 1 - \frac{D-1}{2\times {\rm SNR}^2}
	\end{equation}
	where $D$ is the size of the parameter space~\cite{faithreq}. The inclusion of a Proca mass increases the number of parameters by 1, hence the total parameter space is specified by $\left( M, m, \mu, a, p_0, e_0, x_0, d_L, \theta_s, \phi_s, \theta_K, \phi_K, \Phi_{\theta,0}, \Phi_{\phi,0}, \Phi_{r,0}\right)$, where $M$ is the mass of the SMBH (the primary), $m$ is the mass of the smaller black hole (the secondary), $\mu$ is the mass of the Proca field, $a$ is the spin value of the SMBH, $p_0$, $e_0$, and $x_0$ are the initial semi-latus rectum, eccentricity and inclination, respectively, $d_L$ is the luminosity distance to the system, $\theta_S$ and $\phi_S$ are the barycentric sky location of the system, $\theta_K$ and $\phi_K$ describe the orientation of the EMRI angular momentum vector in the barycentric coordinate system, and $\Phi_{\theta/\phi/r,0}$ are the initial phases for the polar,azimuthal, and radial motion, respectively. Orientation of the spin vector of the SMBH is currently ignored in waveform generation due to limitations of the waveform generator package employed during this study. For signals with SNR on the order of 20, the faithfulness threshold for detectability is $\sim 0.98$. For any pair of signals that produces a faithfulness below this threshold, they will likely be distinguishable with LISA. 
	
	The second measure of detectability we utilize is the number of orbits completed by the binary, as this is easily infered from the waveform directly measured by LISA. The number of orbits completed by the binary at separatrix is directly related to the orbital frequency by
	\begin{equation}
		N_{orbits} = \frac{1}{2 \pi}\int_0^{T_{\text{sep}}} \Omega_{\phi} dt
	\end{equation}
	where $\Omega_{\phi}$ is the azimuthal orbital frequency, which is related to the phase of the trajectory $\Phi_{\phi}$ by $\frac{d \Phi_{\phi}}{dt} = \Omega_{\phi}$. The resolvable deviations in the phase of the trajectory of an EMRI system by LISA can be estimated by a Fisher-matrix anaylsis~\cite{Lindblom_2008, faithreq}. The result is that LISA should be able to distinguish a phase shift of $\Delta \Phi_{\phi} \sim 0.05 $ radians, or in terms of the GW phase $\Delta \Psi \sim 0.1$ radians. This roughly translates to a resolution in the number of completed orbits at the time the trajectory reaches the separatrix as $\Delta N_{orbits} \sim 0.001$. 
	
	Figure \ref{fig:faith_Norbits} shows three example EMRI systems with SMBH masses $M=(10^5, 10^6, 10^7) M_{\odot}$, SMBH dimensionless spin $\chi=0.9$, and initial eccentricity of the orbit $e=0.2$. The initial semi-latus rectum is chosen so that the lifetime of the system approximately equals the LISA mission lifetime. As expected, for higher Proca masses, the dressed waveform increasingly deviates from the GR-in-vacua (undressed) waveform. This is due to the increased coupling between the Proca cloud and the background spacetime. Since the size of the Proca cloud roughly scales as $\frac{1}{\mu}$ (c.f. Eq. \eqref{eq:radialasymptotic}), higher mass translates to a more compact Proca cloud. Thus, as the secondary black hole inspirals, Proca clouds with greater Proca masses have a larger effect on the trajectory due to the compactness (c.f. Eq. \eqref{eq:modifiedfluxes}, in particular $\Gamma(r)$). In addition, the energy and angular momentum fluxes from the Proca cloud are monotonically increasing up until just before the superradiant condition fails. Hence, higher compactness of the cloud together with larger values of the asymptotic fluxes explain the greater deviation from the GR-in-vacua scenario.

    It was found that the initial value of the eccentricity has little effect on the faithfulness, i.e. the difference between the Proca waveform and the GR-in-vacua waveform doesnt change much with eccentricity. This is likely due to the averaging procedure over eccentricity of the effect the Proca flux has on the secondaries trajectory. In other words, the function $\Gamma(r)$ is only a function of the radial distance and not the eccentricity. Though factors such as 
    \begin{equation}
    \left [ \frac{dE_{orbit}}{dE_{Kerr}} + \frac{\mathfrak{m}}{\omega} \frac{dE_{orbit}}{dL_{Kerr}} \right ]
    \end{equation}
    are expected to change with eccentricity, this effect is evidently small. 

    The spin of the SMBH, on the other hand, plays a larger role (cf appendix \ref{app:dataplots}). While the spin doesn't effect the 'knee' of the graph in figure \ref{fig:faith_Norbits}, it does change the minimum value of the faithfulness. In other words, lower spins of black holes are less able to constrain lower Proca masses.

    Its also important to determine the astrophysical relevance of this type of modification to the background. This is achieved by comparing the timescale for gravitational emission from the Proca cloud to the timescale for inspiral of the secondary object. The latter is fixed to be ${\sim } 5$ years, the approximate LISA mission lifetime, and the former depends on the gravitational coupling $\alpha$. In other words, the astrophysical relevance for the study performed in this paper depends on the gravitational coupling. The 'knee' of the faithfulness statistic as a function of $\alpha$ in figure \ref{fig:faith_Norbits} occurs approximately at $\alpha\sim0.07$. The corresponding timescale for gravitational emission from the Proca condensate ranges from $10^3$ - $10^5$ years (cf. appendix \ref{app:proccloud} and figure \ref{fig:gravem}). This means the modification to the gravitational wave signal due to the time-dependent background as discussed here is of astrophysical relevance since the two relevant timescales are separated by three or more orders of magnitude. For higher gravitational couplings, the timescale for gravitational emission shrinks (see figure \ref{fig:gravem}). This translates to less observational relevance since the inspiral phase and gravitational emission must occur coincidentally for LISA to be able to probe the Proca cloud properties using the results of this study. It would be astrophysically remarkable to observe a high gravitational coupling using EMRI's as probes with LISA. Nonetheless, for the purposes of this study, perfect coincidence is assumed. This is a reasonable approximation for lower $\alpha$ values, but becomes unreasonable for $\alpha \gtrsim 0.2$, where the cloud decay timescales roughly equals the inspiral timescale which are both extremely short. However, the faithfulness statistic is much less than unity at this point, so the Proca condensate will already be observable with LISA for the astrophysically-relevant timescales. Hence, there is a range of small gravitational coupling, translating to large cloud decay timescales, where the effect on the inspiral will be observable with LISA and where the cloud is sufficiently long lived. To quantitatively take into account the likeliness of coincidence for these two timescales, a full population synthesis study, analysis, and potentially an N-body simulation would be required, which is beyond the scope of this study.

	\subsubsection{Statistical study} \label{sec:Discoverability}
	To estimate the statistical significance at which we can discover the Proca fields with the EMRI signals, we constructed a likelihood ratio test for our two scenarios: the null hypothesis for the absence of the vector fields and the alternative hypothesis for the presence of them. For the binary simple hypotheses testing problem, the optimal test statistic (TS) is the likelihood ratio of the hypotheses. This is known as the Neyman-Pearson lemma~\cite{10.1098/rsta.1933.0009}. For hypotheses depending on a single parameter, this can be extended to the likelihood ratio for two constant values of that parameter via the Karlin-Rubin theorem~\cite{10.1214/aoms/1177728259} as well. However, beyond these constructions an optimal solution, or in other words the uniformly most powerful test may not, and in general does not exist. The prescription of how the test statistic is constructed may depend on the problem. The most common test statistic for a likelihood ratio test is formed by maximizing the likelihoods of the tested hypotheses separately in their respective parameter spaces and then taking their ratio. The traditional use of maximized likelihood ratios is generally solely due to its similarity to the optimal solution and its generic form. For signals of our interest, this test is also equivalent to using the ratio of highest obtainable SNRs from two hypotheses. After the test statistic is constructed, the significance or the confidence level of the observation is obtained by comparing this value to a distribution of test statistics which arise from the null hypothesis. 
	
\begin{figure}
     \centering
     \includegraphics[width=\columnwidth,clip]{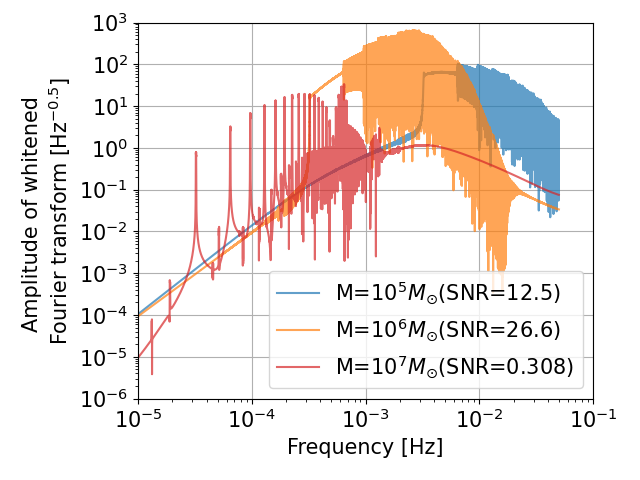}
     \caption{Amplitudes of whitened Fourier transforms of 4 year long polarization averaged EMRI waveforms for different SMBH masses, with all the other parameters specified as below Eq. \eqref{eq:logl} ($p_0=11~M$ for each $M$). The SNR of the waveforms are also given in the figure.}
     \label{fig:fftall}
 \end{figure}

	Here, due to the computational cost of maximization in 15 dimensional parameter space with slow generation of waveforms, we constructed our test statistic by only maximizing in the parameter space of the modified waveforms in the presence of vector fields; and assumed that if the correct waveform model is used, the maximization in the full parameter space gives the parameter values for the EMRI system, which are practically not different than their real physical values. This made our job easier since in our generation of signals, we know the physical parameters and we do not need to perform maximization for them. We calculated the likelihood for the null hypothesis ($\mathcal{L}_{n}$) using the 15 orbital parameters except the Proca mass that we use ($\mathbf{x}$), and for the alternative hypothesis we use these 15 parameters and perform a maximization of the likelihood ($\mathcal{L}_{a}$) over the Proca mass. We assume that these values will maximize $\mathcal{L}_{a}$ in every scenario; since in the low mass limit for the Proca fields, our alternative hypothesis corresponds to the null hypothesis and therefore is inclusive of the null hypothesis too. Consequently, via maximizing $\mathcal{L}_{a}$, we can reach the true $\mathbf{x}$ values in both scenarios. We take the ratio of these two likelihoods as our test statistic.
	\begin{equation}
		{\rm TS}=\frac{\sup_{\mathbf{x},\mu} \mathcal{L}_{a}(\mathbf{x},\mu)}{\mathcal{L}_{n}(\mathbf{x})}
	\end{equation}
 When calculating the likelihoods we assume a Gaussian noise according to the LISA's noise curve. Hence, logarithm of the likelihoods can be written up to a constant, for data $d$ and model prediction $h(\mathbf{\Theta})$ with parameters $\mathbf{\Theta}$ as 
	\begin{equation}
		\log \mathcal{L}=-\frac{1}{2}\langle d-h(\mathbf{\Theta})|d-h(\mathbf{\Theta})\rangle
        \label{eq:logl}
	\end{equation}
 In order to estimate the discovery probability of the effect of Proca clouds on the GWs, we performed sets of waveform generations. In these, we set the orbital parameters as $M=10^6~M_\odot$, $m=10~M_\odot$, $a=0.9$, $p_0=0.53\times10^{-6}~{\rm pc}$ (= 11~$M$ in geometric units $G$=$c$=1), $e_0=0.7$, $\iota_0=0$ ($x_0 = 1$), $\mathbf{\Phi}=\{0,0,0\}$, $qS=0.7$, $\phi_S=0.2$, $qK=0.7$, $\phi_K=0.6$, $d_L=1~{\rm Gpc}$; generated 4 year long waveforms sampled at 0.02 Hz, and added noise according to the LISA noise curve. The central BH mass was chosen as $10^6M_{\odot}$ as it lies in a sweet spot for obtaining high SNRs and being able to constrain low Proca masses. Higher BH masses significantly reduces the SNR, as shown in figure \ref{fig:fftall}; where as lower BH masses are less capable of constraining the boson masses as shown in figure \ref{fig:faith_Norbits}.
 
\begin{figure}
		\centering
		\includegraphics[width=\columnwidth]{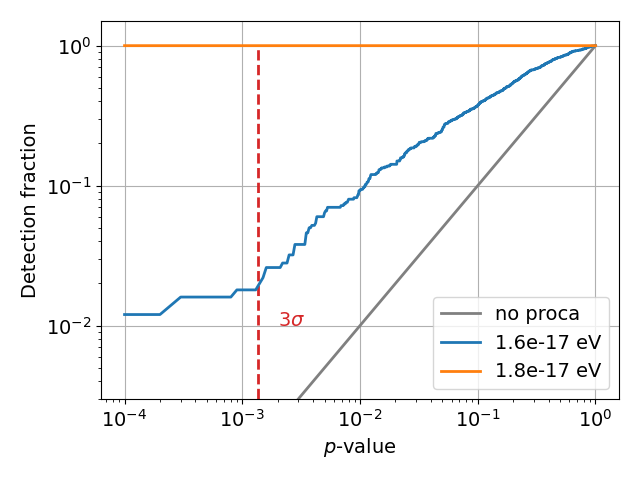}
		\caption{The discovery probability for different boson masses as a function of significance. The gray line indicates the false positive probability as a reference. The dashed red line shows the $p$-value corresponding to one sided 3$\sigma$ significance.}
		\label{fig:disc_ratio}
	\end{figure}

 The first set of waveforms we generated consists of waveforms in the absence of vector fields. TS from this set forms the background distribution we used for computing the significances. In the second and third set, we generated waveforms with Proca masses $\mu=1.6\times10^{-17}$ eV and $\mu=1.8\times10^{-17}$ eV. We found that for the masses above $\mu=1.8\times10^{-17}$ eV, the presence of the vector Proca fields can be discovered at $3\sigma$ significance with full efficiency, where as for lower masses this fraction is lower. The discovery probability from a single EMRI signal for different Proca masses as a function of the $p$-value is shown in figure \ref{fig:disc_ratio}.
	
	\subsubsection{LISA Proca Discovery Potential}
	The combination of the two previous studies suggest LISA should be able to detect superradiantly-generated Proca fields in the mass range $\mu \in $ \ProcaBounds eV. The lower bound comes from both the $\frac{1}{\mu}$ scaling of the Proca cloud radius and the reduced SNR for higher SMBH masses. At lower mass values, the Proca cloud extends over several thousands of gravitational radii and so not enough mass is within the orbital radius to significantly modify the secondary BH's trajectory. The resulting waveforms are thus not 'different' enough to be distinguishable with LISA (as determined through the faithfulness and number of orbits statistics). The upper bound comes from the saturation of the superradiant condition. Above a critical mass, the superradiant condition is no longer satisfied and hence a superradiant bound state is not formed. Since the energy flux from the Proca cloud depends on the combination $M\cdot \mu$, lower mass values of the SMBH allow for increased detection probability for higher mass values of the Proca field (and vice versa).

	\subsection{Possible Effects on Parameter Estimation}

\begin{figure*}
    \centering
    \subfloat[]{\includegraphics[width=0.32\textwidth]{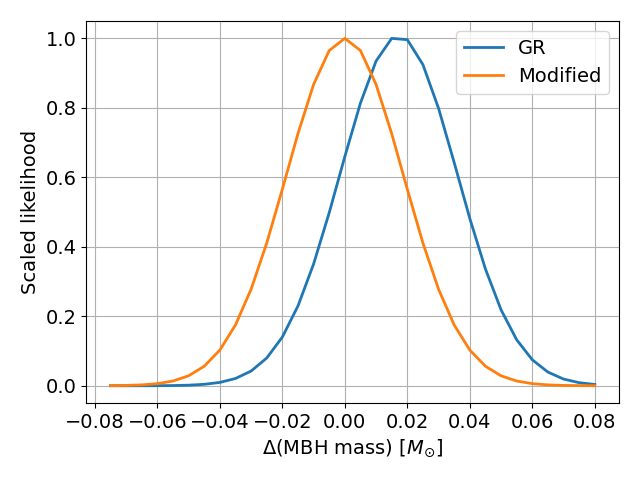}}
 \subfloat[]{\includegraphics[width=0.32\textwidth]{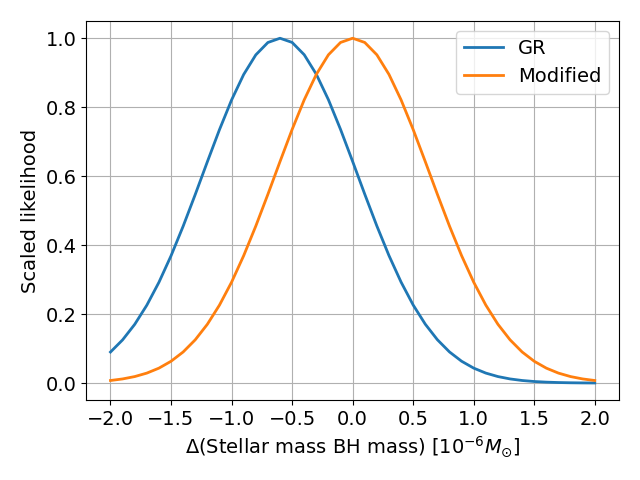}}
 \subfloat[]{\includegraphics[width=0.32\textwidth]{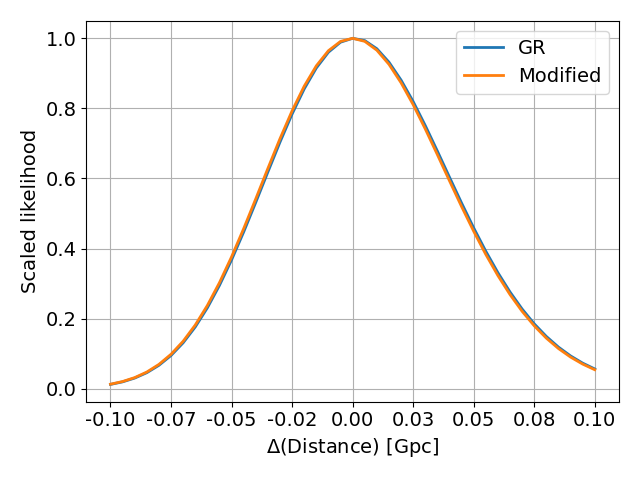}}
    \caption{Possible effects of the use of inaccurate waveforms for parameter estimation for boson mass 1.6$\times10^{-17}$eV. Each figure shows the likelihood for the different values of the parameter on the x-axis, conditioned on the true values of other parameters. The blue (orange) curves show the likelihoods calculated without (with) assuming the Proca fields. Each curve is scaled to have a maximum value of 1. }
    \label{fig:pe}
\end{figure*}

 EMRI signals allow very precise estimation of certain orbital parameters such as the masses of the BHs. However, using inaccurate waveforms in parameter estimation can yield biased likelihoods. Use of canonical GR waveforms in the presence of Proca fields can produce such effects. Here with a simple analysis, we provide such examples for a Proca mass of $1.6\times10^{-17}$ eV. Although modification from such a mass may not be clearly identified, as found in the previous subsection, we found that it can nevertheless bias the parameter estimations comparable to the estimation variances. We calculated the one dimensional conditional likelihoods for the parameters $M$, $m$ and $d_L$, while using the true values for the rest of the parameters. We calculated the likelihoods using the GR templates and the modified templates due to Proca fields, using the true mass of the field. In figure \ref{fig:pe} we show the possible effects of using the inaccurate templates in parameter estimation. It is seen that $M$ and $m$ estimates are more prone to biases than the $d_L$ estimate. We note that in order to capture this effect more accurately and precisely, one should perform a \emph{full} parameter estimation on the full parameter space. Due to the high computational cost of such a study, we performed a one dimensional study with the goal of probing possible effects and tendencies.
 
	\section{Conclusion} \label{sec:conclusion}
	EMRI systems provide a unique arena to study fundamental fields beyond the standard model, most notably dark matter candidates. Due to the feeble interaction between dark matter candidates and the standard model fields, the mass range of such candidates covers tens of orders of magnitudes. The dark photon, in particular, is a well motivated candidate for dark matter. It has several production mechanisms including the misalignment mechanism, quantum flucuations during inflation, tachyonic instabilities arising from couplings to a misaligned axion, and topological defect decays. Various couplings have been proposed for the dark photon. Direct detection searches assume different couplings in the hope of detecting a dark photon-involved process. 
 
    Gravitational interactions, on the other hand, need only assume a minimal coupling between gravity and the dark photon. The study performed here investigated the gravitational instability arising from a perturbation in the Proca field in the vicinity of a rotating uncharged black hole. The superradiant phenomenon is responsible for a build up of a Proca cloud around a Kerr black hole. The resulting cloud modifies the dynamical behavior of an EMRI system during the inspiralling phase, which in turn causes a modification of the measured GW waveform. Depending on the regime of the parameter space, such a modification can be measured by the future space-borne gravitational observatory LISA. 

    After solving the Proca equations on a Kerr background and determining the fluxes of energy and angular momentum from the resulting quasibound state, the modification of the waveform was calculated along with derived statistical quantities that allows one to quantify the difference between modified and unmodified waveforms. Limits on the detectable mass range of the Proca field using LISA were obtained and found that LISA should be able to detect Proca fields in the mass range \ProcaBounds eV.

    Throughout the study, several approximations were leveraged. Primarily, the Einstein-Proca system is linearized with respect to the Kerr background. This separates the Proca field from the gravitational perturbations, greatly aiding analytical developments. Secondly, the secondary black hole is approximated to be travelling adiabatically on a sequence of geodesics and is only described as a point particle, the so-called skeletonized approach. Higher order corrections to the motion due to self-force interactions are neglected. Thirdly, the coupling between the secondary black hole and Proca cloud is minimal. The only coupling between them is through the modification to the background the Proca cloud induces via the integrals of motion. This neglects resonant effects between the Proca cloud and secondary black hole. It should be pointed out that dynamical friction effects, accretion, and resonant transitions have been shown to have drastic effects on the orbital phase for the case of a superradiant scalar field. Its reasonable to suspect such dramatic effects will also be present in the Proca field scenario. For example, resonant transitions can produce floating/sinking orbits which have a drastic effect on the resulting gravitational wave signal. Dynamical friction and accretion produce an additional torque on the binary, gradually dephasing it with respect to the vacuum scenario. The results in this study are thus a conservative treatment since incorporating these effects will provide additional ways in which the signal can be modified, widening the observational potential of LISA (see appendix \ref{app:comp}). Finally, the variation of the local gravitational potential is approximated via the function $\Gamma(r)$ in Eq. \eqref{eq:modifiedfluxes}.

    Future studies planned by the authors involve a more accurate calculation of the Einstein-Proca equations of motion, involving numerical relativity calculations to accurately determine the geodesics, as well as dynamical friction effects and accretion effects on the secondary BH which would provide further modifications to the signal. Additionally, transitions between bound-bound states and bound-unbound states induced by the secondary black hole will likely also have a significant effect. These resonant transitions are planned for future studies as well. Hence, future plans involve more accurate predictions for the interplay between a Proca cloud around an EMRI system with the plan to generate data analysis-read templates for the future LISA mission. These templates will form the foundation for probing the existence of macroscopic Bose-Einstein condensates around EMRI's. We also plan to extend our analysis to the Generalized Proca theories \cite{Heisenberg:2014rta,Heisenberg:2018vsk}, since the presence of derivative interactions will have large implications both for the background dynamics as well as for the perturbations.

    \begin{acknowledgements}
    S.F. and L.H. thank Nils Siemonsen for providing example code used in their studies to solve the Proca equations and discussing the workings of their code. S.F. thanks Sam Dolan for also providing their Proca equations of motion solver. Both code bases aided in developing the custom Proca solver used in this study and greatly advanced the development stage. S.F. also thanks Michael Katz for discussions on the Fast EMRI Waveform code base used in the custom GWform generator built for this study. L.H. is supported by funding from the European Research Council (ERC) under the European Unions Horizon 2020 research and innovation programme grant agreement No 801781 and by the Swiss National Science Foundation grant 179740. L.H. further acknowledges support from the Deutsche Forschungsgemeinschaft (DFG, German Research Foundation) under Germany’s Excellence Strategy EXC 2181/1 -390900948 (the Heidelberg STRUCTURES Excellence Cluster).
    \end{acknowledgements}

	\bibliographystyle{apsrev4-2}
	\bibliography{main.bib}

\begin{thebibliography}{100}%
\makeatletter
\providecommand \@ifxundefined [1]{%
 \@ifx{#1\undefined}
}%
\providecommand \@ifnum [1]{%
 \ifnum #1\expandafter \@firstoftwo
 \else \expandafter \@secondoftwo
 \fi
}%
\providecommand \@ifx [1]{%
 \ifx #1\expandafter \@firstoftwo
 \else \expandafter \@secondoftwo
 \fi
}%
\providecommand \natexlab [1]{#1}%
\providecommand \enquote  [1]{``#1''}%
\providecommand \bibnamefont  [1]{#1}%
\providecommand \bibfnamefont [1]{#1}%
\providecommand \citenamefont [1]{#1}%
\providecommand \href@noop [0]{\@secondoftwo}%
\providecommand \href [0]{\begingroup \@sanitize@url \@href}%
\providecommand \@href[1]{\@@startlink{#1}\@@href}%
\providecommand \@@href[1]{\endgroup#1\@@endlink}%
\providecommand \@sanitize@url [0]{\catcode `\\12\catcode `\$12\catcode
  `\&12\catcode `\#12\catcode `\^12\catcode `\_12\catcode `\%12\relax}%
\providecommand \@@startlink[1]{}%
\providecommand \@@endlink[0]{}%
\providecommand \url  [0]{\begingroup\@sanitize@url \@url }%
\providecommand \@url [1]{\endgroup\@href {#1}{\urlprefix }}%
\providecommand \urlprefix  [0]{URL }%
\providecommand \Eprint [0]{\href }%
\providecommand \doibase [0]{https://doi.org/}%
\providecommand \selectlanguage [0]{\@gobble}%
\providecommand \bibinfo  [0]{\@secondoftwo}%
\providecommand \bibfield  [0]{\@secondoftwo}%
\providecommand \translation [1]{[#1]}%
\providecommand \BibitemOpen [0]{}%
\providecommand \bibitemStop [0]{}%
\providecommand \bibitemNoStop [0]{.\EOS\space}%
\providecommand \EOS [0]{\spacefactor3000\relax}%
\providecommand \BibitemShut  [1]{\csname bibitem#1\endcsname}%
\let\auto@bib@innerbib\@empty
\bibitem [{\citenamefont {Abbott}\ \emph {et~al.}(2016)\citenamefont {Abbott}
  \emph {et~al.}}]{PhysRevLett.116.061102}%
  \BibitemOpen
  \bibfield  {author} {\bibinfo {author} {\bibfnamefont {B.~P.}\ \bibnamefont
  {Abbott}} \emph {et~al.} (\bibinfo {collaboration} {LIGO Scientific
  Collaboration and Virgo Collaboration}),\ }\href
  {https://doi.org/10.1103/PhysRevLett.116.061102} {\bibfield  {journal}
  {\bibinfo  {journal} {Phys. Rev. Lett.}\ }\textbf {\bibinfo {volume} {116}},\
  \bibinfo {pages} {061102} (\bibinfo {year} {2016})}\BibitemShut {NoStop}%
\bibitem [{\citenamefont {{The LIGO Scientific Collaboration and the Virgo
  Collaboration and the KAGRA Collaboration}}\ \emph
  {et~al.}(2021)\citenamefont {{The LIGO Scientific Collaboration and the Virgo
  Collaboration and the KAGRA Collaboration}} \emph {et~al.}}]{LIGO1}%
  \BibitemOpen
  \bibfield  {author} {\bibinfo {author} {\bibnamefont {{The LIGO Scientific
  Collaboration and the Virgo Collaboration and the KAGRA Collaboration}}}
  \emph {et~al.},\ }\href@noop {} {\bibinfo {title} {Tests of general
  relativity with gwtc-3}} (\bibinfo {year} {2021}),\ \Eprint
  {https://arxiv.org/abs/2112.06861} {arXiv:2112.06861 [gr-qc]} \BibitemShut
  {NoStop}%
\bibitem [{\citenamefont {{The LIGO Scientific Collaboration}}\ \emph
  {et~al.}(2022)\citenamefont {{The LIGO Scientific Collaboration}},
  \citenamefont {the Virgo~Collaboration}, \citenamefont {the
  KAGRA~Collaboration} \emph {et~al.}}]{LIGOdarkphoton}%
  \BibitemOpen
  \bibfield  {author} {\bibinfo {author} {\bibnamefont {{The LIGO Scientific
  Collaboration}}}, \bibinfo {author} {\bibnamefont {the Virgo~Collaboration}},
  \bibinfo {author} {\bibnamefont {the KAGRA~Collaboration}}, \emph {et~al.},\
  }\href {https://doi.org/10.1103/PhysRevD.105.063030} {\bibfield  {journal}
  {\bibinfo  {journal} {Physical Review D}\ }\textbf {\bibinfo {volume}
  {105}},\ \bibinfo {pages} {063030} (\bibinfo {year} {2022})},\ \bibinfo
  {note} {arXiv:2105.13085 [astro-ph, physics:gr-qc,
  physics:hep-ph]}\BibitemShut {NoStop}%
\bibitem [{\citenamefont {An}\ \emph {et~al.}(2020)\citenamefont {An},
  \citenamefont {Pospelov}, \citenamefont {Pradler} \emph
  {et~al.}}]{An_Pospelov_Pradler_Ritz_2020}%
  \BibitemOpen
  \bibfield  {author} {\bibinfo {author} {\bibfnamefont {H.}~\bibnamefont
  {An}}, \bibinfo {author} {\bibfnamefont {M.}~\bibnamefont {Pospelov}},
  \bibinfo {author} {\bibfnamefont {J.}~\bibnamefont {Pradler}}, \emph
  {et~al.},\ }\href {https://doi.org/10.1103/PhysRevD.102.115022} {\bibfield
  {journal} {\bibinfo  {journal} {Physical Review D}\ }\textbf {\bibinfo
  {volume} {102}},\ \bibinfo {pages} {115022} (\bibinfo {year}
  {2020})}\BibitemShut {NoStop}%
\bibitem [{\citenamefont {D{\textquotesingle}Eramo}\ \emph
  {et~al.}(2022)\citenamefont {D{\textquotesingle}Eramo}, \citenamefont
  {Valentino}, \citenamefont {Giar{\`{e} }} \emph {et~al.}}]{D_Eramo_2022}%
  \BibitemOpen
  \bibfield  {author} {\bibinfo {author} {\bibfnamefont {F.}~\bibnamefont
  {D{\textquotesingle}Eramo}}, \bibinfo {author} {\bibfnamefont {E.~D.}\
  \bibnamefont {Valentino}}, \bibinfo {author} {\bibfnamefont {W.}~\bibnamefont
  {Giar{\`{e} }}}, \emph {et~al.},\ }\href
  {https://doi.org/10.1088/1475-7516/2022/09/022} {\bibfield  {journal}
  {\bibinfo  {journal} {Journal of Cosmology and Astroparticle Physics}\
  }\textbf {\bibinfo {volume} {2022}}\bibinfo  {number} { (09)},\ \bibinfo
  {pages} {022}}\BibitemShut {NoStop}%
\bibitem [{\citenamefont {Pitjev}\ and\ \citenamefont
  {Pitjeva}(2013)}]{Pitjev_2013}%
  \BibitemOpen
\bibfield  {number} {  }\bibfield  {author} {\bibinfo {author} {\bibfnamefont
  {N.~P.}\ \bibnamefont {Pitjev}}\ and\ \bibinfo {author} {\bibfnamefont
  {E.~V.}\ \bibnamefont {Pitjeva}},\ }\href
  {https://doi.org/10.1134/s1063773713020060} {\bibfield  {journal} {\bibinfo
  {journal} {Astronomy Letters}\ }\textbf {\bibinfo {volume} {39}},\ \bibinfo
  {pages} {141} (\bibinfo {year} {2013})}\BibitemShut {NoStop}%
\bibitem [{\citenamefont {Polisensky}\ and\ \citenamefont
  {Ricotti}(2011)}]{Polisensky_2011}%
  \BibitemOpen
  \bibfield  {author} {\bibinfo {author} {\bibfnamefont {E.}~\bibnamefont
  {Polisensky}}\ and\ \bibinfo {author} {\bibfnamefont {M.}~\bibnamefont
  {Ricotti}},\ }\bibfield  {journal} {\bibinfo  {journal} {Physical Review D}\
  }\textbf {\bibinfo {volume} {83}},\ \href
  {https://doi.org/10.1103/physrevd.83.043506} {10.1103/physrevd.83.043506}
  (\bibinfo {year} {2011})\BibitemShut {NoStop}%
\bibitem [{\citenamefont {Aasi}\ \emph {et~al.}(2015)\citenamefont {Aasi},
  \citenamefont {Abbott}, \citenamefont {Abbott} \emph {et~al.}}]{2015}%
  \BibitemOpen
  \bibfield  {author} {\bibinfo {author} {\bibfnamefont {J.}~\bibnamefont
  {Aasi}}, \bibinfo {author} {\bibfnamefont {B.~P.}\ \bibnamefont {Abbott}},
  \bibinfo {author} {\bibfnamefont {R.}~\bibnamefont {Abbott}}, \emph
  {et~al.},\ }\href {https://doi.org/10.1088/0264-9381/32/7/074001} {\bibfield
  {journal} {\bibinfo  {journal} {Classical and Quantum Gravity}\ }\textbf
  {\bibinfo {volume} {32}},\ \bibinfo {pages} {074001} (\bibinfo {year}
  {2015})}\BibitemShut {NoStop}%
\bibitem [{\citenamefont {Accadia}\ \emph {et~al.}(2012)\citenamefont
  {Accadia}, \citenamefont {Acernese}, \citenamefont {Alshourbagy} \emph
  {et~al.}}]{virgo1}%
  \BibitemOpen
  \bibfield  {author} {\bibinfo {author} {\bibfnamefont {T.}~\bibnamefont
  {Accadia}}, \bibinfo {author} {\bibfnamefont {F.}~\bibnamefont {Acernese}},
  \bibinfo {author} {\bibfnamefont {M.}~\bibnamefont {Alshourbagy}}, \emph
  {et~al.},\ }\href {https://doi.org/10.1088/1748-0221/7/03/P03012} {\bibfield
  {journal} {\bibinfo  {journal} {Journal of Instrumentation}\ }\textbf
  {\bibinfo {volume} {7}}\bibinfo  {number} { (03)},\ \bibinfo {pages}
  {P03012–P03012}}\BibitemShut {NoStop}%
\bibitem [{\citenamefont {Acernese}\ \emph {et~al.}(2015)\citenamefont
  {Acernese}, \citenamefont {Agathos}, \citenamefont {Agatsuma} \emph
  {et~al.}}]{virgo2}%
  \BibitemOpen
\bibfield  {number} {  }\bibfield  {author} {\bibinfo {author} {\bibfnamefont
  {F.}~\bibnamefont {Acernese}}, \bibinfo {author} {\bibfnamefont
  {M.}~\bibnamefont {Agathos}}, \bibinfo {author} {\bibfnamefont
  {K.}~\bibnamefont {Agatsuma}}, \emph {et~al.},\ }\href
  {https://doi.org/10.1088/0264-9381/32/2/024001} {\bibfield  {journal}
  {\bibinfo  {journal} {Classical and Quantum Gravity}\ }\textbf {\bibinfo
  {volume} {32}},\ \bibinfo {pages} {024001} (\bibinfo {year} {2015})},\
  \bibinfo {note} {arXiv:1408.3978 [gr-qc, physics:physics]}\BibitemShut
  {NoStop}%
\bibitem [{\citenamefont {Akutsu}\ \emph {et~al.}(2021)\citenamefont {Akutsu},
  \citenamefont {Ando}, \citenamefont {Arai} \emph {et~al.}}]{kagra1}%
  \BibitemOpen
  \bibfield  {author} {\bibinfo {author} {\bibfnamefont {T.}~\bibnamefont
  {Akutsu}}, \bibinfo {author} {\bibfnamefont {M.}~\bibnamefont {Ando}},
  \bibinfo {author} {\bibfnamefont {K.}~\bibnamefont {Arai}}, \emph {et~al.},\
  }\href {https://doi.org/10.1093/ptep/ptaa125} {\bibfield  {journal} {\bibinfo
   {journal} {Progress of Theoretical and Experimental Physics}\ }\textbf
  {\bibinfo {volume} {2021}},\ \bibinfo {pages} {05A101} (\bibinfo {year}
  {2021})}\BibitemShut {NoStop}%
\bibitem [{\citenamefont {collaboration}\ \emph {et~al.}(2019)\citenamefont
  {collaboration} \emph {et~al.}}]{kagra2}%
  \BibitemOpen
  \bibfield  {author} {\bibinfo {author} {\bibfnamefont {K.}~\bibnamefont
  {collaboration}} \emph {et~al.},\ }\href
  {https://doi.org/10.1038/s41550-018-0658-y} {\bibfield  {journal} {\bibinfo
  {journal} {Nature Astronomy}\ }\textbf {\bibinfo {volume} {3}},\ \bibinfo
  {pages} {35–40} (\bibinfo {year} {2019})}\BibitemShut {NoStop}%
\bibitem [{\citenamefont {Danzmann}(2000)}]{lisa1}%
  \BibitemOpen
  \bibfield  {author} {\bibinfo {author} {\bibfnamefont {K.}~\bibnamefont
  {Danzmann}},\ }\href {https://doi.org/10.1016/S0273-1177(99)00973-4}
  {\bibfield  {journal} {\bibinfo  {journal} {Advances in Space Research}\
  }\textbf {\bibinfo {volume} {25}},\ \bibinfo {pages} {1129–1136} (\bibinfo
  {year} {2000})}\BibitemShut {NoStop}%
\bibitem [{\citenamefont {Amaro-Seoane}\ \emph {et~al.}(2017)\citenamefont
  {Amaro-Seoane}, \citenamefont {Audley}, \citenamefont {Babak} \emph
  {et~al.}}]{lisa2}%
  \BibitemOpen
  \bibfield  {author} {\bibinfo {author} {\bibfnamefont {P.}~\bibnamefont
  {Amaro-Seoane}}, \bibinfo {author} {\bibfnamefont {H.}~\bibnamefont
  {Audley}}, \bibinfo {author} {\bibfnamefont {S.}~\bibnamefont {Babak}}, \emph
  {et~al.},\ }\href {https://doi.org/10.48550/ARXIV.1702.00786} {\bibinfo
  {title} {Laser interferometer space antenna}} (\bibinfo {year}
  {2017})\BibitemShut {NoStop}%
\bibitem [{ET(2023)}]{ET}%
  \BibitemOpen
  \href@noop {} {\bibinfo {title} {et-gw.eu}},\ \bibinfo {howpublished}
  {\url{https://www.et-gw.eu/}} (\bibinfo {year} {2023}),\ \bibinfo {note}
  {[Accessed 05-Apr-2023]}\BibitemShut {NoStop}%
\bibitem [{\citenamefont {Reitze}\ \emph {et~al.}(2019)\citenamefont {Reitze},
  \citenamefont {Adhikari}, \citenamefont {Ballmer} \emph
  {et~al.}}]{reitze2019cosmic}%
  \BibitemOpen
  \bibfield  {author} {\bibinfo {author} {\bibfnamefont {D.}~\bibnamefont
  {Reitze}}, \bibinfo {author} {\bibfnamefont {R.~X.}\ \bibnamefont
  {Adhikari}}, \bibinfo {author} {\bibfnamefont {S.}~\bibnamefont {Ballmer}},
  \emph {et~al.},\ }\href@noop {} {\bibinfo {title} {Cosmic explorer: The u.s.
  contribution to gravitational-wave astronomy beyond ligo}} (\bibinfo {year}
  {2019}),\ \Eprint {https://arxiv.org/abs/1907.04833} {arXiv:1907.04833
  [astro-ph.IM]} \BibitemShut {NoStop}%
\bibitem [{\citenamefont {Dwyer}\ \emph {et~al.}(2015)\citenamefont {Dwyer},
  \citenamefont {Sigg}, \citenamefont {Ballmer} \emph
  {et~al.}}]{PhysRevD.91.082001}%
  \BibitemOpen
  \bibfield  {author} {\bibinfo {author} {\bibfnamefont {S.}~\bibnamefont
  {Dwyer}}, \bibinfo {author} {\bibfnamefont {D.}~\bibnamefont {Sigg}},
  \bibinfo {author} {\bibfnamefont {S.~W.}\ \bibnamefont {Ballmer}}, \emph
  {et~al.},\ }\href {https://doi.org/10.1103/PhysRevD.91.082001} {\bibfield
  {journal} {\bibinfo  {journal} {Phys. Rev. D}\ }\textbf {\bibinfo {volume}
  {91}},\ \bibinfo {pages} {082001} (\bibinfo {year} {2015})}\BibitemShut
  {NoStop}%
\bibitem [{\citenamefont {Seto}\ \emph {et~al.}(2001)\citenamefont {Seto},
  \citenamefont {Kawamura},\ and\ \citenamefont {Nakamura}}]{decigo1}%
  \BibitemOpen
  \bibfield  {author} {\bibinfo {author} {\bibfnamefont {N.}~\bibnamefont
  {Seto}}, \bibinfo {author} {\bibfnamefont {S.}~\bibnamefont {Kawamura}},\
  and\ \bibinfo {author} {\bibfnamefont {T.}~\bibnamefont {Nakamura}},\ }\href
  {https://doi.org/10.1103/PhysRevLett.87.221103} {\bibfield  {journal}
  {\bibinfo  {journal} {Physical Review Letters}\ }\textbf {\bibinfo {volume}
  {87}},\ \bibinfo {pages} {221103} (\bibinfo {year} {2001})}\BibitemShut
  {NoStop}%
\bibitem [{\citenamefont {Kawamura}\ \emph {et~al.}(2019)\citenamefont
  {Kawamura}, \citenamefont {Nakamura}, \citenamefont {Ando} \emph
  {et~al.}}]{decigo2}%
  \BibitemOpen
  \bibfield  {author} {\bibinfo {author} {\bibfnamefont {S.}~\bibnamefont
  {Kawamura}}, \bibinfo {author} {\bibfnamefont {T.}~\bibnamefont {Nakamura}},
  \bibinfo {author} {\bibfnamefont {M.}~\bibnamefont {Ando}}, \emph {et~al.},\
  }\href {https://doi.org/10.1142/S0218271818450013} {\bibfield  {journal}
  {\bibinfo  {journal} {International Journal of Modern Physics D}\ }\textbf
  {\bibinfo {volume} {28}},\ \bibinfo {pages} {1845001} (\bibinfo {year}
  {2019})}\BibitemShut {NoStop}%
\bibitem [{\citenamefont {Antoniadis}\ \emph {et~al.}(2022)\citenamefont
  {Antoniadis}, \citenamefont {Arzoumanian}, \citenamefont {Babak} \emph
  {et~al.}}]{EPTA1}%
  \BibitemOpen
  \bibfield  {author} {\bibinfo {author} {\bibfnamefont {J.}~\bibnamefont
  {Antoniadis}}, \bibinfo {author} {\bibfnamefont {Z.}~\bibnamefont
  {Arzoumanian}}, \bibinfo {author} {\bibfnamefont {S.}~\bibnamefont {Babak}},
  \emph {et~al.},\ }\href {https://doi.org/10.1093/mnras/stab3418} {\bibfield
  {journal} {\bibinfo  {journal} {Monthly Notices of the Royal Astronomical
  Society}\ }\textbf {\bibinfo {volume} {510}},\ \bibinfo {pages} {4873–4887}
  (\bibinfo {year} {2022})}\BibitemShut {NoStop}%
\bibitem [{\citenamefont {Perera}\ \emph {et~al.}(2019)\citenamefont {Perera},
  \citenamefont {DeCesar}, \citenamefont {Demorest} \emph {et~al.}}]{EPTA2}%
  \BibitemOpen
  \bibfield  {author} {\bibinfo {author} {\bibfnamefont {B.~B.~P.}\
  \bibnamefont {Perera}}, \bibinfo {author} {\bibfnamefont {M.~E.}\
  \bibnamefont {DeCesar}}, \bibinfo {author} {\bibfnamefont {P.~B.}\
  \bibnamefont {Demorest}}, \emph {et~al.},\ }\href
  {https://doi.org/10.1093/mnras/stz2857} {\bibfield  {journal} {\bibinfo
  {journal} {Monthly Notices of the Royal Astronomical Society}\ }\textbf
  {\bibinfo {volume} {490}},\ \bibinfo {pages} {4666–4687} (\bibinfo {year}
  {2019})},\ \bibinfo {note} {arXiv:1909.04534 [astro-ph]}\BibitemShut
  {NoStop}%
\bibitem [{\citenamefont {Zhu}\ \emph {et~al.}(2015)\citenamefont {Zhu},
  \citenamefont {Wen}, \citenamefont {Hobbs} \emph {et~al.}}]{nanograv1}%
  \BibitemOpen
  \bibfield  {author} {\bibinfo {author} {\bibfnamefont {X.-J.}\ \bibnamefont
  {Zhu}}, \bibinfo {author} {\bibfnamefont {L.}~\bibnamefont {Wen}}, \bibinfo
  {author} {\bibfnamefont {G.}~\bibnamefont {Hobbs}}, \emph {et~al.},\ }\href
  {https://doi.org/10.1093/mnras/stv381} {\bibfield  {journal} {\bibinfo
  {journal} {Monthly Notices of the Royal Astronomical Society}\ }\textbf
  {\bibinfo {volume} {449}},\ \bibinfo {pages} {1650–1663} (\bibinfo {year}
  {2015})},\ \bibinfo {note} {arXiv:1502.06001 [astro-ph,
  physics:gr-qc]}\BibitemShut {NoStop}%
\bibitem [{\citenamefont {Lam}\ \emph {et~al.}(2019)\citenamefont {Lam},
  \citenamefont {McLaughlin}, \citenamefont {Arzoumanian} \emph
  {et~al.}}]{nanograv2}%
  \BibitemOpen
  \bibfield  {author} {\bibinfo {author} {\bibfnamefont {M.~T.}\ \bibnamefont
  {Lam}}, \bibinfo {author} {\bibfnamefont {M.~A.}\ \bibnamefont {McLaughlin}},
  \bibinfo {author} {\bibfnamefont {Z.}~\bibnamefont {Arzoumanian}}, \emph
  {et~al.},\ }\href {https://doi.org/10.3847/1538-4357/ab01cd} {\bibfield
  {journal} {\bibinfo  {journal} {The Astrophysical Journal}\ }\textbf
  {\bibinfo {volume} {872}},\ \bibinfo {pages} {193} (\bibinfo {year}
  {2019})},\ \bibinfo {note} {arXiv:1809.03058 [astro-ph]}\BibitemShut
  {NoStop}%
\bibitem [{\citenamefont {Manchester}(2006)}]{ppta}%
  \BibitemOpen
  \bibfield  {author} {\bibinfo {author} {\bibfnamefont {R.~N.}\ \bibnamefont
  {Manchester}},\ }\href {https://doi.org/10.1088/1009-9271/6/S2/27} {\bibfield
   {journal} {\bibinfo  {journal} {Chinese Journal of Astronomy and
  Astrophysics}\ }\textbf {\bibinfo {volume} {6}},\ \bibinfo {pages} {139}
  (\bibinfo {year} {2006})}\BibitemShut {NoStop}%
\bibitem [{\citenamefont {Berti}\ \emph
  {et~al.}(2019{\natexlab{a}})\citenamefont {Berti}, \citenamefont {Barausse},
  \citenamefont {Cholis} \emph {et~al.}}]{lisawhite}%
  \BibitemOpen
  \bibfield  {author} {\bibinfo {author} {\bibfnamefont {E.}~\bibnamefont
  {Berti}}, \bibinfo {author} {\bibfnamefont {E.}~\bibnamefont {Barausse}},
  \bibinfo {author} {\bibfnamefont {I.}~\bibnamefont {Cholis}}, \emph
  {et~al.},\ }\href@noop {} {\bibinfo {title} {Tests of general relativity and
  fundamental physics with space-based gravitational wave detectors}} (\bibinfo
  {year} {2019}{\natexlab{a}}),\ \Eprint {https://arxiv.org/abs/1903.02781}
  {arXiv:1903.02781 [astro-ph.HE]} \BibitemShut {NoStop}%
\bibitem [{\citenamefont {Babak}\ \emph {et~al.}(2008)\citenamefont {Babak},
  \citenamefont {Fang}, \citenamefont {Gair} \emph
  {et~al.}}]{Babak_Fang_Gair_Glampedakis_Hughes_2008}%
  \BibitemOpen
  \bibfield  {author} {\bibinfo {author} {\bibfnamefont {S.}~\bibnamefont
  {Babak}}, \bibinfo {author} {\bibfnamefont {H.}~\bibnamefont {Fang}},
  \bibinfo {author} {\bibfnamefont {J.~R.}\ \bibnamefont {Gair}}, \emph
  {et~al.},\ }\bibfield  {journal} {\bibinfo  {journal} {arXiv:gr-qc/0607007}\
  }\href {https://doi.org/10.1103/PhysRevD.77.04990}
  {10.1103/PhysRevD.77.04990} (\bibinfo {year} {2008}),\ \bibinfo {note}
  {arXiv: gr-qc/0607007}\BibitemShut {NoStop}%
\bibitem [{\citenamefont {Barack}\ and\ \citenamefont
  {Cutler}(2004)}]{Barack_Cutler_2004}%
  \BibitemOpen
  \bibfield  {author} {\bibinfo {author} {\bibfnamefont {L.}~\bibnamefont
  {Barack}}\ and\ \bibinfo {author} {\bibfnamefont {C.}~\bibnamefont
  {Cutler}},\ }\href {https://doi.org/10.1103/PhysRevD.69.082005} {\bibfield
  {journal} {\bibinfo  {journal} {Physical Review D}\ }\textbf {\bibinfo
  {volume} {69}},\ \bibinfo {pages} {082005} (\bibinfo {year}
  {2004})}\BibitemShut {NoStop}%
\bibitem [{\citenamefont {Chua}\ \emph {et~al.}(2017)\citenamefont {Chua},
  \citenamefont {Moore},\ and\ \citenamefont {Gair}}]{AAK}%
  \BibitemOpen
  \bibfield  {author} {\bibinfo {author} {\bibfnamefont {A.~J.~K.}\
  \bibnamefont {Chua}}, \bibinfo {author} {\bibfnamefont {C.~J.}\ \bibnamefont
  {Moore}},\ and\ \bibinfo {author} {\bibfnamefont {J.~R.}\ \bibnamefont
  {Gair}},\ }\href {https://doi.org/10.1103/PhysRevD.96.044005} {\bibfield
  {journal} {\bibinfo  {journal} {Phys. Rev. D}\ }\textbf {\bibinfo {volume}
  {96}},\ \bibinfo {pages} {044005} (\bibinfo {year} {2017})}\BibitemShut
  {NoStop}%
\bibitem [{\citenamefont {Sigurdsson}\ and\ \citenamefont
  {Rees}(1997)}]{Sigurdsson_1997}%
  \BibitemOpen
  \bibfield  {author} {\bibinfo {author} {\bibfnamefont {S.}~\bibnamefont
  {Sigurdsson}}\ and\ \bibinfo {author} {\bibfnamefont {M.~J.}\ \bibnamefont
  {Rees}},\ }\href {https://doi.org/10.1093/mnras/284.2.318} {\bibfield
  {journal} {\bibinfo  {journal} {Monthly Notices of the Royal Astronomical
  Society}\ }\textbf {\bibinfo {volume} {284}},\ \bibinfo {pages} {318}
  (\bibinfo {year} {1997})}\BibitemShut {NoStop}%
\bibitem [{\citenamefont {Alexander}(2005)}]{Alexander_2005}%
  \BibitemOpen
  \bibfield  {author} {\bibinfo {author} {\bibfnamefont {T.}~\bibnamefont
  {Alexander}},\ }\href {https://doi.org/10.1016/j.physrep.2005.08.002}
  {\bibfield  {journal} {\bibinfo  {journal} {Physics Reports}\ }\textbf
  {\bibinfo {volume} {419}},\ \bibinfo {pages} {65} (\bibinfo {year}
  {2005})}\BibitemShut {NoStop}%
\bibitem [{\citenamefont {Merritt}(2006)}]{Merritt_2006}%
  \BibitemOpen
  \bibfield  {author} {\bibinfo {author} {\bibfnamefont {D.}~\bibnamefont
  {Merritt}},\ }\href {https://doi.org/10.1088/0034-4885/69/9/r01} {\bibfield
  {journal} {\bibinfo  {journal} {Reports on Progress in Physics}\ }\textbf
  {\bibinfo {volume} {69}},\ \bibinfo {pages} {R01} (\bibinfo {year}
  {2006})}\BibitemShut {NoStop}%
\bibitem [{\citenamefont {Bortolas}\ and\ \citenamefont
  {Mapelli}(2019)}]{Bortolas_2019}%
  \BibitemOpen
  \bibfield  {author} {\bibinfo {author} {\bibfnamefont {E.}~\bibnamefont
  {Bortolas}}\ and\ \bibinfo {author} {\bibfnamefont {M.}~\bibnamefont
  {Mapelli}},\ }\href {https://doi.org/10.1093/mnras/stz440} {\bibfield
  {journal} {\bibinfo  {journal} {Monthly Notices of the Royal Astronomical
  Society}\ }\textbf {\bibinfo {volume} {485}},\ \bibinfo {pages} {2125}
  (\bibinfo {year} {2019})}\BibitemShut {NoStop}%
\bibitem [{\citenamefont {Amaro-Seoane}\ \emph {et~al.}(2007)\citenamefont
  {Amaro-Seoane}, \citenamefont {Gair}, \citenamefont {Freitag} \emph
  {et~al.}}]{Amaro_Seoane_2007}%
  \BibitemOpen
  \bibfield  {author} {\bibinfo {author} {\bibfnamefont {P.}~\bibnamefont
  {Amaro-Seoane}}, \bibinfo {author} {\bibfnamefont {J.~R.}\ \bibnamefont
  {Gair}}, \bibinfo {author} {\bibfnamefont {M.}~\bibnamefont {Freitag}}, \emph
  {et~al.},\ }\href {https://doi.org/10.1088/0264-9381/24/17/r01} {\bibfield
  {journal} {\bibinfo  {journal} {Classical and Quantum Gravity}\ }\textbf
  {\bibinfo {volume} {24}},\ \bibinfo {pages} {R113} (\bibinfo {year}
  {2007})}\BibitemShut {NoStop}%
\bibitem [{\citenamefont {Berry}\ \emph {et~al.}(2019)\citenamefont {Berry},
  \citenamefont {Hughes}, \citenamefont {Sopuerta} \emph {et~al.}}]{lisaemri}%
  \BibitemOpen
  \bibfield  {author} {\bibinfo {author} {\bibfnamefont {C.~P.~L.}\
  \bibnamefont {Berry}}, \bibinfo {author} {\bibfnamefont {S.~A.}\ \bibnamefont
  {Hughes}}, \bibinfo {author} {\bibfnamefont {C.~F.}\ \bibnamefont
  {Sopuerta}}, \emph {et~al.},\ }\href@noop {} {\bibinfo {title} {The unique
  potential of extreme mass-ratio inspirals for gravitational-wave astronomy}}
  (\bibinfo {year} {2019}),\ \Eprint {https://arxiv.org/abs/1903.03686}
  {arXiv:1903.03686 [astro-ph.HE]} \BibitemShut {NoStop}%
\bibitem [{\citenamefont {Gair}\ \emph {et~al.}(2004)\citenamefont {Gair},
  \citenamefont {Barack}, \citenamefont {Creighton} \emph
  {et~al.}}]{Gair_2004}%
  \BibitemOpen
  \bibfield  {author} {\bibinfo {author} {\bibfnamefont {J.~R.}\ \bibnamefont
  {Gair}}, \bibinfo {author} {\bibfnamefont {L.}~\bibnamefont {Barack}},
  \bibinfo {author} {\bibfnamefont {T.}~\bibnamefont {Creighton}}, \emph
  {et~al.},\ }\href {https://doi.org/10.1088/0264-9381/21/20/003} {\bibfield
  {journal} {\bibinfo  {journal} {Classical and Quantum Gravity}\ }\textbf
  {\bibinfo {volume} {21}},\ \bibinfo {pages} {S1595} (\bibinfo {year}
  {2004})}\BibitemShut {NoStop}%
\bibitem [{\citenamefont {Babak}\ \emph {et~al.}(2017)\citenamefont {Babak},
  \citenamefont {Gair}, \citenamefont {Sesana} \emph {et~al.}}]{Babak_2017}%
  \BibitemOpen
  \bibfield  {author} {\bibinfo {author} {\bibfnamefont {S.}~\bibnamefont
  {Babak}}, \bibinfo {author} {\bibfnamefont {J.}~\bibnamefont {Gair}},
  \bibinfo {author} {\bibfnamefont {A.}~\bibnamefont {Sesana}}, \emph
  {et~al.},\ }\bibfield  {journal} {\bibinfo  {journal} {Physical Review D}\
  }\textbf {\bibinfo {volume} {95}},\ \href
  {https://doi.org/10.1103/physrevd.95.103012} {10.1103/physrevd.95.103012}
  (\bibinfo {year} {2017})\BibitemShut {NoStop}%
\bibitem [{\citenamefont {Babak}\ \emph {et~al.}(2010)\citenamefont {Babak},
  \citenamefont {Baker}, \citenamefont {Benacquista} \emph
  {et~al.}}]{Babak_2010}%
  \BibitemOpen
  \bibfield  {author} {\bibinfo {author} {\bibfnamefont {S.}~\bibnamefont
  {Babak}}, \bibinfo {author} {\bibfnamefont {J.~G.}\ \bibnamefont {Baker}},
  \bibinfo {author} {\bibfnamefont {M.~J.}\ \bibnamefont {Benacquista}}, \emph
  {et~al.},\ }\href {https://doi.org/10.1088/0264-9381/27/8/084009} {\bibfield
  {journal} {\bibinfo  {journal} {Classical and Quantum Gravity}\ }\textbf
  {\bibinfo {volume} {27}},\ \bibinfo {pages} {084009} (\bibinfo {year}
  {2010})}\BibitemShut {NoStop}%
\bibitem [{\citenamefont {Gair}\ \emph {et~al.}(2013)\citenamefont {Gair},
  \citenamefont {Vallisneri}, \citenamefont {Larson} \emph
  {et~al.}}]{Gair_Vallisneri_Larson_Baker_2013}%
  \BibitemOpen
  \bibfield  {author} {\bibinfo {author} {\bibfnamefont {J.~R.}\ \bibnamefont
  {Gair}}, \bibinfo {author} {\bibfnamefont {M.}~\bibnamefont {Vallisneri}},
  \bibinfo {author} {\bibfnamefont {S.~L.}\ \bibnamefont {Larson}}, \emph
  {et~al.},\ }\href {https://doi.org/10.12942/lrr-2013-7} {\bibfield  {journal}
  {\bibinfo  {journal} {Living Reviews in Relativity}\ }\textbf {\bibinfo
  {volume} {16}},\ \bibinfo {pages} {7} (\bibinfo {year} {2013})},\ \bibinfo
  {note} {arXiv:1212.5575 [gr-qc]}\BibitemShut {NoStop}%
\bibitem [{\citenamefont {Gair}\ \emph {et~al.}(2017)\citenamefont {Gair},
  \citenamefont {Babak}, \citenamefont {Sesana} \emph {et~al.}}]{Gair_2017}%
  \BibitemOpen
  \bibfield  {author} {\bibinfo {author} {\bibfnamefont {J.~R.}\ \bibnamefont
  {Gair}}, \bibinfo {author} {\bibfnamefont {S.}~\bibnamefont {Babak}},
  \bibinfo {author} {\bibfnamefont {A.}~\bibnamefont {Sesana}}, \emph
  {et~al.},\ }\href {https://doi.org/10.1088/1742-6596/840/1/012021} {\bibfield
   {journal} {\bibinfo  {journal} {Journal of Physics: Conference Series}\
  }\textbf {\bibinfo {volume} {840}},\ \bibinfo {pages} {012021} (\bibinfo
  {year} {2017})}\BibitemShut {NoStop}%
\bibitem [{\citenamefont {Hui}\ \emph {et~al.}(2017)\citenamefont {Hui},
  \citenamefont {Ostriker}, \citenamefont {Tremaine},\ and\ \citenamefont
  {Witten}}]{Hui_2017}%
  \BibitemOpen
  \bibfield  {author} {\bibinfo {author} {\bibfnamefont {L.}~\bibnamefont
  {Hui}}, \bibinfo {author} {\bibfnamefont {J.~P.}\ \bibnamefont {Ostriker}},
  \bibinfo {author} {\bibfnamefont {S.}~\bibnamefont {Tremaine}},\ and\
  \bibinfo {author} {\bibfnamefont {E.}~\bibnamefont {Witten}},\ }\href
  {https://doi.org/10.1103/physrevd.95.043541} {\bibfield  {journal} {\bibinfo
  {journal} {Physical Review D}\ }\textbf {\bibinfo {volume} {95}},\ \bibinfo
  {pages} {043541} (\bibinfo {year} {2017})},\ \bibinfo {note}
  {arXiv:1610.08297 [astro-ph, physics:hep-ph, physics:hep-th]}\BibitemShut
  {NoStop}%
\bibitem [{\citenamefont {Miller}\ \emph {et~al.}(2021)\citenamefont {Miller},
  \citenamefont {Clesse}, \citenamefont {De~Lillo} \emph
  {et~al.}}]{Miller_Clesse_De}%
  \BibitemOpen
  \bibfield  {author} {\bibinfo {author} {\bibfnamefont {A.~L.}\ \bibnamefont
  {Miller}}, \bibinfo {author} {\bibfnamefont {S.}~\bibnamefont {Clesse}},
  \bibinfo {author} {\bibfnamefont {F.}~\bibnamefont {De~Lillo}}, \emph
  {et~al.},\ }\href {https://doi.org/10.1016/j.dark.2021.100836} {\bibfield
  {journal} {\bibinfo  {journal} {Physics of the Dark Universe}\ }\textbf
  {\bibinfo {volume} {32}},\ \bibinfo {pages} {100836} (\bibinfo {year}
  {2021})},\ \bibinfo {note} {arXiv:2012.12983 [astro-ph,
  physics:gr-qc]}\BibitemShut {NoStop}%
\bibitem [{\citenamefont {Chen}\ \emph {et~al.}(2017)\citenamefont {Chen},
  \citenamefont {Schive},\ and\ \citenamefont
  {Chiueh}}]{Chen_Schive_Chiueh_2017}%
  \BibitemOpen
  \bibfield  {author} {\bibinfo {author} {\bibfnamefont {S.-R.}\ \bibnamefont
  {Chen}}, \bibinfo {author} {\bibfnamefont {H.-Y.}\ \bibnamefont {Schive}},\
  and\ \bibinfo {author} {\bibfnamefont {T.}~\bibnamefont {Chiueh}},\ }\href
  {https://doi.org/10.1093/mnras/stx449} {\bibfield  {journal} {\bibinfo
  {journal} {Monthly Notices of the Royal Astronomical Society}\ }\textbf
  {\bibinfo {volume} {468}},\ \bibinfo {pages} {1338–1348} (\bibinfo {year}
  {2017})},\ \bibinfo {note} {arXiv:1606.09030 [astro-ph]}\BibitemShut
  {NoStop}%
\bibitem [{\citenamefont {Nelson}\ and\ \citenamefont
  {Scholtz}(2011)}]{Nelson_Scholtz_2011}%
  \BibitemOpen
  \bibfield  {author} {\bibinfo {author} {\bibfnamefont {A.~E.}\ \bibnamefont
  {Nelson}}\ and\ \bibinfo {author} {\bibfnamefont {J.}~\bibnamefont
  {Scholtz}},\ }\href {https://doi.org/10.1103/PhysRevD.84.103501} {\bibfield
  {journal} {\bibinfo  {journal} {Physical Review D}\ }\textbf {\bibinfo
  {volume} {84}},\ \bibinfo {pages} {103501} (\bibinfo {year} {2011})},\
  \bibinfo {note} {arXiv:1105.2812 [astro-ph, physics:hep-ph]}\BibitemShut
  {NoStop}%
\bibitem [{\citenamefont {Goodsell}\ \emph {et~al.}(2009)\citenamefont
  {Goodsell}, \citenamefont {Jaeckel}, \citenamefont {Redondo} \emph
  {et~al.}}]{Goodsell_Jaeckel_Redondo_Ringwald_2009}%
  \BibitemOpen
  \bibfield  {author} {\bibinfo {author} {\bibfnamefont {M.}~\bibnamefont
  {Goodsell}}, \bibinfo {author} {\bibfnamefont {J.}~\bibnamefont {Jaeckel}},
  \bibinfo {author} {\bibfnamefont {J.}~\bibnamefont {Redondo}}, \emph
  {et~al.},\ }\href {https://doi.org/10.1088/1126-6708/2009/11/027} {\bibfield
  {journal} {\bibinfo  {journal} {Journal of High Energy Physics}\ }\textbf
  {\bibinfo {volume} {2009}},\ \bibinfo {pages} {027–027} (\bibinfo {year}
  {2009})},\ \bibinfo {note} {arXiv:0909.0515 [hep-ph,
  physics:hep-th]}\BibitemShut {NoStop}%
\bibitem [{\citenamefont {Co}\ \emph {et~al.}(2019)\citenamefont {Co},
  \citenamefont {Pierce}, \citenamefont {Zhang} \emph {et~al.}}]{Co_2019}%
  \BibitemOpen
  \bibfield  {author} {\bibinfo {author} {\bibfnamefont {R.~T.}\ \bibnamefont
  {Co}}, \bibinfo {author} {\bibfnamefont {A.}~\bibnamefont {Pierce}}, \bibinfo
  {author} {\bibfnamefont {Z.}~\bibnamefont {Zhang}}, \emph {et~al.},\
  }\bibfield  {journal} {\bibinfo  {journal} {Physical Review D}\ }\textbf
  {\bibinfo {volume} {99}},\ \href {https://doi.org/10.1103/physrevd.99.075002}
  {10.1103/physrevd.99.075002} (\bibinfo {year} {2019})\BibitemShut {NoStop}%
\bibitem [{\citenamefont {Graham}\ \emph {et~al.}(2016)\citenamefont {Graham},
  \citenamefont {Mardon},\ and\ \citenamefont {Rajendran}}]{Graham_2016}%
  \BibitemOpen
  \bibfield  {author} {\bibinfo {author} {\bibfnamefont {P.~W.}\ \bibnamefont
  {Graham}}, \bibinfo {author} {\bibfnamefont {J.}~\bibnamefont {Mardon}},\
  and\ \bibinfo {author} {\bibfnamefont {S.}~\bibnamefont {Rajendran}},\
  }\bibfield  {journal} {\bibinfo  {journal} {Physical Review D}\ }\textbf
  {\bibinfo {volume} {93}},\ \href {https://doi.org/10.1103/physrevd.93.103520}
  {10.1103/physrevd.93.103520} (\bibinfo {year} {2016})\BibitemShut {NoStop}%
\bibitem [{\citenamefont {Su}\ \emph {et~al.}(1994)\citenamefont {Su},
  \citenamefont {Heckel}, \citenamefont {Adelberger} \emph
  {et~al.}}]{PhysRevD.50.3614}%
  \BibitemOpen
  \bibfield  {author} {\bibinfo {author} {\bibfnamefont {Y.}~\bibnamefont
  {Su}}, \bibinfo {author} {\bibfnamefont {B.~R.}\ \bibnamefont {Heckel}},
  \bibinfo {author} {\bibfnamefont {E.~G.}\ \bibnamefont {Adelberger}}, \emph
  {et~al.},\ }\href {https://doi.org/10.1103/PhysRevD.50.3614} {\bibfield
  {journal} {\bibinfo  {journal} {Phys. Rev. D}\ }\textbf {\bibinfo {volume}
  {50}},\ \bibinfo {pages} {3614} (\bibinfo {year} {1994})}\BibitemShut
  {NoStop}%
\bibitem [{\citenamefont {Schlamminger}\ \emph {et~al.}(2008)\citenamefont
  {Schlamminger}, \citenamefont {Choi}, \citenamefont {Wagner} \emph
  {et~al.}}]{Schlamminger_2008}%
  \BibitemOpen
  \bibfield  {author} {\bibinfo {author} {\bibfnamefont {S.}~\bibnamefont
  {Schlamminger}}, \bibinfo {author} {\bibfnamefont {K.-Y.}\ \bibnamefont
  {Choi}}, \bibinfo {author} {\bibfnamefont {T.~A.}\ \bibnamefont {Wagner}},
  \emph {et~al.},\ }\bibfield  {journal} {\bibinfo  {journal} {Physical Review
  Letters}\ }\textbf {\bibinfo {volume} {100}},\ \href
  {https://doi.org/10.1103/physrevlett.100.041101}
  {10.1103/physrevlett.100.041101} (\bibinfo {year} {2008})\BibitemShut
  {NoStop}%
\bibitem [{\citenamefont {Williams}\ \emph {et~al.}(2004)\citenamefont
  {Williams}, \citenamefont {Turyshev},\ and\ \citenamefont
  {Boggs}}]{PhysRevLett.93.261101}%
  \BibitemOpen
  \bibfield  {author} {\bibinfo {author} {\bibfnamefont {J.~G.}\ \bibnamefont
  {Williams}}, \bibinfo {author} {\bibfnamefont {S.~G.}\ \bibnamefont
  {Turyshev}},\ and\ \bibinfo {author} {\bibfnamefont {D.~H.}\ \bibnamefont
  {Boggs}},\ }\href {https://doi.org/10.1103/PhysRevLett.93.261101} {\bibfield
  {journal} {\bibinfo  {journal} {Phys. Rev. Lett.}\ }\textbf {\bibinfo
  {volume} {93}},\ \bibinfo {pages} {261101} (\bibinfo {year}
  {2004})}\BibitemShut {NoStop}%
\bibitem [{\citenamefont {TURYSHEV}\ and\ \citenamefont
  {WILLIAMS}(2007)}]{TURYSHEV_2007}%
  \BibitemOpen
  \bibfield  {author} {\bibinfo {author} {\bibfnamefont {S.~G.}\ \bibnamefont
  {TURYSHEV}}\ and\ \bibinfo {author} {\bibfnamefont {J.~G.}\ \bibnamefont
  {WILLIAMS}},\ }\href {https://doi.org/10.1142/s0218271807011838} {\bibfield
  {journal} {\bibinfo  {journal} {International Journal of Modern Physics D}\
  }\textbf {\bibinfo {volume} {16}},\ \bibinfo {pages} {2165} (\bibinfo {year}
  {2007})}\BibitemShut {NoStop}%
\bibitem [{\citenamefont {Baryakhtar}\ \emph {et~al.}(2017)\citenamefont
  {Baryakhtar}, \citenamefont {Lasenby},\ and\ \citenamefont
  {Teo}}]{Baryakhtar_Lasenby_Teo_2017}%
  \BibitemOpen
  \bibfield  {author} {\bibinfo {author} {\bibfnamefont {M.}~\bibnamefont
  {Baryakhtar}}, \bibinfo {author} {\bibfnamefont {R.}~\bibnamefont
  {Lasenby}},\ and\ \bibinfo {author} {\bibfnamefont {M.}~\bibnamefont {Teo}},\
  }\href {https://doi.org/10.1103/PhysRevD.96.035019} {\bibfield  {journal}
  {\bibinfo  {journal} {Physical Review D}\ }\textbf {\bibinfo {volume} {96}},\
  \bibinfo {pages} {035019} (\bibinfo {year} {2017})},\ \bibinfo {note} {arXiv:
  1704.05081}\BibitemShut {NoStop}%
\bibitem [{\citenamefont {East}\ and\ \citenamefont
  {Pretorius}(2017)}]{East_Pretorius_2017}%
  \BibitemOpen
  \bibfield  {author} {\bibinfo {author} {\bibfnamefont {W.~E.}\ \bibnamefont
  {East}}\ and\ \bibinfo {author} {\bibfnamefont {F.}~\bibnamefont
  {Pretorius}},\ }\href {https://doi.org/10.1103/PhysRevLett.119.041101}
  {\bibfield  {journal} {\bibinfo  {journal} {Physical Review Letters}\
  }\textbf {\bibinfo {volume} {119}},\ \bibinfo {pages} {041101} (\bibinfo
  {year} {2017})}\BibitemShut {NoStop}%
\bibitem [{\citenamefont {East}(2017)}]{East_2017}%
  \BibitemOpen
  \bibfield  {author} {\bibinfo {author} {\bibfnamefont {W.~E.}\ \bibnamefont
  {East}},\ }\href {https://doi.org/10.1103/PhysRevD.96.024004} {\bibfield
  {journal} {\bibinfo  {journal} {Physical Review D}\ }\textbf {\bibinfo
  {volume} {96}},\ \bibinfo {pages} {024004} (\bibinfo {year} {2017})},\
  \bibinfo {note} {arXiv:1705.01544 [astro-ph, physics:gr-qc,
  physics:hep-ph]}\BibitemShut {NoStop}%
\bibitem [{\citenamefont {Pierce}\ \emph {et~al.}(2018)\citenamefont {Pierce},
  \citenamefont {Riles},\ and\ \citenamefont {Zhao}}]{Pierce_Riles_Zhao_2018}%
  \BibitemOpen
  \bibfield  {author} {\bibinfo {author} {\bibfnamefont {A.}~\bibnamefont
  {Pierce}}, \bibinfo {author} {\bibfnamefont {K.}~\bibnamefont {Riles}},\ and\
  \bibinfo {author} {\bibfnamefont {Y.}~\bibnamefont {Zhao}},\ }\href
  {https://doi.org/10.1103/PhysRevLett.121.061102} {\bibfield  {journal}
  {\bibinfo  {journal} {Physical Review Letters}\ }\textbf {\bibinfo {volume}
  {121}},\ \bibinfo {pages} {061102} (\bibinfo {year} {2018})},\ \bibinfo
  {note} {arXiv:1801.10161 [astro-ph, physics:gr-qc,
  physics:hep-ph]}\BibitemShut {NoStop}%
\bibitem [{\citenamefont {Maselli}\ \emph {et~al.}(2020)\citenamefont
  {Maselli}, \citenamefont {Franchini}, \citenamefont {Gualtieri},\ and\
  \citenamefont {Sotiriou}}]{Maselli_Franchini_Gualtieri_Sotiriou_2020}%
  \BibitemOpen
  \bibfield  {author} {\bibinfo {author} {\bibfnamefont {A.}~\bibnamefont
  {Maselli}}, \bibinfo {author} {\bibfnamefont {N.}~\bibnamefont {Franchini}},
  \bibinfo {author} {\bibfnamefont {L.}~\bibnamefont {Gualtieri}},\ and\
  \bibinfo {author} {\bibfnamefont {T.~P.}\ \bibnamefont {Sotiriou}},\ }\href
  {https://doi.org/10.1103/PhysRevLett.125.141101} {\bibfield  {journal}
  {\bibinfo  {journal} {Physical Review Letters}\ }\textbf {\bibinfo {volume}
  {125}},\ \bibinfo {pages} {141101} (\bibinfo {year} {2020})},\ \bibinfo
  {note} {arXiv:2004.11895 [astro-ph, physics:gr-qc,
  physics:hep-th]}\BibitemShut {NoStop}%
\bibitem [{\citenamefont {Maselli}\ \emph {et~al.}(2022)\citenamefont
  {Maselli}, \citenamefont {Franchini}, \citenamefont {Gualtieri},
  \citenamefont {Sotiriou}, \citenamefont {Barsanti},\ and\ \citenamefont
  {Pani}}]{Maselli_Franchini_Gualtieri_Sotiriou_Barsanti_Pani_2022}%
  \BibitemOpen
  \bibfield  {author} {\bibinfo {author} {\bibfnamefont {A.}~\bibnamefont
  {Maselli}}, \bibinfo {author} {\bibfnamefont {N.}~\bibnamefont {Franchini}},
  \bibinfo {author} {\bibfnamefont {L.}~\bibnamefont {Gualtieri}}, \bibinfo
  {author} {\bibfnamefont {T.~P.}\ \bibnamefont {Sotiriou}}, \bibinfo {author}
  {\bibfnamefont {S.}~\bibnamefont {Barsanti}},\ and\ \bibinfo {author}
  {\bibfnamefont {P.}~\bibnamefont {Pani}},\ }\href
  {https://doi.org/10.1038/s41550-021-01589-5} {\bibfield  {journal} {\bibinfo
  {journal} {Nature Astronomy}\ }\textbf {\bibinfo {volume} {6}},\ \bibinfo
  {pages} {464–470} (\bibinfo {year} {2022})},\ \bibinfo {note}
  {arXiv:2106.11325 [astro-ph, physics:gr-qc, physics:hep-th]}\BibitemShut
  {NoStop}%
\bibitem [{\citenamefont {Siemonsen}\ \emph {et~al.}(2022)\citenamefont
  {Siemonsen}, \citenamefont {May},\ and\ \citenamefont {East}}]{superrad}%
  \BibitemOpen
  \bibfield  {author} {\bibinfo {author} {\bibfnamefont {N.}~\bibnamefont
  {Siemonsen}}, \bibinfo {author} {\bibfnamefont {T.}~\bibnamefont {May}},\
  and\ \bibinfo {author} {\bibfnamefont {W.~E.}\ \bibnamefont {East}},\ }\href
  {https://doi.org/10.48550/ARXIV.2211.03845} {\bibinfo {title} {Superrad: A
  black hole superradiance gravitational waveform model}} (\bibinfo {year}
  {2022})\BibitemShut {NoStop}%
\bibitem [{\citenamefont {Baumann}\ \emph {et~al.}(2020)\citenamefont
  {Baumann}, \citenamefont {Chia}, \citenamefont {Porto} \emph
  {et~al.}}]{Baumann_Chia_Porto_Stout_2020}%
  \BibitemOpen
  \bibfield  {author} {\bibinfo {author} {\bibfnamefont {D.}~\bibnamefont
  {Baumann}}, \bibinfo {author} {\bibfnamefont {H.~S.}\ \bibnamefont {Chia}},
  \bibinfo {author} {\bibfnamefont {R.~A.}\ \bibnamefont {Porto}}, \emph
  {et~al.},\ }\href {https://doi.org/10.1103/PhysRevD.101.083019} {\bibfield
  {journal} {\bibinfo  {journal} {Physical Review D}\ }\textbf {\bibinfo
  {volume} {101}},\ \bibinfo {pages} {083019} (\bibinfo {year} {2020})},\
  \bibinfo {note} {arXiv:1912.04932 [gr-qc, physics:hep-ph,
  physics:hep-th]}\BibitemShut {NoStop}%
\bibitem [{\citenamefont {Frolov}\ \emph {et~al.}(2018)\citenamefont {Frolov},
  \citenamefont {Krtou\v{s}}, \citenamefont {Kubiz\v{n}\'{a}k} \emph
  {et~al.}}]{fkks}%
  \BibitemOpen
  \bibfield  {author} {\bibinfo {author} {\bibfnamefont {V.~P.}\ \bibnamefont
  {Frolov}}, \bibinfo {author} {\bibfnamefont {P.}~\bibnamefont {Krtou\v{s}}},
  \bibinfo {author} {\bibfnamefont {D.}~\bibnamefont {Kubiz\v{n}\'{a}k}}, \emph
  {et~al.},\ }\href {https://doi.org/10.1103/PhysRevLett.120.231103} {\bibfield
   {journal} {\bibinfo  {journal} {Physical Review Letters}\ }\textbf {\bibinfo
  {volume} {120}},\ \bibinfo {pages} {231103} (\bibinfo {year} {2018})},\
  \bibinfo {note} {arXiv: 1804.00030}\BibitemShut {NoStop}%
\bibitem [{\citenamefont {Dolan}(2018)}]{Dolan_2018}%
  \BibitemOpen
  \bibfield  {author} {\bibinfo {author} {\bibfnamefont {S.~R.}\ \bibnamefont
  {Dolan}},\ }\bibfield  {journal} {\bibinfo  {journal} {Physical Review D}\
  }\textbf {\bibinfo {volume} {98}},\ \href
  {https://doi.org/10.1103/physrevd.98.104006} {10.1103/physrevd.98.104006}
  (\bibinfo {year} {2018})\BibitemShut {NoStop}%
\bibitem [{\citenamefont {Baumann}\ \emph
  {et~al.}(2019{\natexlab{a}})\citenamefont {Baumann}, \citenamefont {Chia},\
  and\ \citenamefont {Porto}}]{Baumann_Chia_Porto_2019}%
  \BibitemOpen
  \bibfield  {author} {\bibinfo {author} {\bibfnamefont {D.}~\bibnamefont
  {Baumann}}, \bibinfo {author} {\bibfnamefont {H.~S.}\ \bibnamefont {Chia}},\
  and\ \bibinfo {author} {\bibfnamefont {R.~A.}\ \bibnamefont {Porto}},\ }\href
  {https://doi.org/10.1103/PhysRevD.99.044001} {\bibfield  {journal} {\bibinfo
  {journal} {Physical Review D}\ }\textbf {\bibinfo {volume} {99}},\ \bibinfo
  {pages} {044001} (\bibinfo {year} {2019}{\natexlab{a}})},\ \bibinfo {note}
  {arXiv:1804.03208 [astro-ph, physics:gr-qc, physics:hep-ph,
  physics:hep-th]}\BibitemShut {NoStop}%
\bibitem [{\citenamefont {Baumann}\ \emph
  {et~al.}(2019{\natexlab{b}})\citenamefont {Baumann}, \citenamefont {Chia},
  \citenamefont {Stout} \emph {et~al.}}]{gravatom}%
  \BibitemOpen
  \bibfield  {author} {\bibinfo {author} {\bibfnamefont {D.}~\bibnamefont
  {Baumann}}, \bibinfo {author} {\bibfnamefont {H.~S.}\ \bibnamefont {Chia}},
  \bibinfo {author} {\bibfnamefont {J.}~\bibnamefont {Stout}}, \emph {et~al.},\
  }\href {https://doi.org/10.1088/1475-7516/2019/12/006} {\bibfield  {journal}
  {\bibinfo  {journal} {Journal of Cosmology and Astroparticle Physics}\
  }\textbf {\bibinfo {volume} {2019}}\bibfield  {number} {\bibinfo  {number} {
  (12)},\ \bibinfo {pages} {006–006}},\ }\bibinfo {note} {arXiv:
  1908.10370}\BibitemShut {NoStop}%
\bibitem [{\citenamefont {Siemonsen}\ and\ \citenamefont
  {East}(2020)}]{Siemonsen_2020}%
  \BibitemOpen
  \bibfield  {author} {\bibinfo {author} {\bibfnamefont {N.}~\bibnamefont
  {Siemonsen}}\ and\ \bibinfo {author} {\bibfnamefont {W.~E.}\ \bibnamefont
  {East}},\ }\bibfield  {journal} {\bibinfo  {journal} {Physical Review D}\
  }\textbf {\bibinfo {volume} {101}},\ \href
  {https://doi.org/10.1103/physrevd.101.024019} {10.1103/physrevd.101.024019}
  (\bibinfo {year} {2020})\BibitemShut {NoStop}%
\bibitem [{\citenamefont {Inc.}(2023)}]{Mathematica}%
  \BibitemOpen
  \bibfield  {author} {\bibinfo {author} {\bibfnamefont {W.~R.}\ \bibnamefont
  {Inc.}},\ }\href {https://www.wolfram.com/mathematica} {\bibinfo {title}
  {Mathematica, {V}ersion 13.2}} (\bibinfo {year} {2023}),\ \bibinfo {note}
  {champaign, IL, 2022}\BibitemShut {NoStop}%
\bibitem [{\citenamefont {{Teukolsky}}\ and\ \citenamefont
  {{Press}}(1974)}]{Teukolsky3}%
  \BibitemOpen
  \bibfield  {author} {\bibinfo {author} {\bibfnamefont {S.~A.}\ \bibnamefont
  {{Teukolsky}}}\ and\ \bibinfo {author} {\bibfnamefont {W.~H.}\ \bibnamefont
  {{Press}}},\ }\href {https://doi.org/10.1086/153180} {\bibfield  {journal}
  {\bibinfo  {journal} {\apj}\ }\textbf {\bibinfo {volume} {193}},\ \bibinfo
  {pages} {443} (\bibinfo {year} {1974})}\BibitemShut {NoStop}%
\bibitem [{\citenamefont {Brito}\ \emph {et~al.}(2015)\citenamefont {Brito},
  \citenamefont {Cardoso},\ and\ \citenamefont
  {Pani}}]{Brito_Cardoso_Pani_2015}%
  \BibitemOpen
  \bibfield  {author} {\bibinfo {author} {\bibfnamefont {R.}~\bibnamefont
  {Brito}}, \bibinfo {author} {\bibfnamefont {V.}~\bibnamefont {Cardoso}},\
  and\ \bibinfo {author} {\bibfnamefont {P.}~\bibnamefont {Pani}},\ }\href
  {https://doi.org/10.1007/978-3-319-19000-6} {\emph {\bibinfo {title}
  {Superradiance}}},\ \bibinfo {series} {Lecture Notes in Physics}, Vol.\
  \bibinfo {volume} {906}\ (\bibinfo  {publisher} {Springer International
  Publishing},\ \bibinfo {address} {Cham},\ \bibinfo {year} {2015})\BibitemShut
  {NoStop}%
\bibitem [{\citenamefont {Unruh}(1974)}]{unruh}%
  \BibitemOpen
  \bibfield  {author} {\bibinfo {author} {\bibfnamefont {W.~G.}\ \bibnamefont
  {Unruh}},\ }\href {https://doi.org/10.1103/PhysRevD.10.3194} {\bibfield
  {journal} {\bibinfo  {journal} {Phys. Rev. D}\ }\textbf {\bibinfo {volume}
  {10}},\ \bibinfo {pages} {3194} (\bibinfo {year} {1974})}\BibitemShut
  {NoStop}%
\bibitem [{\citenamefont {Starobinskii}(1973)}]{starobinsky}%
  \BibitemOpen
  \bibfield  {author} {\bibinfo {author} {\bibfnamefont {A.~A.}\ \bibnamefont
  {Starobinskii}},\ }\href {https://www.osti.gov/biblio/4580095} {\bibfield
  {journal} {\bibinfo  {journal} {Zh. Eksp. Teor. Fiz. 64: No. 1, 48-57(Jan
  1973).}\ } (\bibinfo {year} {1973})}\BibitemShut {NoStop}%
\bibitem [{\citenamefont {Balakumar}\ \emph {et~al.}(2020)\citenamefont
  {Balakumar}, \citenamefont {Winstanley}, \citenamefont {Bernar} \emph
  {et~al.}}]{BALAKUMAR2020135904}%
  \BibitemOpen
  \bibfield  {author} {\bibinfo {author} {\bibfnamefont {V.}~\bibnamefont
  {Balakumar}}, \bibinfo {author} {\bibfnamefont {E.}~\bibnamefont
  {Winstanley}}, \bibinfo {author} {\bibfnamefont {R.~P.}\ \bibnamefont
  {Bernar}}, \emph {et~al.},\ }\href
  {https://doi.org/https://doi.org/10.1016/j.physletb.2020.135904} {\bibfield
  {journal} {\bibinfo  {journal} {Physics Letters B}\ }\textbf {\bibinfo
  {volume} {811}},\ \bibinfo {pages} {135904} (\bibinfo {year}
  {2020})}\BibitemShut {NoStop}%
\bibitem [{\citenamefont {Richartz}\ \emph {et~al.}(2009)\citenamefont
  {Richartz}, \citenamefont {Weinfurtner}, \citenamefont {Penner},\ and\
  \citenamefont {Unruh}}]{richartz2009generalised}%
  \BibitemOpen
  \bibfield  {author} {\bibinfo {author} {\bibfnamefont {M.}~\bibnamefont
  {Richartz}}, \bibinfo {author} {\bibfnamefont {S.}~\bibnamefont
  {Weinfurtner}}, \bibinfo {author} {\bibfnamefont {A.~J.}\ \bibnamefont
  {Penner}},\ and\ \bibinfo {author} {\bibfnamefont {W.~G.}\ \bibnamefont
  {Unruh}},\ }\href@noop {} {\bibinfo {title} {Generalised superradiant
  scattering}} (\bibinfo {year} {2009}),\ \Eprint
  {https://arxiv.org/abs/0909.2317} {arXiv:0909.2317 [gr-qc]} \BibitemShut
  {NoStop}%
\bibitem [{\citenamefont {Berti}\ \emph
  {et~al.}(2019{\natexlab{b}})\citenamefont {Berti}, \citenamefont {Brito},
  \citenamefont {Macedo} \emph {et~al.}}]{Berti_Brito_Macedo_Raposo_Rosa_2019}%
  \BibitemOpen
  \bibfield  {author} {\bibinfo {author} {\bibfnamefont {E.}~\bibnamefont
  {Berti}}, \bibinfo {author} {\bibfnamefont {R.}~\bibnamefont {Brito}},
  \bibinfo {author} {\bibfnamefont {C.~F.~B.}\ \bibnamefont {Macedo}}, \emph
  {et~al.},\ }\href {https://doi.org/10.1103/PhysRevD.99.104039} {\bibfield
  {journal} {\bibinfo  {journal} {Physical Review D}\ }\textbf {\bibinfo
  {volume} {99}},\ \bibinfo {pages} {104039} (\bibinfo {year}
  {2019}{\natexlab{b}})},\ \bibinfo {note} {arXiv:1904.03131
  [gr-qc]}\BibitemShut {NoStop}%
\bibitem [{\citenamefont {Traykova}\ \emph {et~al.}(2021)\citenamefont
  {Traykova}, \citenamefont {Clough}, \citenamefont {Helfer}, \citenamefont
  {Berti}, \citenamefont {Ferreira},\ and\ \citenamefont
  {Hui}}]{Traykova_2021}%
  \BibitemOpen
  \bibfield  {author} {\bibinfo {author} {\bibfnamefont {D.}~\bibnamefont
  {Traykova}}, \bibinfo {author} {\bibfnamefont {K.}~\bibnamefont {Clough}},
  \bibinfo {author} {\bibfnamefont {T.}~\bibnamefont {Helfer}}, \bibinfo
  {author} {\bibfnamefont {E.}~\bibnamefont {Berti}}, \bibinfo {author}
  {\bibfnamefont {P.~G.}\ \bibnamefont {Ferreira}},\ and\ \bibinfo {author}
  {\bibfnamefont {L.}~\bibnamefont {Hui}},\ }\bibfield  {journal} {\bibinfo
  {journal} {Physical Review D}\ }\textbf {\bibinfo {volume} {104}},\ \href
  {https://doi.org/10.1103/physrevd.104.103014} {10.1103/physrevd.104.103014}
  (\bibinfo {year} {2021}),\ \bibinfo {note} {arXiv: 2106.08280}\BibitemShut
  {NoStop}%
\bibitem [{\citenamefont {Katz}\ \emph {et~al.}(2021)\citenamefont {Katz},
  \citenamefont {Chua}, \citenamefont {Speri} \emph {et~al.}}]{few}%
  \BibitemOpen
  \bibfield  {author} {\bibinfo {author} {\bibfnamefont {M.~L.}\ \bibnamefont
  {Katz}}, \bibinfo {author} {\bibfnamefont {A.~J.}\ \bibnamefont {Chua}},
  \bibinfo {author} {\bibfnamefont {L.}~\bibnamefont {Speri}}, \emph {et~al.},\
  }\bibfield  {journal} {\bibinfo  {journal} {Physical Review D}\ }\textbf
  {\bibinfo {volume} {104}},\ \href
  {https://doi.org/10.1103/physrevd.104.064047} {10.1103/physrevd.104.064047}
  (\bibinfo {year} {2021})\BibitemShut {NoStop}%
\bibitem [{\citenamefont {Peters}\ and\ \citenamefont
  {Mathews}(1963)}]{PeterMatthew}%
  \BibitemOpen
  \bibfield  {author} {\bibinfo {author} {\bibfnamefont {P.~C.}\ \bibnamefont
  {Peters}}\ and\ \bibinfo {author} {\bibfnamefont {J.}~\bibnamefont
  {Mathews}},\ }\href {https://doi.org/10.1103/PhysRev.131.435} {\bibfield
  {journal} {\bibinfo  {journal} {Phys. Rev.}\ }\textbf {\bibinfo {volume}
  {131}},\ \bibinfo {pages} {435} (\bibinfo {year} {1963})}\BibitemShut
  {NoStop}%
\bibitem [{\citenamefont {Yoshino}\ and\ \citenamefont
  {Kodama}(2014)}]{Yoshino_Kodama_2014}%
  \BibitemOpen
  \bibfield  {author} {\bibinfo {author} {\bibfnamefont {H.}~\bibnamefont
  {Yoshino}}\ and\ \bibinfo {author} {\bibfnamefont {H.}~\bibnamefont
  {Kodama}},\ }\bibfield  {journal} {\bibinfo  {journal} {Progress of
  Theoretical and Experimental Physics}\ }\textbf {\bibinfo {volume} {2014}},\
  \href {https://doi.org/10.1093/ptep/ptu029} {10.1093/ptep/ptu029} (\bibinfo
  {year} {2014}),\ \bibinfo {note} {arXiv:1312.2326 [gr-qc, physics:hep-ph,
  physics:hep-th]}\BibitemShut {NoStop}%
\bibitem [{\citenamefont {East}(2018)}]{East_2018}%
  \BibitemOpen
  \bibfield  {author} {\bibinfo {author} {\bibfnamefont {W.~E.}\ \bibnamefont
  {East}},\ }\href {https://doi.org/10.1103/PhysRevLett.121.131104} {\bibfield
  {journal} {\bibinfo  {journal} {Physical Review Letters}\ }\textbf {\bibinfo
  {volume} {121}},\ \bibinfo {pages} {131104} (\bibinfo {year} {2018})},\
  \bibinfo {note} {arXiv:1807.00043 [astro-ph, physics:gr-qc,
  physics:hep-ph]}\BibitemShut {NoStop}%
\bibitem [{\citenamefont {Pani}\ \emph {et~al.}(2012)\citenamefont {Pani},
  \citenamefont {Cardoso}, \citenamefont {Gualtieri} \emph
  {et~al.}}]{Pani_Cardoso_Gualtieri_Berti_Ishibashi_2012}%
  \BibitemOpen
  \bibfield  {author} {\bibinfo {author} {\bibfnamefont {P.}~\bibnamefont
  {Pani}}, \bibinfo {author} {\bibfnamefont {V.}~\bibnamefont {Cardoso}},
  \bibinfo {author} {\bibfnamefont {L.}~\bibnamefont {Gualtieri}}, \emph
  {et~al.},\ }\href {https://doi.org/10.1103/PhysRevD.86.104017} {\bibfield
  {journal} {\bibinfo  {journal} {Physical Review D}\ }\textbf {\bibinfo
  {volume} {86}},\ \bibinfo {pages} {104017} (\bibinfo {year} {2012})},\
  \bibinfo {note} {arXiv: 1209.0773}\BibitemShut {NoStop}%
\bibitem [{\citenamefont {Robson}\ \emph {et~al.}(2019)\citenamefont {Robson},
  \citenamefont {Cornish},\ and\ \citenamefont {Liu}}]{lisapsd}%
  \BibitemOpen
  \bibfield  {author} {\bibinfo {author} {\bibfnamefont {T.}~\bibnamefont
  {Robson}}, \bibinfo {author} {\bibfnamefont {N.~J.}\ \bibnamefont
  {Cornish}},\ and\ \bibinfo {author} {\bibfnamefont {C.}~\bibnamefont {Liu}},\
  }\href {https://doi.org/10.1088/1361-6382/ab1101} {\bibfield  {journal}
  {\bibinfo  {journal} {Classical and Quantum Gravity}\ }\textbf {\bibinfo
  {volume} {36}},\ \bibinfo {pages} {105011} (\bibinfo {year}
  {2019})}\BibitemShut {NoStop}%
\bibitem [{\citenamefont {Finn}(1992)}]{Finn_1992}%
  \BibitemOpen
  \bibfield  {author} {\bibinfo {author} {\bibfnamefont {L.~S.}\ \bibnamefont
  {Finn}},\ }\href {https://doi.org/10.1103/PhysRevD.46.5236} {\bibfield
  {journal} {\bibinfo  {journal} {Physical Review D}\ }\textbf {\bibinfo
  {volume} {46}},\ \bibinfo {pages} {5236–5249} (\bibinfo {year} {1992})},\
  \bibinfo {note} {arXiv:gr-qc/9209010}\BibitemShut {NoStop}%
\bibitem [{\citenamefont {Damour}\ \emph {et~al.}(1998)\citenamefont {Damour},
  \citenamefont {Iyer},\ and\ \citenamefont
  {Sathyaprakash}}]{Damour_Iyer_Sathyaprakash_1998}%
  \BibitemOpen
  \bibfield  {author} {\bibinfo {author} {\bibfnamefont {T.}~\bibnamefont
  {Damour}}, \bibinfo {author} {\bibfnamefont {B.~R.}\ \bibnamefont {Iyer}},\
  and\ \bibinfo {author} {\bibfnamefont {B.~S.}\ \bibnamefont
  {Sathyaprakash}},\ }\href {https://doi.org/10.1103/PhysRevD.57.885}
  {\bibfield  {journal} {\bibinfo  {journal} {Physical Review D}\ }\textbf
  {\bibinfo {volume} {57}},\ \bibinfo {pages} {885–907} (\bibinfo {year}
  {1998})},\ \bibinfo {note} {arXiv:gr-qc/9708034}\BibitemShut {NoStop}%
\bibitem [{\citenamefont {Harry}\ \emph {et~al.}(2016)\citenamefont {Harry},
  \citenamefont {Privitera}, \citenamefont {Boh\'{e}} \emph {et~al.}}]{HPBB}%
  \BibitemOpen
  \bibfield  {author} {\bibinfo {author} {\bibfnamefont {I.}~\bibnamefont
  {Harry}}, \bibinfo {author} {\bibfnamefont {S.}~\bibnamefont {Privitera}},
  \bibinfo {author} {\bibfnamefont {A.}~\bibnamefont {Boh\'{e}}}, \emph
  {et~al.},\ }\href {https://doi.org/10.1103/PhysRevD.94.024012} {\bibfield
  {journal} {\bibinfo  {journal} {Physical Review D}\ }\textbf {\bibinfo
  {volume} {94}},\ \bibinfo {pages} {024012} (\bibinfo {year} {2016})},\
  \bibinfo {note} {arXiv:1603.02444 [astro-ph, physics:gr-qc]}\BibitemShut
  {NoStop}%
\bibitem [{\citenamefont {Chatziioannou}\ \emph {et~al.}(2017)\citenamefont
  {Chatziioannou}, \citenamefont {Klein}, \citenamefont {Yunes} \emph
  {et~al.}}]{faithreq}%
  \BibitemOpen
  \bibfield  {author} {\bibinfo {author} {\bibfnamefont {K.}~\bibnamefont
  {Chatziioannou}}, \bibinfo {author} {\bibfnamefont {A.}~\bibnamefont
  {Klein}}, \bibinfo {author} {\bibfnamefont {N.}~\bibnamefont {Yunes}}, \emph
  {et~al.},\ }\href {https://doi.org/10.1103/PhysRevD.95.104004} {\bibfield
  {journal} {\bibinfo  {journal} {Physical Review D}\ }\textbf {\bibinfo
  {volume} {95}},\ \bibinfo {pages} {104004} (\bibinfo {year} {2017})},\
  \bibinfo {note} {arXiv:1703.03967 [astro-ph, physics:gr-qc]}\BibitemShut
  {NoStop}%
\bibitem [{\citenamefont {Lindblom}\ \emph {et~al.}(2008)\citenamefont
  {Lindblom}, \citenamefont {Owen},\ and\ \citenamefont
  {Brown}}]{Lindblom_2008}%
  \BibitemOpen
  \bibfield  {author} {\bibinfo {author} {\bibfnamefont {L.}~\bibnamefont
  {Lindblom}}, \bibinfo {author} {\bibfnamefont {B.~J.}\ \bibnamefont {Owen}},\
  and\ \bibinfo {author} {\bibfnamefont {D.~A.}\ \bibnamefont {Brown}},\
  }\bibfield  {journal} {\bibinfo  {journal} {Physical Review D}\ }\textbf
  {\bibinfo {volume} {78}},\ \href {https://doi.org/10.1103/physrevd.78.124020}
  {10.1103/physrevd.78.124020} (\bibinfo {year} {2008})\BibitemShut {NoStop}%
\bibitem [{\citenamefont {{Neyman}}\ and\ \citenamefont
  {{Pearson}}(1933)}]{10.1098/rsta.1933.0009}%
  \BibitemOpen
  \bibfield  {author} {\bibinfo {author} {\bibfnamefont {J.}~\bibnamefont
  {{Neyman}}}\ and\ \bibinfo {author} {\bibfnamefont {E.~S.}\ \bibnamefont
  {{Pearson}}},\ }\href {https://doi.org/10.1098/rsta.1933.0009} {\bibfield
  {journal} {\bibinfo  {journal} {Philosophical Transactions of the Royal
  Society of London Series A}\ }\textbf {\bibinfo {volume} {231}},\ \bibinfo
  {pages} {289} (\bibinfo {year} {1933})}\BibitemShut {NoStop}%
\bibitem [{\citenamefont {Karlin}\ and\ \citenamefont
  {Rubin}(1956)}]{10.1214/aoms/1177728259}%
  \BibitemOpen
  \bibfield  {author} {\bibinfo {author} {\bibfnamefont {S.}~\bibnamefont
  {Karlin}}\ and\ \bibinfo {author} {\bibfnamefont {H.}~\bibnamefont {Rubin}},\
  }\href {https://doi.org/10.1214/aoms/1177728259} {\bibfield  {journal}
  {\bibinfo  {journal} {The Annals of Mathematical Statistics}\ }\textbf
  {\bibinfo {volume} {27}},\ \bibinfo {pages} {272 } (\bibinfo {year}
  {1956})}\BibitemShut {NoStop}%
\bibitem [{\citenamefont {Heisenberg}(2014)}]{Heisenberg:2014rta}%
  \BibitemOpen
  \bibfield  {author} {\bibinfo {author} {\bibfnamefont {L.}~\bibnamefont
  {Heisenberg}},\ }\href {https://doi.org/10.1088/1475-7516/2014/05/015}
  {\bibfield  {journal} {\bibinfo  {journal} {JCAP}\ }\textbf {\bibinfo
  {volume} {05}},\ \bibinfo {pages} {015}},\ \Eprint
  {https://arxiv.org/abs/1402.7026} {arXiv:1402.7026 [hep-th]} \BibitemShut
  {NoStop}%
\bibitem [{\citenamefont {Heisenberg}(2019)}]{Heisenberg:2018vsk}%
  \BibitemOpen
  \bibfield  {author} {\bibinfo {author} {\bibfnamefont {L.}~\bibnamefont
  {Heisenberg}},\ }\href {https://doi.org/10.1016/j.physrep.2018.11.006}
  {\bibfield  {journal} {\bibinfo  {journal} {Phys. Rept.}\ }\textbf {\bibinfo
  {volume} {796}},\ \bibinfo {pages} {1} (\bibinfo {year} {2019})},\ \Eprint
  {https://arxiv.org/abs/1807.01725} {arXiv:1807.01725 [gr-qc]} \BibitemShut
  {NoStop}%
\bibitem [{\citenamefont {Vicente}\ and\ \citenamefont
  {Cardoso}(2022)}]{Vicente_2022}%
  \BibitemOpen
  \bibfield  {author} {\bibinfo {author} {\bibfnamefont {R.}~\bibnamefont
  {Vicente}}\ and\ \bibinfo {author} {\bibfnamefont {V.}~\bibnamefont
  {Cardoso}},\ }\bibfield  {journal} {\bibinfo  {journal} {Physical Review D}\
  }\textbf {\bibinfo {volume} {105}},\ \href
  {https://doi.org/10.1103/physrevd.105.083008} {10.1103/physrevd.105.083008}
  (\bibinfo {year} {2022})\BibitemShut {NoStop}%
\bibitem [{\citenamefont {Traykova}\ \emph {et~al.}(2023)\citenamefont
  {Traykova}, \citenamefont {Vicente}, \citenamefont {Clough}, \citenamefont
  {Helfer}, \citenamefont {Berti}, \citenamefont {Ferreira},\ and\
  \citenamefont {Hui}}]{traykova2023relativistic}%
  \BibitemOpen
  \bibfield  {author} {\bibinfo {author} {\bibfnamefont {D.}~\bibnamefont
  {Traykova}}, \bibinfo {author} {\bibfnamefont {R.}~\bibnamefont {Vicente}},
  \bibinfo {author} {\bibfnamefont {K.}~\bibnamefont {Clough}}, \bibinfo
  {author} {\bibfnamefont {T.}~\bibnamefont {Helfer}}, \bibinfo {author}
  {\bibfnamefont {E.}~\bibnamefont {Berti}}, \bibinfo {author} {\bibfnamefont
  {P.~G.}\ \bibnamefont {Ferreira}},\ and\ \bibinfo {author} {\bibfnamefont
  {L.}~\bibnamefont {Hui}},\ }\href@noop {} {\bibinfo {title} {Relativistic
  drag forces on black holes from scalar dark matter clouds of all sizes}}
  (\bibinfo {year} {2023}),\ \Eprint {https://arxiv.org/abs/2305.10492}
  {arXiv:2305.10492 [gr-qc]} \BibitemShut {NoStop}%
\bibitem [{\citenamefont {Zhang}\ and\ \citenamefont
  {Yang}(2020)}]{Zhang_2020}%
  \BibitemOpen
  \bibfield  {author} {\bibinfo {author} {\bibfnamefont {J.}~\bibnamefont
  {Zhang}}\ and\ \bibinfo {author} {\bibfnamefont {H.}~\bibnamefont {Yang}},\
  }\bibfield  {journal} {\bibinfo  {journal} {Physical Review D}\ }\textbf
  {\bibinfo {volume} {101}},\ \href
  {https://doi.org/10.1103/physrevd.101.043020} {10.1103/physrevd.101.043020}
  (\bibinfo {year} {2020})\BibitemShut {NoStop}%
\bibitem [{\citenamefont {Baumann}\ \emph
  {et~al.}(2022{\natexlab{a}})\citenamefont {Baumann}, \citenamefont {Bertone},
  \citenamefont {Stout},\ and\ \citenamefont {Tomaselli}}]{Baumann_2022}%
  \BibitemOpen
  \bibfield  {author} {\bibinfo {author} {\bibfnamefont {D.}~\bibnamefont
  {Baumann}}, \bibinfo {author} {\bibfnamefont {G.}~\bibnamefont {Bertone}},
  \bibinfo {author} {\bibfnamefont {J.}~\bibnamefont {Stout}},\ and\ \bibinfo
  {author} {\bibfnamefont {G.~M.}\ \bibnamefont {Tomaselli}},\ }\bibfield
  {journal} {\bibinfo  {journal} {Physical Review Letters}\ }\textbf {\bibinfo
  {volume} {128}},\ \href {https://doi.org/10.1103/physrevlett.128.221102}
  {10.1103/physrevlett.128.221102} (\bibinfo {year}
  {2022}{\natexlab{a}})\BibitemShut {NoStop}%
\bibitem [{\citenamefont {Buehler}\ and\ \citenamefont
  {Desjacques}(2023)}]{Buehler_2023}%
  \BibitemOpen
  \bibfield  {author} {\bibinfo {author} {\bibfnamefont {R.}~\bibnamefont
  {Buehler}}\ and\ \bibinfo {author} {\bibfnamefont {V.}~\bibnamefont
  {Desjacques}},\ }\bibfield  {journal} {\bibinfo  {journal} {Physical Review
  D}\ }\textbf {\bibinfo {volume} {107}},\ \href
  {https://doi.org/10.1103/physrevd.107.023516} {10.1103/physrevd.107.023516}
  (\bibinfo {year} {2023})\BibitemShut {NoStop}%
\bibitem [{\citenamefont {Cole}\ \emph {et~al.}(2022)\citenamefont {Cole},
  \citenamefont {Bertone}, \citenamefont {Coogan}, \citenamefont {Gaggero},
  \citenamefont {Karydas}, \citenamefont {Kavanagh}, \citenamefont {Spieksma},\
  and\ \citenamefont {Tomaselli}}]{cole2022disks}%
  \BibitemOpen
  \bibfield  {author} {\bibinfo {author} {\bibfnamefont {P.~S.}\ \bibnamefont
  {Cole}}, \bibinfo {author} {\bibfnamefont {G.}~\bibnamefont {Bertone}},
  \bibinfo {author} {\bibfnamefont {A.}~\bibnamefont {Coogan}}, \bibinfo
  {author} {\bibfnamefont {D.}~\bibnamefont {Gaggero}}, \bibinfo {author}
  {\bibfnamefont {T.}~\bibnamefont {Karydas}}, \bibinfo {author} {\bibfnamefont
  {B.~J.}\ \bibnamefont {Kavanagh}}, \bibinfo {author} {\bibfnamefont
  {T.~F.~M.}\ \bibnamefont {Spieksma}},\ and\ \bibinfo {author} {\bibfnamefont
  {G.~M.}\ \bibnamefont {Tomaselli}},\ }\href@noop {} {\bibinfo {title} {Disks,
  spikes, and clouds: distinguishing environmental effects on bbh gravitational
  waveforms}} (\bibinfo {year} {2022}),\ \Eprint
  {https://arxiv.org/abs/2211.01362} {arXiv:2211.01362 [gr-qc]} \BibitemShut
  {NoStop}%
\bibitem [{\citenamefont {Tomaselli}\ \emph {et~al.}(2023)\citenamefont
  {Tomaselli}, \citenamefont {Spieksma},\ and\ \citenamefont
  {Bertone}}]{tomaselli2023dynamical}%
  \BibitemOpen
  \bibfield  {author} {\bibinfo {author} {\bibfnamefont {G.~M.}\ \bibnamefont
  {Tomaselli}}, \bibinfo {author} {\bibfnamefont {T.~F.~M.}\ \bibnamefont
  {Spieksma}},\ and\ \bibinfo {author} {\bibfnamefont {G.}~\bibnamefont
  {Bertone}},\ }\href@noop {} {\bibinfo {title} {Dynamical friction in
  gravitational atoms}} (\bibinfo {year} {2023}),\ \Eprint
  {https://arxiv.org/abs/2305.15460} {arXiv:2305.15460 [gr-qc]} \BibitemShut
  {NoStop}%
\bibitem [{\citenamefont {Baumann}\ \emph
  {et~al.}(2022{\natexlab{b}})\citenamefont {Baumann}, \citenamefont {Bertone},
  \citenamefont {Stout},\ and\ \citenamefont {Tomaselli}}]{Baumann_2022_ion}%
  \BibitemOpen
  \bibfield  {author} {\bibinfo {author} {\bibfnamefont {D.}~\bibnamefont
  {Baumann}}, \bibinfo {author} {\bibfnamefont {G.}~\bibnamefont {Bertone}},
  \bibinfo {author} {\bibfnamefont {J.}~\bibnamefont {Stout}},\ and\ \bibinfo
  {author} {\bibfnamefont {G.~M.}\ \bibnamefont {Tomaselli}},\ }\bibfield
  {journal} {\bibinfo  {journal} {Physical Review D}\ }\textbf {\bibinfo
  {volume} {105}},\ \href {https://doi.org/10.1103/physrevd.105.115036}
  {10.1103/physrevd.105.115036} (\bibinfo {year}
  {2022}{\natexlab{b}})\BibitemShut {NoStop}%
\bibitem [{\citenamefont {Unruh}(1976)}]{Unrun1976}%
  \BibitemOpen
  \bibfield  {author} {\bibinfo {author} {\bibfnamefont {W.~G.}\ \bibnamefont
  {Unruh}},\ }\href {https://doi.org/10.1103/PhysRevD.14.3251} {\bibfield
  {journal} {\bibinfo  {journal} {Phys. Rev. D}\ }\textbf {\bibinfo {volume}
  {14}},\ \bibinfo {pages} {3251} (\bibinfo {year} {1976})}\BibitemShut
  {NoStop}%
\bibitem [{\citenamefont {Zhang}\ and\ \citenamefont
  {Yang}(2019)}]{Zhang_2019}%
  \BibitemOpen
  \bibfield  {author} {\bibinfo {author} {\bibfnamefont {J.}~\bibnamefont
  {Zhang}}\ and\ \bibinfo {author} {\bibfnamefont {H.}~\bibnamefont {Yang}},\
  }\bibfield  {journal} {\bibinfo  {journal} {Physical Review D}\ }\textbf
  {\bibinfo {volume} {99}},\ \href {https://doi.org/10.1103/physrevd.99.064018}
  {10.1103/physrevd.99.064018} (\bibinfo {year} {2019})\BibitemShut {NoStop}%
\bibitem [{\citenamefont {Cao}\ and\ \citenamefont
  {Tang}(2023)}]{cao2023signatures}%
  \BibitemOpen
  \bibfield  {author} {\bibinfo {author} {\bibfnamefont {Y.}~\bibnamefont
  {Cao}}\ and\ \bibinfo {author} {\bibfnamefont {Y.}~\bibnamefont {Tang}},\
  }\href@noop {} {\bibinfo {title} {Signatures of ultralight bosons in compact
  binary inspiral and outspiral}} (\bibinfo {year} {2023}),\ \Eprint
  {https://arxiv.org/abs/2307.05181} {arXiv:2307.05181 [gr-qc]} \BibitemShut
  {NoStop}%
\bibitem [{\citenamefont {Ferreira}\ \emph {et~al.}(2017)\citenamefont
  {Ferreira}, \citenamefont {Macedo},\ and\ \citenamefont
  {Cardoso}}]{Ferreira_2017}%
  \BibitemOpen
  \bibfield  {author} {\bibinfo {author} {\bibfnamefont {M.~C.}\ \bibnamefont
  {Ferreira}}, \bibinfo {author} {\bibfnamefont {C.~F.}\ \bibnamefont
  {Macedo}},\ and\ \bibinfo {author} {\bibfnamefont {V.}~\bibnamefont
  {Cardoso}},\ }\bibfield  {journal} {\bibinfo  {journal} {Physical Review D}\
  }\textbf {\bibinfo {volume} {96}},\ \href
  {https://doi.org/10.1103/physrevd.96.083017} {10.1103/physrevd.96.083017}
  (\bibinfo {year} {2017})\BibitemShut {NoStop}%
\bibitem [{\citenamefont {Hannuksela}\ \emph {et~al.}(2019)\citenamefont
  {Hannuksela}, \citenamefont {Wong}, \citenamefont {Brito}, \citenamefont
  {Berti},\ and\ \citenamefont {Li}}]{Hannuksela_2019}%
  \BibitemOpen
  \bibfield  {author} {\bibinfo {author} {\bibfnamefont {O.~A.}\ \bibnamefont
  {Hannuksela}}, \bibinfo {author} {\bibfnamefont {K.~W.~K.}\ \bibnamefont
  {Wong}}, \bibinfo {author} {\bibfnamefont {R.}~\bibnamefont {Brito}},
  \bibinfo {author} {\bibfnamefont {E.}~\bibnamefont {Berti}},\ and\ \bibinfo
  {author} {\bibfnamefont {T.~G.~F.}\ \bibnamefont {Li}},\ }\href
  {https://doi.org/10.1038/s41550-019-0712-4} {\bibfield  {journal} {\bibinfo
  {journal} {Nature Astronomy}\ }\textbf {\bibinfo {volume} {3}},\ \bibinfo
  {pages} {447} (\bibinfo {year} {2019})}\BibitemShut {NoStop}%
\end{thebibliography}%

    \appendix

    \onecolumngrid

    \section{Decomposition of Proca Equations} \label{app:decomp}

    \subsubsection*{\bf Angular Equation} \label{sec:angularequation}
	Multiply the angular equation Eq. \eqref{eq:ProcaFKKStheta} by $q_{\theta}$ and define $\Lambda = \frac{\mu^2}{\nu^2} - \frac{\sigma}{\nu} + 2a \omega m - a^2 \omega^2$ and $\gamma^2 = \omega^2 - \mu^2$, where we've denoted the mode number by $m$ and no longer by $\mathfrak{m}$ as in the main text. Rearranging the terms, one finds

    \begin{equation} \label{eq:angular}
		q_{\theta} \left( \partial_{\theta}^2  + \cot{\theta} \partial_{\theta}  - \frac{m^2}{\sin{\theta}^2} + \Lambda \right) S + \left( \left( \gamma^2 - 2 \sigma \nu \right) a^2 \cos{\theta}^2 - \gamma^2 \nu^2 a^4 \cos{\theta}^4 - 2 a^2 \nu^2 \cos{\theta} \sin{\theta} \partial_{\theta} \right) S = 0
	\end{equation}
	Now we expand the angular variable $S$ in terms of functions proportional to the associated Legendre polynomials as
	\begin{equation}
	S = \underset{l^{\prime} = |m|}{\Sigma} b_{l^{\prime}} Y^m_{2l^{\prime}-|m|+\eta}(\theta)
	\end{equation}
	where $Y^m_{l^{\prime}}(\theta) = Y^m_{l^{\prime}}(\theta,0)$, and we expand the function in a basis with definite parity as the angular equation respects parity. We then insert this into the angular equation above. 
	
	Using various relations of the associated legendre polynomials in the first term of Eq. \eqref{eq:angular}, one can easily show
	\begin{equation}
	q_{\theta} \left( \partial_{\theta}^2  + \cot{\theta} \partial_{\theta}  - \frac{m^2}{\sin{\theta}^2} + \Lambda \right) S \rightarrow q_{\theta} (-l(l+1) + \Lambda) Y^m_{l}
	\end{equation}
	where we dropped the $b_l$ coefficient. We operate on Eq. \eqref{eq:angular} with $\int d\Omega \bar{Y}^m_{l}$ and define the quantity $\langle lm | X | l^{\prime} m \rangle \equiv \int d\Omega \bar{Y}^m_l X Y^m_{l^{\prime}}$. Then we have the expression
	\begin{align*}
		0 = & (-l^{\prime}(l^{\prime}+1) + \Lambda) \langle lm | l^{\prime}m \rangle + \\
		& ((\gamma^2 - 2 \sigma \nu) a^2 - a^2 \nu^2 (-l^{\prime}(l^{\prime}+1) + \Lambda)) \langle lm| \cos{\theta}^2 | l^{\prime}m \rangle -\\
		& 2a^2 \nu^2 \langle lm | \cos{\theta} \sin{\theta} \partial_{\theta} | l^{\prime}m \rangle -\\
		& \gamma^2 \nu^2 a^4 \langle lm | \cos{\theta}^4 | l^{\prime}m \rangle
	\end{align*}
	Representing the trigonometric functions in a spherical harmonic basis, we can easily calculate 
	\begin{align*}
		\langle lm | l^{\prime}m \rangle &= \delta_{l l^{\prime}} \\
		a_{l l^{\prime}} \equiv \langle lm| \cos{\theta}^2 | l^{\prime}m \rangle &= \frac{1}{3} \sqrt{\frac{16 \pi}{5}} \langle l | 2 | l^{\prime} \rangle  + \frac{\sqrt{4 \pi}}{3} \langle l | 0 | l^{\prime} \rangle \\
		b_{l l^{\prime}} \equiv \langle lm | \cos{\theta}^4 | l^{\prime}m \rangle &= \frac{16 \sqrt{\pi}}{105} \langle l | 4 | l^{\prime} \rangle + \frac{7 \sqrt{4 \pi}}{35} \langle l | 0 | l^{\prime} \rangle + \frac{10}{35} \sqrt{\frac{16 \pi}{5}} \langle l | 2 | l^{\prime} \rangle \\
		d_{l l^{\prime}} \equiv \langle lm | \cos{\theta} \sin{\theta} \partial_{\theta} | l^{\prime}m \rangle &= \sqrt{\frac{4 \pi}{3}} \frac{l^{\prime} \sqrt{(l^{\prime}+1)^2 - m^2}}{\sqrt{(2l^{\prime}+1)(2l+3)}} \langle l | 1 | l^{\prime} + 1 \rangle - \sqrt{\frac{4 \pi}{3}} \frac{(l^{\prime}+m)(\sqrt{l^{2 \prime} - m^2}}{\sqrt{(2l^{\prime}+1)(2l^{\prime}-1)}} \langle l | 1 | l^{\prime} - 1 \rangle
	\end{align*}
	where $\langle l1, l2, l3 \rangle$ denotes the triple product integral $\int d\Omega Y^m_{l1} Y^m_{l2} Y^m_{l3}$ and can be represented in terms of the 3J-symbols. Thus, the angular equation, after reinserting the $b_{l^{\prime}}$ coefficients, becomes
	\begin{equation}
	\mathcal{M}_{l l^{\prime}} b_{l^{\prime}} = 0
	\end{equation}
	where 
	\begin{equation}
	\mathcal{M}_{l l^{\prime}} = \left( \Lambda - l^{\prime}(l^{\prime}+1) \right) \delta_{l l^{\prime}} + \left( \nu^2 (l^{\prime}(l^{\prime}+1) - \Lambda) - 2 \sigma \nu + \gamma^2 \right) a^2 a_{l l^{\prime}} - \gamma^2 \nu^2 a^4 b_{l l^{\prime}} - 2a^2 \nu^2 d_{l l^{\prime}}
	\end{equation}
	
	This implies that $b_{l^{\prime}}$ lives in the kernel of the map $\mathcal{M}$. For there to be non-trivial solutions to this eigenvalue equation, we require $Det(\mathcal{M}) = 0$, which places restrictions on the complex-valued eigenvalue $\nu$. In general, the complex eigenvalue $\nu$ depends on $\mu$, $\omega$, $l$, and $a$. A generic solution can be obtained numerically, though insight can be found by taking suitable limits.

	\paragraph*{static limit:}
	In the limit of staticity ($a \rightarrow 0)$, the map takes the diagonal form
	\begin{equation}
	\mathcal{M} = (\Lambda - l^{\prime}(l^{\prime}+1))\delta_{l l^{\prime}}
	\end{equation}
	The non-trivial solution constraint then enforces
	\begin{equation}
	\Lambda = l^{\prime} (l^{\prime}+1)
	\end{equation}
	which, in terms of the eigenvalue $\nu$, yields
	\begin{equation}
	\mu^2 - \omega \nu - l^{\prime}(l^{\prime}+1) \nu^2 = 0
	\end{equation}
	We thus find the solutions
	\begin{equation}
	\nu = \begin{cases}
		\frac{\mu^2}{\omega}&  l^{\prime}=0 \\ 
		- \frac{\omega \pm \sqrt{\omega^2 + 4 \mu^2 l^{\prime} (l^{\prime} + 1)}}{2 l^{\prime}(l^{\prime}+1)} & l^{\prime} >0
	\end{cases}
	\end{equation}
	We have a pair of modes for $l^{\prime}>0$ and a single (even-parity) mode for $l^{\prime}=0$. For the latter, the associated eigenvector is a single spherical harmonic and is thus even-parity under a parity transformation, while for the former we have a pair of modes. As the angular equation respects parity, this pair have the same parity. Thus, we've found the even-parity solutions

	\paragraph*{Marginally-bound case: $\gamma^2=0$}
	Consider now the case $\omega^2 = \mu^2$. This is the threshold between a quasibound mode and an unbound state. The matrix $\mathcal{M}$ is tridiagonal, as can be seen using properties of the 3J-symbols. Consider the truncated series
	\begin{equation}
	S = Y^m_l + b_1 Y^m_{l+2} + 0*Y^m_{l+4}
	\end{equation}
	where $l=|m|+\eta$ and $\eta = 0,1$ denotes the parity.
	
	In general, we have three equations coming from
	\begin{equation}
	\mathcal{M} \cdot b
	\end{equation}
	with $b=(1,b_1,0)^T$.

	In the $\eta=0$ case, we have $m = \pm l$. The solution $S=Y^{\pm l}_{l}$ with $\nu = \frac{\pm \omega}{m - a \omega}$ is an exact solution since $\Lambda = l(l+1)$, $\sigma = \pm m \nu$. It can be shown this solution together with this eigenvalue solves the angular equation by direct insertion. It can also be shown this eigenvalue corresponds to the $S=-1$ polarization state of the Proca field.

    In the case $\eta=1$ and $b_1=0$, we now have two non-trivial equations, with $m = \pm (l-1)$. It can be shown by direct computation that the corresponding eigenvalues are
	\begin{equation}
	\nu = \frac{1}{2a} \left( \pm l - a*\omega + \epsilon \sqrt{ (\mp l + a \omega)^2 + 4 a \omega} \right)
	\end{equation}
	These correspond to both parity-odd and parity-even polarization states, as shown in~\cite{Dolan_2018}.

    For $\eta=0$ and $m=1$, the last polarization state can be recovered by finding the middle root of the following cubic expression~\cite{Dolan_2018}
	\begin{equation}
	a \nu^3 (m-a \omega) - \nu^2 ((m+1)(m+2) - a \omega (2m - a \omega)) + \omega \nu + \omega^2 = 0
	\end{equation}

	\subsubsection*{\bf Radial equation} \label{sec:radialequation}
	The radial equation Eq. \eqref{eq:ProcaFKKSradial} at asymptotic infinity can be shown to reduce to
	\begin{equation}
	\left(\left(1-\frac{r_s}{r}\right) \partial_r \partial_r  + \frac{\omega^2}{1-\frac{r_s}{r}} - \mu^2 \right)R(r)=0
	\end{equation}
	Remarkably, there is an exact solution in terms of Whittaker M and W functions.
	The exact solution is
	\begin{equation}
	R(r) = c_1 M_{\xi, \chi} \left[2\left(r_s - r \right)Q\right] + c_2 W_{\xi, \chi} \left[ 2\left(r_s - r \right)Q\right]
	\end{equation}
	where $\xi = \frac{r_s(\mu^2 - 2 \omega^2)}{2 Q}$, $\chi = \frac{- i \sqrt{-1 + 4 r_s^2 \omega^2}}{2}$, $c_{1,2}$ are constants, $Q \equiv \sqrt{\mu^2 - \omega^2}$, and $W_{\xi, \chi}[x]$, $M_{\xi \chi}[x]$ are the Whittaker W and M functions. Asymptotically, the solution takes the form
	\begin{equation}
	R(r\rightarrow \infty) = r^{\frac{(2\omega^2 - \mu^2)M}{Q}} e^{-Qr},
	\end{equation}
	assuming vanishing boundary conditions at infinity, as employed later.

    Now, we rewrite the derivatives of the radial equation in terms of the tortoise co-ordinate, $\frac{\partial r^*}{dr} = \frac{r^2 + a^2}{\Delta}$. One finds
	\begin{equation}
	\left( \frac{(r^2 + a^2)^2}{q_r \Delta} \partial_{r^*}^2 + \frac{K_r^2}{q_r \Delta} \right) R(r) + \left( \frac{2-q_r}{q_r^2} \frac{\sigma}{\nu} - \frac{\mu^2}{\nu^2} + \frac{2r(2-q_r)}{q_r^2} \partial_{r^*} \right) R(r)=0
	\end{equation}
	Multiplying by $q_r \Delta$,
	\begin{equation}
	\left( (r^2 + a^2)^2 \partial_{r^*}^2 + K_r^2 \right) R(r) + q_r \Delta\left( \frac{2-q_r}{q_r^2} \frac{\sigma}{\nu} - \frac{\mu^2}{\nu^2} + \frac{2r(2-q_r)}{q_r^2} \partial_{r^*} \right) R(r)=0
	\end{equation}
	In the limit $r\rightarrow r_+$, $\Delta = (r-r_+)(r-r_-) \rightarrow 0$, and so the above differential equation reduces to
	\begin{equation}
	\left( (r_+^2 + a^2)^2 \partial_{r^*}^2 + K_{r_+}^2 \right) R(r)=0
	\end{equation}
	assuming $\partial_{r^*} R(r)$ is finite at the outer horizon.
	Taking $m=0$, this reduces $K_{r_+}$ to $-(r_+^2+a^2)\omega$. The ODE is readily solved, yielding
	\begin{equation}
	R(r\rightarrow r_+) = e^{-i \omega r^*}
	\end{equation}

	If we take a generic m, then the asymptotic form is
	\begin{equation}
	R(r\rightarrow r_+) \sim e^{-r^* \sqrt{ \frac{m(a \omega -m)}{r_s r_+} + \frac{1}{r_s^2} (m-r_s \omega)(m+r_s \omega)}} = e^{-i r^* \frac{K_{r_+}}{(r_{+}^2 +a^2)}}
	\end{equation}
	
	\paragraph*{Asymptotic form}
	The two asymptotic forms of the radial equation are then
	\begin{equation} \label{eq:radialasymptotic}
		R(r) = 
		\begin{cases}
			e^{-i \omega r^*} & r \rightarrow r_+ \\
			r^{\frac{(2\omega^2 - \mu^2)M}{Q}} e^{-Qr} & r \rightarrow \infty 
		\end{cases}
	\end{equation}
	with $Q \equiv \sqrt{\mu^2 - \omega^2}$. 
	
	Write out the tortoise coordinate in terms of the Boyer-lindquist radial co-ordinate to find
	\begin{equation}
	r^* = r + \frac{r_s r_+}{r_+ - r_-} \ln{\frac{r-r_+}{r_s}} - \frac{r_s r_-}{r_+ - r_-} \ln{\frac{r-r_-}{r_s}}
	\end{equation}
	with $r_s = 2*M$ and $r_{\pm}$ the inner and outer horizon radii in Boyer-Lindguist co-ords. Then we rewrite the asymptotic form near the horizon as
	\begin{equation}
	e^{-i \omega r^*} = e^{-i \omega r} \left( \frac{r - r_+}{r_s} \right)^{-i \omega \frac{r_s r_+}{r_+ - r_-}} \left( \frac{r - r_-}{r_s} \right)^{i \omega \frac{r_s r_-}{r_+ - r_-}}
	\end{equation}
	We see there is a pole at $r=r_+$. We can then expand this function in terms of a generalized power series as
	\begin{equation}
	R(r) = x^{-i \kappa} (r_0 + r_1 x + r_2 x^2 + ...)
	\end{equation}
	with $x \equiv \frac{r-r_+}{r_+ - r_-}$ and $\kappa = \frac{\omega r_s r_+}{r_+ - r_-}$, or more concisely,
	\begin{equation} \label{eq:frobeniusseries}
		R(r) = \overset{\infty}{\underset{n=0}{\Sigma}} r_n x^{n-i \kappa}
	\end{equation}
	The coefficients $r_n$ can be determined by inserting the above into the radial equation. We first recast the radial equation into the form
	\begin{equation}\label{eq:recastradial}
		\partial_r^2 R(r) +  \left(\frac{q_r}{\Delta} \left( \frac{2r-r_s}{q_r} - \frac{2r\nu^2 \Delta}{q_r^2} \right) \right)\partial_r R(r) +  \left( \frac{q_r}{\Delta} \left( - \frac{\mu^2}{\nu^2} + \frac{K_r^2}{\Delta q_r} + \frac{(2-q_r) \sigma}{ q_r^2 \nu} \right) \right)R(r) = 0
	\end{equation}
	which is of the form 
	\begin{equation}
	\partial_r^2 R(r) + P(r; \lambda) \partial_r R(r) + Q(r; \lambda) R(r) = 0
	\end{equation}
	where $\lambda$ are all the other parameters, i.e. $\nu$, $\mu$, etc, and 
	\begin{align*}
		P(r; \lambda) &= \frac{1}{(r-r_+)}\left(\frac{q_r}{(r-r_-)} \left( \frac{2r-r_s}{q_r} - \frac{2r\nu^2 \Delta}{q_r^2} \right) \right) = \frac{1}{(r-r_+)}*\tilde{P}(r;\lambda)\\
		Q(r; \lambda) &= \frac{1}{(r-r_+)^2}\left( \frac{q_r}{(r-r_-)^2} \left( - \frac{\mu^2}{\nu^2}\Delta + \frac{K_r^2}{ q_r} + \Delta\frac{(2-q_r) \sigma}{ q_r^2 \nu} \right) \right) = \frac{1}{(r-r_+)^2} \tilde{Q}(r;\lambda)
	\end{align*}
	Its clear then that, as $\Delta = (r-r_+)(r-r_-)$, then the $P(r; \lambda)$ diverges like $\frac{1}{r-r_+}$ and $Q(r;\lambda)$ diverges as $\frac{1}{(r-r_+)^2}$. Hence, $r-r_+$ is a pole of order 1 for $P(r;\lambda)$ and a pole of order 2 for $Q(r;\lambda)$ and both $\tilde{P}(r;\lambda)$ and $\tilde{Q}(r;\lambda)$ are regular at $r=r_+$. Thus, by Fuch's theorem, we can apply the method of Frobenius to find a power series solution of the form Eq. \eqref{eq:frobeniusseries} for the radial function near the outer horizon. Define the following functions
	\begin{align*}
		\tilde{P}_1(r;\lambda) &= \frac{2r-r_s}{(r-r_-)} \\
		\tilde{P}_2(r;\lambda) &= - \frac{2r \nu^2}{q_r}\\
		\tilde{Q}_1(r;\lambda) &= \frac{K_r^2}{(r-r_-)^2}\\
		\tilde{Q}_2(r;\lambda) &= \frac{q_r}{(r-r_-)} \left( \frac{2-q_r}{q_r^2} \frac{\sigma}{\nu} - \frac{\mu^2}{\nu^2} \right)
	\end{align*}
	so that 
	\begin{align*}
		\tilde{P} &= \tilde{P}_1  + (r-r_+) \tilde{P}_2 \\
		\tilde{Q} &= \tilde{Q}_1 + (r-r_+) \tilde{Q}_2
	\end{align*}
	The differential equation then becomes 
	\begin{equation}
	\partial_r^2 R(r) + \frac{\tilde{P}_1}{r-r_+} \partial_r R(r) + \tilde{P_2} \partial_r R(r) + \frac{\tilde{Q}_1}{(r-r_+)^2} R(r) + \frac{\tilde{Q}_2}{(r-r_+)} R(r) = 0
	\end{equation}

	Inserting the expansion Eq. \eqref{eq:frobeniusseries}, one finds
	\begin{align*}
		0&=\sum_{n=0} \left( r_n (n-\kappa)(n-\kappa-1) x^{n-2-\kappa} + \tilde{P}_1 r_n (n-\kappa) x^{n-2-\kappa} + \tilde{Q}_1 r_n x^{n-\kappa-2} + \tilde{P}_2 r_n (n-\kappa) x^{n-\kappa -1} + \tilde{Q}_2 r_n x^{n-\kappa-1} \right) \\
		0&=\sum_{n=0} \left( r_n (n-\kappa)(n-\kappa-1) x^{n-2-\kappa} + \tilde{P}_1 r_n (n-\kappa) x^{n-2-\kappa} + \tilde{Q}_1 r_n x^{n-\kappa-2} \right) + \sum_{n=0} \left( \tilde{P}_2 r_n (n-\kappa) x^{n-\kappa -1} + \tilde{Q}_2 r_n x^{n-\kappa-1} \right) \\
		0&=\sum_{n=0} \left(  (n-\kappa)(n-\kappa-1)  + \tilde{P}_1 (n-\kappa)  + \tilde{Q}_1 \right)r_n x^{n-2-\kappa} + \sum_{n=1} \left( \tilde{P}_2 (n-1-\kappa)  + \tilde{Q}_2  \right)r_{n-1} x^{n-\kappa-2} \\
		0&= \left((-\kappa)(-\kappa-1)  + \tilde{P}_1 (-\kappa)  + \tilde{Q}_1 \right)r_0x^{-2-\kappa} +  \sum_{n=1} \left[\left(  (n-\kappa)(n-\kappa-1)  + \tilde{P}_1 (n-\kappa)  + \tilde{Q}_1 \right)r_n \right.\\
		&\;\;\;\left. + \left( \tilde{P}_2 (n-1-\kappa)  + \tilde{Q}_2  \right)r_{n-1} \right]x^{n-\kappa-2} \\
	\end{align*}
	The indicial equation can be read off as
	\begin{equation}
	\left((-\kappa)(-\kappa-1)  + \tilde{P}_1(r_+;\lambda) (-\kappa)  + \tilde{Q}_1(r_+;\lambda) \right)=0
	\end{equation}
	Solving for $\kappa$,
	\begin{equation}
	\kappa = \frac{\tilde{P}_1(r_+)-1 \pm \sqrt{(1-\tilde{P}_1(r_+))^2 - 4\tilde{Q}_1(r_+)}}{2}
	\end{equation}
	Evaluating the tilde functions, one finds
	\begin{align*}
		\tilde{P}_1(r_+) &= 1\\
		\tilde{Q}_1(r_+) &= \left( \frac{ \omega r_+ r_s - a m}{r_+ - r_-} \right)^2
	\end{align*}
	Hence, we find $\kappa$ to be 
    \begin{equation} \label{eq:indicialroots}
		\kappa = \pm i  \left( \frac{ \omega r_+ r_s - am}{r_+ - r_-} \right)
	\end{equation}

	To find the recursion relation between the coefficients, we reindex the sum to
	\begin{equation}
	\sum_{n=0} \left( \left( (n-k)(n-k-1) + \tilde{P}_1(r) (n-k) + \tilde{Q}_1(r) \right) r_n + \left( \tilde{P}_2(r) (n-1-k) + \tilde{Q}_2(r) \right) r_{n-1} \right) x^{n-k-2} 
	\end{equation}
	Now expand the tilde functions in a taylor series around the outer horizon radius. After an application of the Cauchy product formula, one finds
	\begin{equation}
	\sum_{n=0} \left[ (n-\kappa)(n-\kappa-1) r_n + \sum_{j=0}^{n} \bigg( p_{1,j} (n-j-\kappa) r_{n-j} + q_{1,j}r_{n-j} + p_{2,j} (n-j-1-\kappa) r_{n-j-1} + q_{2,j}r_{n-1-j} \bigg) \right]x^{n-\kappa-2}
	\end{equation}
	where $p_{1/2,j}$ and $q_{1/2,j}$ denote the j'th coefficient in the taylor series expansion of the four tilde functions. Vanishing of the coefficients enforces
	\begin{equation}
	r_n = -\frac{1}{(n-\kappa)(n-\kappa-1)}\sum_{j=0}^{n} \bigg( p_{1,j} (n-j-\kappa) r_{n-j} + q_{1,j}r_{n-j} + p_{2,j} (n-j-1-\kappa) r_{n-j-1} + q_{2,j}r_{n-1-j} \bigg)
	\end{equation}

	Solving for $r_n$, we find the recursion relation for the Frobenius coefficients
	\begin{equation} \label{eq:recursionrelation}
		r_n = \frac{-1}{n(n-2*\kappa)} \bigg[ (p_{2,0}(n-\kappa-1) + q_{2,0}) r_{n-1} + \sum_{j=1} \left((p_{1,j}(n-j-\kappa) + q_{1,j})r_{n-j} + (p_{2,j}(n-j-1-\kappa) + q_{2,j})r_{n-1-j} \right) \bigg]
	\end{equation}
	Taking $r_0=1$ and calculating the taylor series expansion of the four tilde functions, together with the value for $\kappa$ derived from the indicial equation, we have completed our asymptotic expansion of the radial function. By Fuchs theorem, we have a fundamental set of solutions given by Eq. \eqref{eq:frobeniusseries}, together with Eq. \eqref{eq:indicialroots} and Eq. \eqref{eq:recursionrelation}. This solution will be used to calculate the boundary condition at the outer horizon as required for numerically solving Eq. \eqref{eq:recastradial}.
	
	Now the boundary conditions at infinity can be enforced by minimizing the quantity $\ln{R(r_{max})}^2$ over the complex-frequency space for a specified large value of $r_{max}$. This not only enforces the asymptotic boundary conditions at infinity, namely $R(r \rightarrow \infty) \sim e^{-Qr}$, but also will yield the complex-frequency parameter $\omega$. 
	
	With these two boundary conditions enforced, we can numerically solve the radial equation by integrating from $r=r_+ + \epsilon$ to $r_{max}$, with initial conditions determined by the Frobenius expansion at the outer horizon and the asymptotic boundary condition enforced by the minimization of $\ln{R(r_{max})}^2$, which yields the complex-frequency $\omega$. 
	
	After solving the radial and angular equation for the complex eigenvalues, we've determined the Proca field in terms of the parameters $(m,a,M,n,S, \mu)$, where $m$ is the total angular momentum projection, $a$ is the dimensionless spin of the BH, $M$ is the mass of the BH, $n$ is the overtone number specifying the number of zero crossings of the radial function and comes from imposing boundary conditions on a schrodinger-like equation in the non-relativistic regime, $S$ is the spin of the Proca field, taking values $S=-1,0,+1$, and $\mu = \frac{m_{A}}{\hbar}$ is the mass parameter of the Proca field.

    \section{Secular Evolution of Proca Cloud} \label{app:proccloud}
    Here, we display various figures for certain orbital and system parameters. These aid in visualizing the relation between a dressed EMRI system and the vacuum case. The variation of the mass of the cloud over time can be determined from $\frac{ d E_{c}}{dt} \propto E_{c}^2$, which follows from the Teukolsky formalism for the Proca field on a Kerr background. In particular, our choice of normalization is the statement of energy conservation, i.e. $M_{c} = M_{0,bh} - M_{f,bh}$, where $M_{c}$ is the mass of the cloud at saturation, $M_{0,bh}$ and $M_{f,bh}$ are the initial and final masses of the black hole at saturation, respectively. Using \Cref{eq:energyintegral}, and the fact $\mathfrak{T}_{\mu \nu} \sim (A^{\mu})^2$, it then follows that the normalization coefficient of the Proca field is 
    \begin{equation}
        C = \sqrt{\frac{M_{0,bh} - M_{f,bh}}{\bar{E}_c}}
    \end{equation}
    where $\bar{E}_c$ is the unnormalized energy calculated directly from \Cref{eq:energyintegral}. The final mass of the black hole can be determined directly from \Cref{eq:fracchange} and the requirement the saturation condition it met, $\omega = m \Omega_H$.

\begin{figure}
        \centering
        \includegraphics[width=\linewidth]{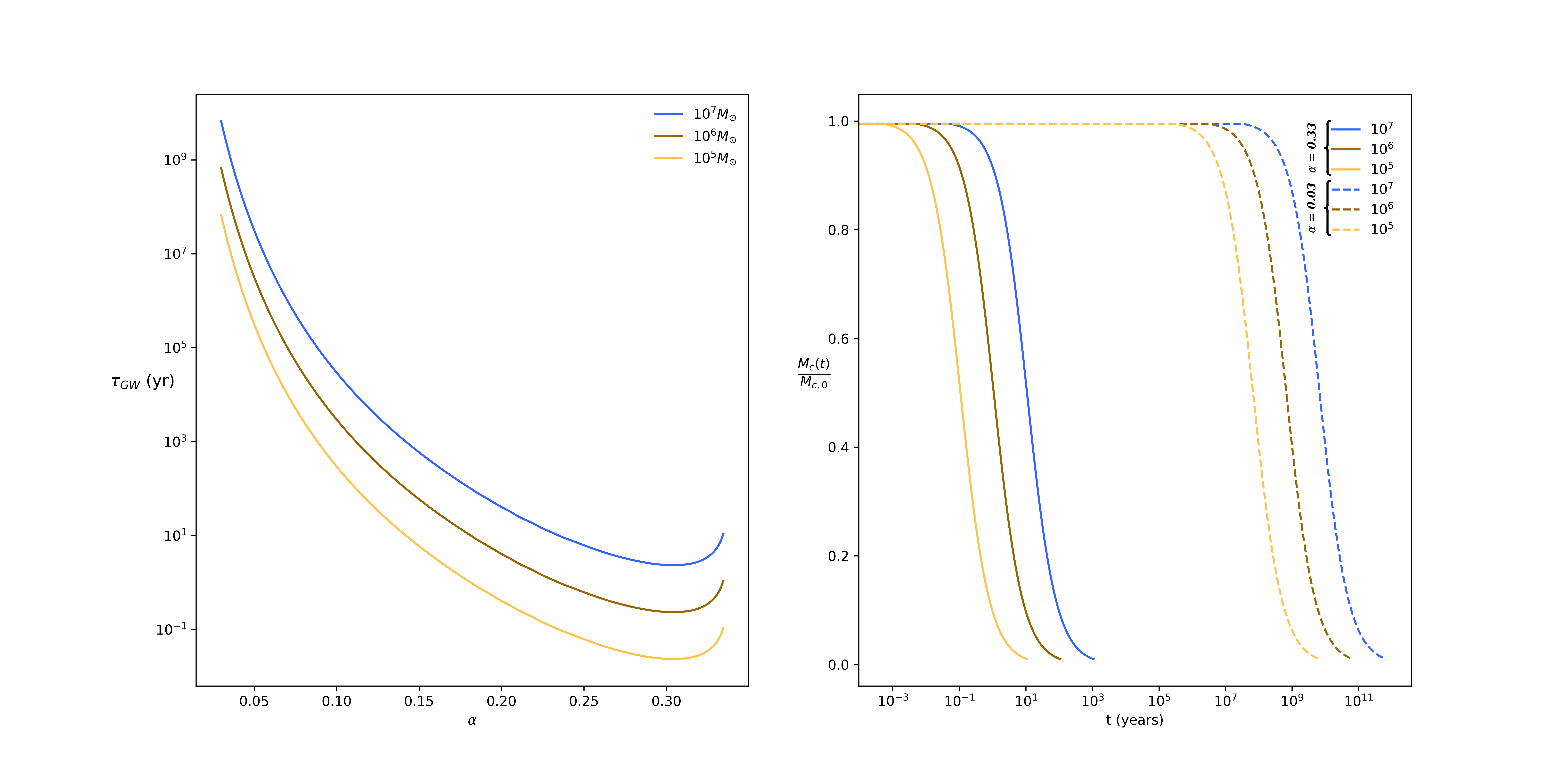}
        \caption{Gravitational emission timescale (left) and relative depletion of the mass of the radiating Proca cloud (right) versus gravitational coupling for various choices of the SMBH mass. The emission timescale tends to shrink for higher couplings, up until the superradiant condition is violated. It's also apparent that lower gravitational couplings permit longer lived clouds.}
        \label{fig:gravem}
    \end{figure}

\begin{figure}
        \centering
        \includegraphics[width=\linewidth]{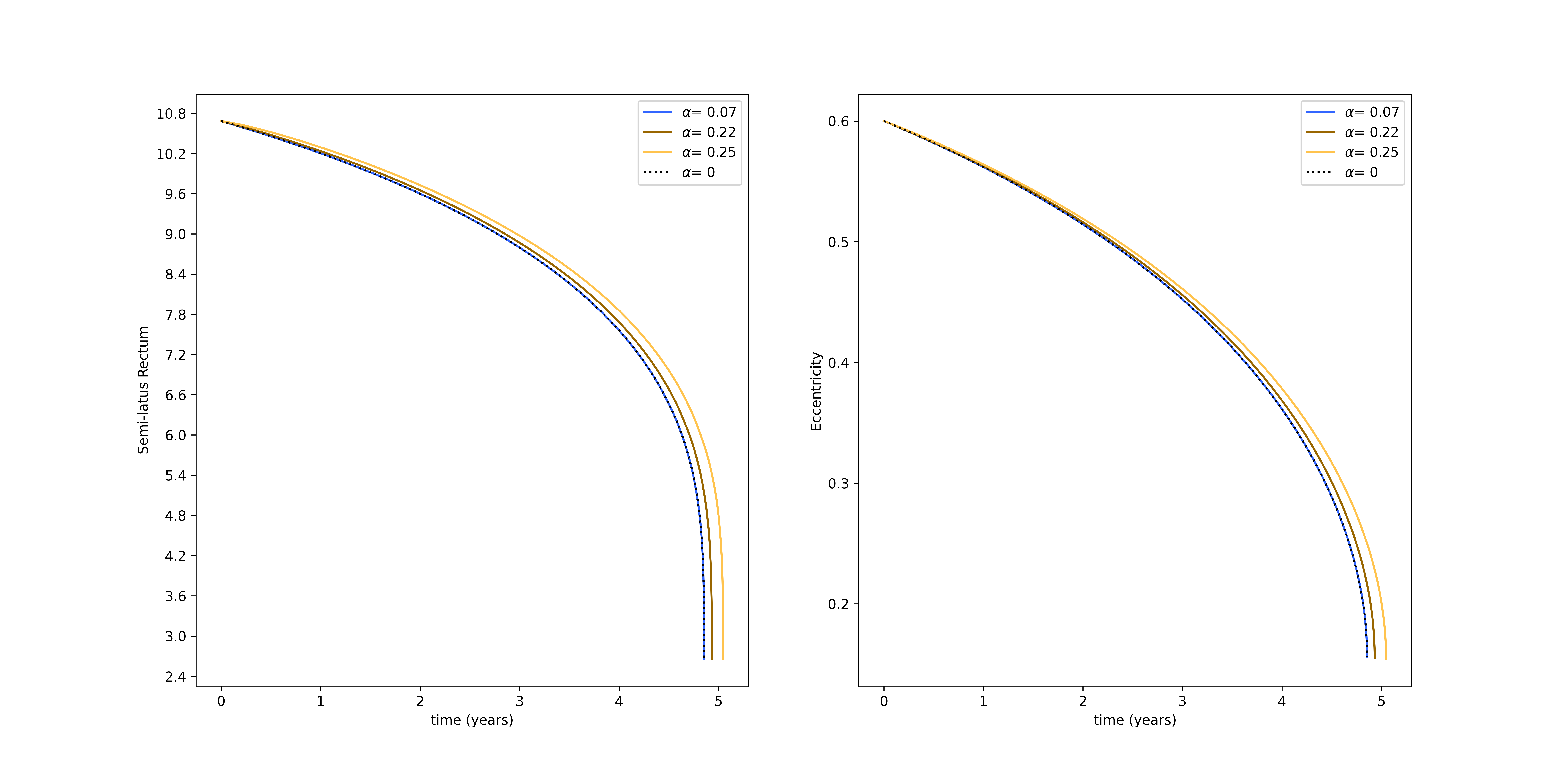}
        \caption{Variation of the semi-latus rectum (left) and eccentricity (right) over the course of the inspiral, for various choices of the gravitational coupling $\alpha$. Here, the SMBH mass is taken to be $M=10^6 M_{\odot}$. The semi-latus rectum is expressed in gravitational units, $\frac{p}{M}$, where $p$ is the semi-latus rectum in geometrized units.}
        \label{fig:orbparams}
    \end{figure}

    The true normalized energy of the cloud is then $E_c = C^2 \bar{E}_c$. This implies the amplitude of the Proca field scales likes $\sim \sqrt{M_{c}}$. From the Teukolsky formalism, the asymptotic energy flux due to perturbations of the Kerr spacetime obey the scaling relations $\dot{E}_{c} \propto \vert Z \vert^2 \propto C^4 = E_c^2$. In other words, $\frac{ d E_{c}}{dt} \propto E_{c}^2$. This relation yields
    \begin{equation}
        M_c(t) = \frac{M_0}{1 + t/\tau}
    \end{equation}
    where $\tau$ is the gravitational emission timescale and is determined from the relation 
    \begin{equation}
        \tau = \frac{M_0}{\frac{d M}{dt}(0)}
    \end{equation}
    The initial time is taken to be the time when the superradiant instability saturates and the cloud reaches a quasibound state. The gravitational emission during the instability is neglected due to the clear separation of timescales. The timescale for depletion via gravitational emission and evolution of the Proca cloud mass is shown in figure \ref{fig:gravem}. The evolution of the orbital parameters for an example system of $M=10^6 M_{\odot}$ and $\chi = 0.9$ due to the secular change in the mass and angular momentum of the background is shown in figure \ref{fig:orbparams}. The strength of the deviation from the vacuum scenario (labeled $\alpha=0$ in the figure) as a function of the gravitational coupling is clearly visible. Stronger values of $\alpha$ correspond to greater deviations from the vacuum scenario, producing waveforms that differ by a larger amount from the undresssed counterpart.
    
    Its should be noted that lower values of $\alpha$ use analytic expressions for the energy flux, while higher values of $\alpha$ use fits to numerical relativity calculations \cite{superrad}. Moreover, for higher values of $\alpha$, $\tau$ diverges near the superradiant threshold as the Proca cloud mass drops off in this region and $\tau \sim \frac{1}{M_0}$.

    \section{Comparison to Other Effects} \label{app:comp}

    As stated in the main text, several affects have been neglected when computing the potential Proca mass range observable with the LISA mission. These include dynamical friction~\cite{Traykova_2021, Vicente_2022, traykova2023relativistic, Hui_2017, Zhang_2020, Baumann_2022, Buehler_2023, cole2022disks, tomaselli2023dynamical}, accretion of the Proca cloud onto the secondary black hole~\cite{Baumann_2022_ion, Unrun1976}, and resonant transitions between Proca states~\cite{Baumann_2022, Baumann_Chia_Porto_Stout_2020, Baumann_Chia_Porto_2019, Zhang_2019}. Within a limited scope, each effect is studied here to understand the role it plays in the potential Proca mass range observable with LISA. 

    First, its been shown that dynamical friction has the same order of magnitude effect as that arising from transitions between states of the cloud. Indeed, the transition from bound to unbound states has been suggested to be interpreted as dynamical friction~\cite{tomaselli2023dynamical} of the Proca cloud on the secondary black hole. However, the full phenomenon of state transitions has either three effects on the orbital trajectory. The orbit either floats, sinks, or is kicked, depending on the initial and final states of the transition. Either of these three effects individually yield a reduction in the faithfulness, with respect to the vacuum case. When including the modification to the background, a sinking orbit may counteract the effect of the modification to the background discussed in the main text. However, a floating orbit has the opposite effect, it enhances the deviation from the vacuum scenario. This can be seen from the expression for the waveform inner product Eq. \eqref{eq:wvinner}. Consider a signal in the detector, whose functional form is $h(f) = A(f) e^{i \phi(f)}$, where $A(f)$ and $\phi(f)$ are the amplitude and phase as a function of frequency, respectively. Let $A_{0}(f)$ and $\phi_{0}(f)$ be the amplitude and phase, respectively, for the vacuum inspiral. Let $\delta \phi_{1}(f)$ denote the deviation from the vacuum scenario due to the modification of the background, as discussed in the main text, and let $\delta \phi_{2}(f)$ denote the additional phase deviation due to the floating or sinking orbits. At lowest order, the amplitude remains unchanged, so the waveform of the perturbed spacetime is
    \begin{equation}
    h_{\text{Proca}}(f) = A_{0}(f) e^{i (\phi_{0}(f) + \delta \phi_{1}(f) + \delta \phi(2)(f)} = h_{\text{vacuum}}(f)e^{i( \delta \phi_{1}(f) + \delta \phi_{2}(f))}
    \end{equation}
    where we've defined $h_{\text{vacuum}}(f) = A_{0}(f) e^{i \phi_{0}(f)}$. The waveform inner product then takes the form
    \begin{equation}
    \langle h_{\text{vacuum}} | h_{\text{Proca}} \rangle = 4*Re\int \frac{h_{\text{vacuum}} h_{\text{Proca}}^* }{S_n} df = 4*Re \int \frac{|A_{0}|^2}{S_n} e^{i( \delta \phi_{1}(f) + \delta \phi_{2}(f))} df
    \end{equation}
    Assuming a small deviation of the phase \footnote{ For state transitions, this is typically a good approximation since the deviation in frequency scales as $q$ for $q\ll 1$ \cite{Baumann_Chia_Porto_Stout_2020}.}, then the inner products, at lowest order, becomes
    \begin{equation}
    \langle h_{\text{vacuum}} | h_{\text{Proca}} \rangle = 4*Re\int \frac{|A_{0}|^2}{S_n} df  - 2*Re  \int \frac{|A_{0}|^2}{S_n} \left(  \delta \phi_{1}(f) + \delta \phi_{2}(f) \right)^2 df 
    \end{equation}
    where we've dropped purely imaginary terms. For floating orbits, $\delta \phi_{1}$ and $\delta \phi_{2}$ have the same sign, so the second term is purely positive, reducing the value of the faithfulness. This implies that the inclusion of state transitions in the cloud will increase the observable mass range of the Proca cloud with LISA. To further elucidate this point, consider for example a $10^7 M_{\odot}$ primary black hole surrounded by a Proca cloud with mass $\mu=9.35*10^{-19} eV$. The results of this study suggest this Proca mass would be unobservable with LISA. However, inclusion of floating reduces the faithfulness even further, potentially pushing it below the critical faithfulness threshold for observability. Over the entire parameter, the observable mass range of the Proca field with LISA is then further extended beyond the range suggested in the main text. This is owed to the fact that floating orbits contribute scenarios where the Proca mass is observable, while sinking orbits, that partially cancel the effect of this study, do not. These estimates were confirmed with explicit numerical computation for an example scenario. It was found that for $\mu=9.35*10^{-19} eV$, $M=10^7 M_{\odot}$, $e0=0.2$, and $\chi=0.9$, inclusion of either a sinking or floating orbit reduced the faithfulness from values above the critical threshold for observability (see figure \ref{fig:faith_Norbits_all}) to values below it. Despite the fact there are regions of the parameter space where the effects can partially cancel, over the entire range of the parameter space, the observable Proca mass range will increase due to those regions which enhance the background modification and thus push the faithfulness statistic below the critical threshold for observability.

    The second additional affect neglected is accretion of the Proca cloud onto the secondary black hole. Its been suggested in~\cite{cao2023signatures} that accretion is strongly $\alpha$-suppressed relative to dynamical friction (and hence state transitions discussed above) and the change in the cloud mass is even further suppressed \footnote{The estimations in~\cite{cao2023signatures} were calculated assuming a scalar cloud, but then suggested to be within the same order of magnitude for the Proca cloud as well.}. Hence, its reasonable to neglect the effect of accretion at this order in the perturbative expansion. Nonetheless, accretion acts as an additional force that can either enhance or impair the radial inspiral~\cite{Baumann_2022_ion, Unrun1976}. In a similar argument as above, over the entire parameter space, the region that enhances the modification to the background will widen the observable Proca mass range with LISA by pushing the faithfulness below the critical threshold. Thus, including accretion will also further expand the potentially observable Proca mass range of LISA.

    The inclusion of these two effects are thus shown to further widen the mass range observable with LISA, further suggesting the results of this study are a conservative estimate. It should also be noted that self-gravity is another effect commonly studied in the literature \cite{Ferreira_2017,Hannuksela_2019}. Since this study uses the \textsc{superrad}~\cite{superrad} package to compute the asymptotic energy and angular momentum fluxes from the Proca cloud, which uses numerical relativity-fitted formulas for large $\alpha$-values and a Newtonian treatment for $\alpha \ll 1$, the self-gravity of the cloud is automatically included in this study.

    \section{Data Plots} \label{app:dataplots}
\begin{figure}
		\centering
		\includegraphics[width=0.88\textwidth,trim=0 0 0 4cm]{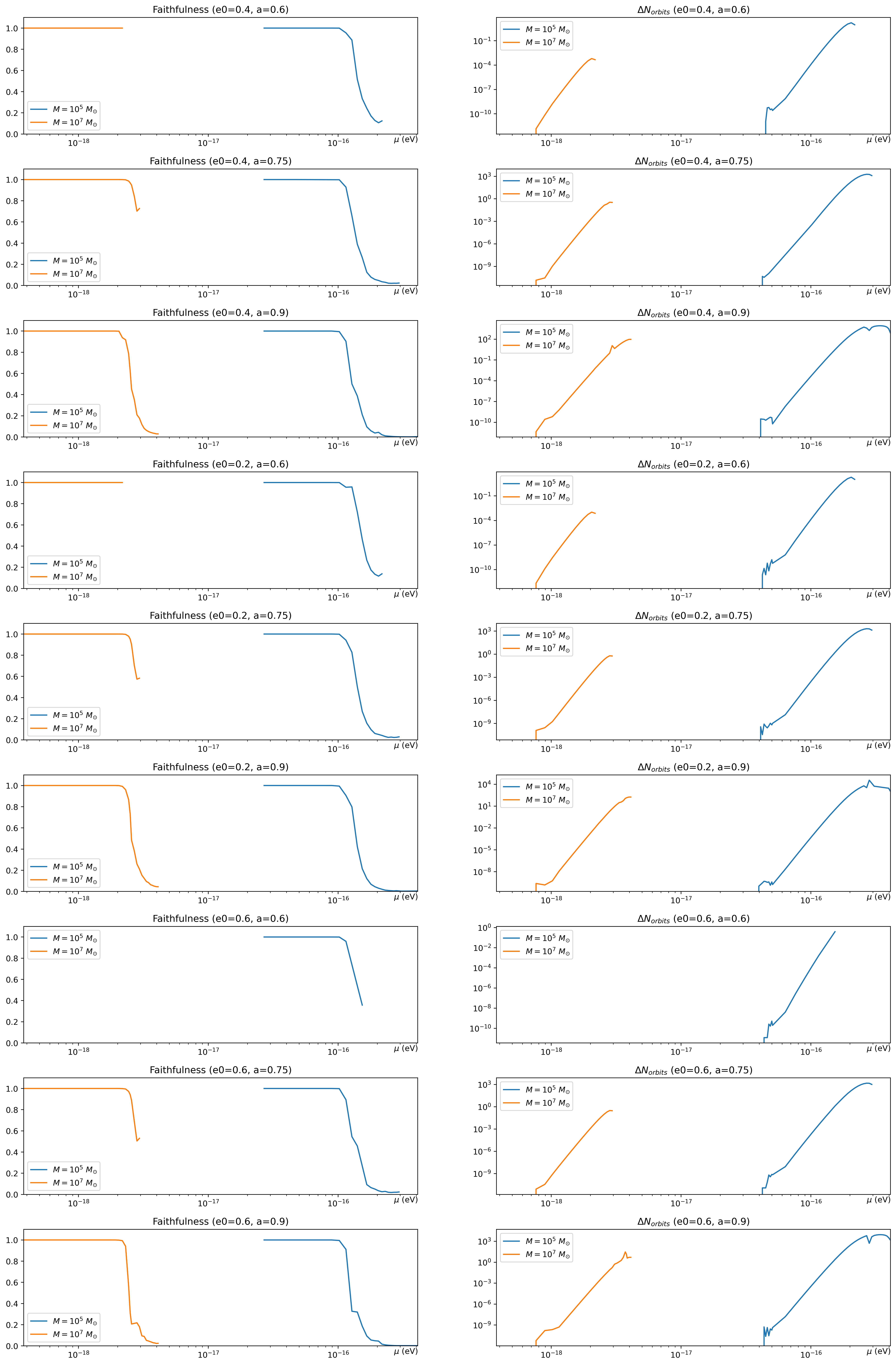}
		\caption{The difference in number of orbits and faithfulness as a function of Proca mass, respectively. The total spacetime dimensionless spin takes values  $\chi=[0.6,0.75,0.9]$ and the initial eccentricity takes values $e0=[0.2,0.4,0.6]$. The difference in number of orbits is the absolute difference between the number of orbits completed by the dressed and undressed waveform at separatrix. The data for $M=10^7 M_{\odot}$, $e0=0.6$, and $\chi=0.6$ is omitted due to numerical instability.}
		\label{fig:faith_Norbits_all}
	\end{figure}

    Displayed here (see figure \ref{fig:faith_Norbits_all}) is the generated data from this study with parameters $e0=[0.2,0.4,0.6]$, $\chi = [0.6,0.75,0.9]$, and $M = [10^5, 10^6, 10^7]$. It is observed that eccentricity plays little role in the approximations of this study and lower black hole spin is less able to constrain the Proca mass.

\end{document}